\documentclass[a4paper, 12pt]{article}
\pdfoutput=1

\usepackage{latexsym,amsmath,amsfonts,amssymb}
\usepackage{tikz}
\usetikzlibrary{decorations.pathmorphing,cd,decorations.markings,calc}
\usepackage{mathrsfs}
\usepackage[american]{babel}
\usepackage{graphicx}
\usepackage{bbm}
\usepackage{cite}
\usepackage{tcolorbox}
\usepackage{cancel}
\usepackage{appendix}

\usepackage[colorlinks=true, citecolor=blue!90!black, linkcolor=blue!90!black, linktocpage=true, urlcolor=red!70!black]{hyperref}

\renewcommand{\baselinestretch}{1.2}
\newcommand{\slashed}{{\bf\not}}

\newcommand{\matht}[1]{{\ensuremath{\boldsymbol{#1}}}}

\numberwithin{equation}{section}

\newcommand{\mat}[1]{\begin{pmatrix} #1 \end{pmatrix}}

\newcommand{\be}{\begin{equation}} \newcommand{\ee}{\end{equation}}
\newcommand{\bea}{\begin{equation} \begin{aligned}} \newcommand{\eea}{\end{aligned} \end{equation}}

\newcommand{\tps}[2]{\texorpdfstring{#1}{#2}}

\newcommand{\cA}{\mathcal{A}}

\newcommand{\cD}{\mathcal{D}}
\newcommand{\cE}{\mathcal{E}}

\newcommand{\cJ}{\mathcal{J}}

\newcommand{\cM}{\mathcal{M}}

\newcommand{\cN}{\mathcal{N}}
\newcommand{\cO}{\mathcal{O}}

\newcommand{\cT}{\mathcal{T}}

\newcommand{\cW}{\mathcal{W}}

\newcommand{\bZ}{\mathbb{Z}}

\newcommand{\unit}{\mathbbm{1}}

\newcommand{\ed}{\;.}
\newcommand{\ec}{\;,}
\newcommand{\one}{^{(1)}}
\newcommand{\two}{^{(2)}}

\def\Spin{\mathrm{Spin}}

\def\repa{\raise4pt\hbox{$\square$}\mkern-14mu\raise-4pt\hbox{$\square$}}
\def\repab{\overline{\raise4pt\hbox{$\square$}\mkern-14mu\raise-4pt\hbox{$\square$}\mkern-1mu}}

\DeclareMathOperator{\Tr}{Tr}

\newcommand{\YM}{\textrm{YM}}
\newcommand{\0}{\mathbf{0}}
\newcommand{\vv}{\mathbf{v}}
\newcommand{\s}{\mathbf{s}}
\newcommand{\cc}{\mathbf{c}}

\begin{document}
\thispagestyle{empty}
	\fontsize{12pt}{20pt}	
	\vspace{13mm}  
	\begin{center}
		{\huge  Symmetries, Universes and Phases of QCD$_2$ \\[5pt] with an Adjoint Dirac Fermion  
  }
		\\[13mm]
		{\large Jeremias Aguilera Damia, \, Giovanni Galati,\, and Luigi Tizzano }	
				\bigskip
				
				{\it
						 Physique Th\'eorique et Math\'ematique and International Solvay Institutes\\
Universit\'e Libre de Bruxelles, C.P. 231, 1050 Brussels, Belgium  \\[.2em]
						
					}
		
	\end{center}

\begin{abstract}

We study 2d $SU(N)$ QCD with an adjoint Dirac fermion. Assuming that the IR limit of the massless theory is captured by a WZW coset CFT, we show that this CFT can be decomposed into a sum of distinct CFTs, each representing a superselection sector (universe) of the gauge theory corresponding to different flux tube sectors. The CFTs describing each universe are related by non-invertible topological lines that exhibit a mixed anomaly with the $\mathbb{Z}^{(1)}_N$ 1-form symmetry. These symmetries exists along the entire RG flow thereby implying deconfinement of the massless theory. We begin by outlining the general features of the model for arbitrary $N$ and then provide a detailed analysis for $N=2$ and $N=3$. In these specific cases, we explicitly determine the IR partition function, identify the symmetries, and explore relevant deformations. Based on these findings and in alignment with various previous studies, we propose a phase diagram for the massive $SU(2)$ gauge theory and calculate its confining string tension. 
\end{abstract}

\newpage
\pagenumbering{arabic}
\setcounter{page}{1}
\setcounter{footnote}{0}
\renewcommand{\thefootnote}{\arabic{footnote}}

{\renewcommand{\baselinestretch}{.88} \parskip=0pt
	\setcounter{tocdepth}{2}
	\tableofcontents}


\section{Introduction}

The study of confinement in gauge theories is a problem of fundamental importance that remains only partially understood. Despite significant progress made through numerical lattice simulations, no analytical tools currently exist that provide a complete understanding.

A sharp criterion for confinement in quantum field theory is based on the presence of an unbroken 1-form symmetry \cite{Aharony:2013hda, Gaiotto:2014kfa}. This is equivalent to the condition that Wilson loops, in a suitable representation, follow an area law and that the quark-antiquark potential exhibits a linear behavior at large separations. Conversely, if the one-form symmetry is spontaneously broken, the system is deconfined (or color screened), Wilson loops follow a perimeter law, and external quarks are screened. 

Beyond symmetry considerations, conventional wisdom suggests that in the presence of confinement, the system develops a mass gap, the vacuum becomes featureless, and the only remaining particles are massive bound states. Indeed, many real-world examples of confining gauge theories display both confinement and a mass gap \emph{at the same time}. This perspective is supported by lattice simulations \cite{Wilson:1974sk} and studies of supersymmetric theories \cite{Seiberg:1994rs}.

However, it is important to emphasize that, although confinement and the formation of a mass gap are typically observed together, they are two distinct phenomena that do not necessarily occur simultaneously. The goal of this work is to analyze specific examples of gauge theories where these two phenomena can manifest independently. We will focus on gauge theories in two spacetime dimensions, where an important feature is that gluons are non-dynamical. These theories are known to possess a rich dynamics and have attracted significant interest since the original work \cite{tHooft:1974pnl}. In particular, we will analyze QCD$_2$ based on the gauge group $SU(N)$ coupled to $N_f=2$ Majorana fermions, equivalently a single Dirac fermion, in the adjoint representation, for low values of $N$. 

Two-dimensional QCD with gauge group $SU(N)$ and two adjoint Majorana fermions belongs to a larger class of theories with $SU(N)$ gauge group and $N_f$ massless adjoint Majorana fermions
that can be broadly characterized as follows:
\begin{itemize}
    \item The case $N_f=0$ corresponds to pure $SU(N)$ Yang-Mills theory in two dimensions (YM$_2$). This theory has no propagating degrees of freedom and no free parameters since $g_{\YM}$ has dimensions of mass; it is solvable and invariant under area-preserving diffeomorphisms. It is also known to exhibit confinement, so that all Wilson loops obey an area law for any value of $N$ \cite{Migdal:1975zg}. This theory was also extensively studied in the large $N$ limit in \cite{Gross:1993hu, Gross:1993yt} due to its proposed connection to string worldsheet theory. For a recent exploration of this connection, see \cite{Aharony:2023tam}.
    
    \item The case $N_f=1$ corresponds to $SU(N)$ YM$_2$ coupled to a single adjoint Majorana fermion, a theory that has received a lot of attention over the years. It was first studied in the large-$N$ limit using discretized light-cone quantization (DLCQ) in \cite{Dalley:1992yy, Bhanot:1993xp} or using light-cone Hamiltonian truncation \cite{Katz:2013qua}. The massless theory is known to develop a mass gap \emph{and} exhibit screening behavior at long distance \cite{Gross:1995bp, Cherman:2019hbq, Komargodski:2020mxz, Dempsey:2021xpf, Dempsey:2022uie}. By applying bosonization techniques \cite{Witten:1983ar} and considering the deep infrared limit ($g_\YM \to \infty$), which is naively associated to removing the YM$_2$ kinetic term, one expects to find a coset CFT at long distances:
    \be\label{cosetNF1}
    \frac{\textrm{Spin}(N^2-1)_1}{SU(N)_N}\ec
    \ee
    which has vanishing central charge and should be understood as a non-trivial TQFT. A priori, it is not immediately clear that the infrared limit of $N_f = 1$ adjoint QCD$_2$ is equivalent to the description obtained by removing the $SU(N)$ kinetic term. Nonetheless, the authors of \cite{Komargodski:2020mxz} have demonstrated that, irrespective of this assumption, $N_f=1$ adjoint QCD$_2$ has a large set of \emph{non-invertible} symmetries associated with the chiral algebra underlying \eqref{cosetNF1}. These lead to deconfinement of the fundamental Wilson loop. Perturbing the massless theory by a small mass $m_{\psi} >0$ for the adjoint Majorana fermion breaks all these non-invertible symmetries, and the system becomes gapped and confined.
    \item The case $N_f = 2$ corresponds to $SU(N)$ YM$_2$ coupled to $N_f = 2$ adjoint Majorana fermions, a theory whose candidate IR dynamics differ significantly from the previous examples. The initial study of this model appeared in \cite{Gopakumar:2012gd}. See also \cite{Isachenkov:2014zua} for subsequent developments. Adopting ideas similar to those discussed around \eqref{cosetNF1}, one expects that the long-distance behavior of this system is captured by the following coset:
    \be\label{cosetNF2}
    \frac{\textrm{Spin}(2(N^2-1))_1}{SU(N)_{2N}}\ec
    \ee
    which has a central charge $c = (N^2 - 1)/3$ and is therefore a non-trivial CFT whose detailed properties will be thoroughly explored in Section \ref{sec: coset generalities}. Furthermore, we will systematically analyze the candidate IR description of this theory for $N = 2, 3$. Using techniques similar to those in \cite{Komargodski:2020mxz}, we will demonstrate that adjoint QCD$_2$ with $N_f = 2$ has a large set of non-invertible symmetries associated with the chiral algebra underlying \eqref{cosetNF2}, which again establishes the deconfinement of the fundamental Wilson loop. The massless theory can be perturbed by two independent mass operators, whose coefficients we denote by $m_1$ and $m_2$. For generic positive values $m_{1,2} > 0$, all non-invertible symmetries are broken, and the theory becomes confined; for large $m_{1,2}$, it flows to pure YM$_2$. When a Majorana fermion remains massless, a subset of these non-invertible symmetries is preserved, allowing us to reproduce the gapped and deconfined IR phase discussed above for $N_f = 1$ adjoint QCD. Thanks to our symmetry-based perspective, we are also able to revisit various aspects that were missing in the study \cite{Gopakumar:2012gd}. In particular, we discuss the consequences of the 1-form symmetry, clarify the decoupling of relevant operators at long distances, and explain why certain potentially dangerous quartic deformations are not generated along the RG flow.

    \item To the best of our knowledge, the study of $SU(N)$ YM$_2$ coupled to $N_f \geq 3$ adjoint Majorana fermions is poorly understood. Building on the intuition gathered from the previous cases, we expect that the long-distance behavior of the model is captured by the coset:
    \be
        \frac{\textrm{Spin}(N_f(N^2-1))_1}{SU(N)_{N_fN}}\ec
        \ee
    with a central charge of $c = \frac{N_f(N_f-1)(N^2-1)}{2(N_f+1)}$, thus describing a non-trivial CFT. We believe that this class of theories should exhibit a large set of associated non-invertible symmetries and a rich massive phase structure. In this work, we will not further analyze these models, but we hope that the methods presented here will inspire others to undertake this study.
   \end{itemize}

The physical intuition behind confinement involves the presence of finite-tension electric strings, which lead to a linear quark-antiquark potential at large distances. Therefore, a key quantity to examine in a gapped and confining phase is the effective tension of these confining strings. This analysis can be approached in mass-deformed adjoint QCD$_2$, where these strings, referred to as $k$-strings, have an effective tension that offers insights into the behavior of a Wilson loop in a representation whose Young tableau has $k$ boxes. In the $N_f = 1$ theory, an exact formula for the string tension has been proposed in \cite{Komargodski:2020mxz}. In this work, we have taken initial steps toward addressing this question by providing an explicit calculation of the only existing confining string tension for $N=2$. However, the complexity of the IR dynamics for $N > 3$ currently prevents us from identifying a universal formula for the effective tension that applies to all values of $N$ and $k$. 

Finally, we would like to mention that several new approaches to studying adjoint QCD have recently been discussed.\footnote{See also \cite{Ambrosino:2023dik,Ambrosino:2024prz} for recent studies on the connection between other examples of 2d gauge theories and integrability.} The work \cite{Dempsey:2023fvm} has proposed a lattice Hamiltonian for $SU(N)$ adjoint QCD$_2$ with $N_f = 1$, see also the lattice analysis in \cite{Bergner:2024ttq}, while \cite{Dempsey:2024ofo} has discussed the same theory on a finite-size circle of length $L$ with periodic boundary conditions.\footnote{The finite-temperature analysis of this theory was conducted long ago in \cite{Kutasov:1993gq}, where it was argued that a Hagedorn transition is expected to occur when the adjoint fermions are massive, leading to a deconfined phase for circle radii smaller than a critical length, $L_{\textrm{crit}}$.} It would be interesting to extend both of these methods to the study of $N_f = 2$ adjoint QCD$_2$ as discussed in this work, to provide further evidence for the candidate IR dynamics we have presented. In particular, these complementary methods may be better suited for larger values of $N$.

The paper is organized as follows. In Section \ref{sec: generalities}, we explore the general features of $SU(N)$ QCD$_2$ with two Majorana fermions (which we denote by $\cT_{SU(N)}$), focusing on the symmetries of the model, and its decomposition into distinct universes due to the presence of a $\bZ\one_N$ 1-form symmetry. We then analyze the general properties of the IR coset \eqref{cosetNF2}, providing a general expression for the partition function as a sum of distinct CFTs. In Section \ref{sec: CS general}, we discuss the Chern-Simons representation of the IR model and show how topological Verlinde lines can be extracted from this framework. Section \ref{sec: N=2} is dedicated to a detailed analysis of $\cT_{SU(2)}$, where we identify the two IR CFTs corresponding to each universe of the theory and examine the generalized (non-invertible) symmetries. Moreover, in Section \ref{sec: mass def}, we investigate massive deformations of the theory, proposing a candidate phase diagram as a function of $m_1$ and $m_2$ and demonstrating how various 0-form and 1-form symmetries act on the vacuum. Section \ref{sec:tension} is devoted to the explicit computation of the string tension for the confining string in $\cT_{SU(2)}$. In Section \ref{sec: N=3}, we turn to $\cT_{SU(3)}$, where we determine its explicit IR partition function and identify three CFTs corresponding to each universe. These are specific modular invariants of the $\cN=(2,2)$ super-minimal models with $c=8/3$. Finally, we describe the generalized symmetries from the Chern-Simons perspective and explore mass and quartic deformations of the theory. Additional detailed computations are provided in several appendices.

\section{Generalities on QCD\tps{\matht{_2}}{2} with an Adjoint Dirac Fermion}\label{sec: generalities}
In this work, we consider a two-dimensional $SU(N)$ Yang-Mills theory coupled to a single massless Dirac fermion in the adjoint representation, which we will often denote by  $\mathcal{T}_{SU(N)}$. In Euclidean signature its action is:
\be\label{eq: Adj QCD action}
S_{\cT_{SU(N)}}=\int d^2x \left(-\frac{1}{4g^2_{\YM}} \Tr (F^2) +i\sum^{2}_{j=1}\Tr\psi_j^{T}\slashed{D}\psi_j\right)\ec
\ee 
where $\psi_{1,2}$ are two (non-chiral) Majorana fermions in the adjoint representation of the $\mathfrak{su}(N)$ Lie algebra. 

In 2d, the gauge coupling is classically relevant and gluons are non-propagating, so the UV limit of the theory \eqref{eq: Adj QCD action} consists of $2(N^2-1)$ free Majorana fermions. Following \cite{Komargodski:2020mxz}, we now review some useful properties of the duality between an \emph{even} number $n$ of free Majorana fermions with gauged $(-1)^{\textrm{F}}$ symmetry 
and the $\Spin(n)_1$ WZW model \cite{Witten:1983ar}.\footnote{We follow the same convention for the $\Spin(n)_1$ theory as outlined in \cite{Ji:2019ugf}.} The global symmetry group of $n$ free Majorana fermions is 
\be
G_f = O(n)_L \times O(n)_R = (SO(n)_L \times SO(n)_R) \rtimes (\bZ^C_2 \times \bZ^{C_L}_2)\ec
\ee
where $\bZ^C_2$ is the diagonal charge conjugation while $\bZ_2^{C_L}$ acts as charge conjugation only on left-moving fermions. The fermion parity symmetry $\bZ^{\textrm{F}}_2$ that we would like to gauge is the diagonal subgroup of $\bZ^{F_L}_2 \times \bZ^{F_R}_2 = Z(SO(n)_L\times SO(n)_R)$.\footnote{Note that, the gauging procedure of $\bZ^{\textrm{F}}_2$ is not unique as one always has the freedom to stack a given fermionic theory with an invertible spin-TQFT in 1+1d whose action is the Arf invariant. See Appendix \ref{app: bosonization} for additional details.} Therefore, after bosonization the global symmetry group of the bosonic model becomes
\be
G_b = \frac{\Spin(n)_L\times \Spin(n)_R}{Z(\Spin(n)_{\textrm{diag}})} \rtimes \bZ^C_2\ed
\ee
The $\bZ^\vee_2$ symmetry dual to $\bZ^{\textrm{F}}_2$ is now a $\bZ_2$ subgroup of $Z(G_b)$ and gives rise to a non-trivial extension of $SO(n)_L\times SO(n)_R$.

To make contact with adjoint QCD, we consider $n=2(N^2-1)$ free Majorana fermions and gauge a ${SU(N) \subset \Spin(2(N^2-1))}$ subgroup on both sides of the duality. The global symmetry group of the gauged theory is the subgroup of $G_b$ that acts non-trivially on gauge-invariant operators and commutes with the gauged symmetry. Its continuous component is $ U(1)_V \times U(1)_A$,
acting naturally on the adjoint Dirac fermion.

Under the assumption that the correct IR limit of $\cT_{SU(N)}$ is reached by formally sending $g^2_{\rm{YM}}\to \infty$, the resulting low-energy effective description is the gauged WZW model for the coset \cite{Kutasov:1994xq,Gopakumar:2012gd,Isachenkov:2014zua}:
\be\label{IRcoset}
\cT_{SU(N),\rm{IR}} = \frac{\Spin(2(N^2-1))_1}{SU(N)_{2N}}\ed
\ee
The above theory describes what is commonly known as Goddard-Kent-Olive (GKO) coset model \cite{Goddard:1984vk}, we will review some relevant aspects of this construction in Appendix \ref{app: coset}. In this study, we will extensively use the coset description \eqref{IRcoset} along with its corresponding chiral algebra.\footnote{The property that adjoint QCD \eqref{eq: Adj QCD action} has an associated $\Spin(2(N^2-1))/SU(N)_{2N}$ chiral algebra does not depend on the value of the gauge coupling $g_{\rm{YM}}$. This can be established trough an Hamiltonian analysis as in \cite{Kutasov:1994xq,Delmastro:2021otj} or by considering the CS/WZW correspondence with suitable boundary conditions. The latter approach will be further analyzed in section \ref{sec: CS general}.} This approach will be instrumental in revealing new generalized (categorical) symmetries within the theory \eqref{eq: Adj QCD action}. Additionally, it will allow us to study the IR mapping for several UV relevant deformations that will be discussed in sections \ref{sec: N=2} and \ref{sec: N=3}.

\subsection{1-form Symmetry, Universes and Decomposition}  \label{subsec: universes}

To fully specify a QCD theory based on the Lie algebra $\mathfrak{g}=\mathfrak{su}(N)$ it is necessary to choose the global form of the gauge group. In the previous section we choose the global form $SU(N)$, therefore $\cT_{SU(N)}$ has a one-form global symmetry $\bZ\one_N$ \cite{Aharony:2013hda,Gaiotto:2014kfa}. 

A $2d$ QFT with a one-form symmetry $\Gamma\one$ is characterized by the existence of topological sectors labeled by a representation $\rho \in (\Gamma\one)^{\vee}$, where $(\Gamma\one)^{\vee}$ is the Pontryagin dual group to $\Gamma\one$. Such topological sectors labeled by $\rho$ describe the theory in presence of a flux tube, a well defined Lorentz invariant configuration in $2d$. By its very definition, a one-form symmetry $\Gamma\one$ implies the existence of a topological local operator $U_g$, with $g\in \Gamma\one$ acting on line operators $L$ as follows:
\be\label{eq: 1form action}
U_gL\,U^{-1}_g = \chi_\rho(g)L\ec
\ee
where $\rho\in (\Gamma\one)^{\vee}$ is an irreducible representation of $\Gamma\one$ and $\chi_g(\rho)$ denotes its associated character. Therefore, the spectrum of line operators in a $2d$ QFT with one-form symmetry can be organized in terms of their charges under $\Gamma\one$. Moreover, the action of $U_g$ on line operators induces the following \emph{decomposition} of the Hilbert space:
\be\label{decomposed}
\mathcal{H}= \bigoplus_{\rho \in (\Gamma\one)^\vee}\mathcal{H}_{\rho}\ec
\ee
where $|\psi\rangle \in \mathcal{H}_{\rho}$ if $U_g|\psi\rangle = \chi_\rho(g)|\psi\rangle$. To summarize, the effect of a one-form symmetry in two dimensions is to break up the theory into distinct superselection sectors commonly known as \emph{universes} \cite{Hellerman:2006zs, Aminov:2019hwg, Sharpe:2019ddn, Tanizaki:2019rbk, Komargodski:2020mxz,Nguyen:2021naa} (see \cite{Sharpe:2022ene} for a recent review). The full Hilbert space of the theory is thus decomposed into the direct sum of Hilbert spaces of different universes. Moreover, the topological nature of $U_g$ precludes the existence of finite energy field configurations interpolating between states of different charge, remarkably even at finite volume. This should be contrasted with the typical scenario that holds for spontaneously broken discrete zero-form symmetries, where such field configurations are generically present at finite volume.

The finite volume partition function of adjoint QCD is indeed subject to decomposition. To see this, let us activate a background two-form gauge field $B\two \in H^2(X, \bZ_N)$ for the one-form symmetry $\bZ_N\one$ and impose $\int_X B\two = k \mod N$. Inserting the $B\two$-flux is equivalent to inserting the topological local operator $U_g$, therefore the structure \eqref{decomposed} implies that we obtain a sum of partition functions for each universe weighted with different phases:
\be
Z_{\mathrm{QCD}} = \sum^{N-1}_{p=0} e^{2\pi i p k/N} Z_p\ec
\ee
where $Z_p$ is the partiton function of the universe $p$. The projection into a given universe is simply achieved by gauging the 1-form symmetry $\bZ\one_N$ which in the above sum would be obtained by summing over $k$. However, there are multiple ways to gauge a $1$-form symmetry parametrized by the discrete torsion $\exp{\left(-\frac{2\pi p_0}{N}\int B\two \right)}$, where $p_0$ is an integer in $\bZ_N$. The resulting partition function is
\be
Z_{\mathrm{QCD}/\bZ_N\one} = \frac{1}{N}\sum^{N-1}_{p,k =0} e^{2\pi i (p-p_0)k/N}Z_p\ec
\ee
and thus performing the sum over $k$ projects the theory into a single universe $Z_{p_0}$.

An important set of observables in the $SU(N)$ theory are (non-topological) Wilson lines in a given representation of the gauge group. These are charged under the 1-form symmetry $\bZ_N\one$. Due to \eqref{eq: 1form action}, the eigenvalue of the topological operator $U_g$ must jump when it crosses such line, thereby serving as a static domain wall separating distinct universes. Moreover, 0-form global symmetries in 2d are generated by topological line operators which can also be charged under the 1-form symmetry. Their charge signals a mixed 't Hooft anomaly between the 0-form and 1-form symmetry \cite{Komargodski:2020mxz,Cherman:2019hbq}. Therefore, 0-form symmetry lines with non-trivial 1-form symmetry charge connect different universes. In this case, the structure of distinct universes is determined solely by the physics of a single one.\footnote{Note that when theories are projected into fixed universes by gauging the 1-form symmetry, all charged lines are projected out, consistent with the absence of further decomposition.}

\subsection{Properties of the IR Coset}\label{sec: coset generalities}
As anticipated, the candidate IR description for $\cT_{SU(N)}$ is the coset CFT
\be\label{eq: so coset}
\cT_{SU(N),\textrm{IR}} = \frac{\Spin(2(N^2-1))_1}{SU(N)_{2N}}\ec
\ee
which has central charge
\be\label{eq: SO/SU central charge}
c_{\cT_\textrm{SU(N),IR}} = \frac{N^2-1}{3}\ed
\ee

Furthermore, the above coset CFT is expected to possess distinct universes whose structure should reproduce the discussion of Section \ref{subsec: universes}. To deepen our understanding of the physics within each universe, it is necessary to calculate the torus partition function of \eqref{eq: so coset}. We will now examine a well-established systematic approach to this calculation (see also Appendix \ref{app: coset} for an extensive review).

Let us first list the spectrum of highest weight representations $\Lambda$ and associated dimensions $h_{\Lambda}$ for the chiral algebra $\mathfrak{so}(n)_1$ with \emph{even} $n$:
\be
\Lambda=\{\mathbf{0},\mathbf{v},\s,\cc\}\ec\qquad h_{\Lambda}=\left\{0,\frac{1}{2},\frac{n}{16},\frac{n}{16}\right\}\ed
\ee
where $\0, \vv,\s,\cc$ denote respectively the identity, vector, spinor and co-spinor representations. The modular invariant partition function for the $\Spin(n)_1$ CFT appearing in the numerator of \eqref{eq: so coset} is
\be\label{eq: Z SO(2n)}
Z_{\Spin(n)_1}=|\chi_{\0}|^2+|\chi_{\vv}|^2+|\chi_{\s}|^2+|\chi_{\cc}|^2=\frac12\left(|\chi_{\0+\vv}|^2+|\chi_{\0-\vv}|^2+|\chi_{\s+\cc}|^2+|\chi_{\s-\cc}|^2\right)\ec
\ee
where $\chi_\Lambda$ are characters associated with the corresponding representation $\Lambda$. In the second equality above, we just expressed the partition function in a different basis as it is going to prove useful below. 

Note that the torus characters of $\mathfrak{so}(n)_1$ are known explicitly by means of the free fermion representation \cite{Isachenkov:2014zua}. An important remark is in order at this point. Because of the $\bZ_2$ outer automorphism of the $\mathfrak{so}(n)$ algebra that exchanges the spinor and the co-spinor representations, these two are actually isomorphic. Accordingly, their characters are identical in absence of fugacities. In particular, this implies that $\chi_{\s-\cc}$ is trivial.\footnote{Explicit forms for this characters can be found for instance in \cite{Isachenkov:2014zua}. In particular, Eq. (3.13) there proves that $\chi_{\s-\cc}=0$ for trivial fugacities.} 

Let us now review all the data required to describe the coset denominator theory. The highest weight representations for the algebra $\mathfrak{su}(N)_{2N}$ will be denoted by $\Lambda'$, these are accounted by arrays of fundamental weights
\be
\Lambda'=[\lambda_1,\ldots ,\lambda_{N-1}]\ec\quad \lambda_i\geq 0\ec\quad \sum^{N-1}_{i=1}\lambda_i\leq 2N \ec
\ee
with associated dimensions
\be\label{eq: SU(N) dimensions}
h_{\Lambda'}=\frac{{\cal C}(\Lambda')}{2(k+h^{\vee})}=\frac{(\Lambda',\Lambda'+2\rho)}{6N}
\ee
where ${\cal C}(\Lambda')$ denotes the quadratic Casimir of the representation $\Lambda'$, $(\Lambda',\Lambda'')=F_{ij}\lambda'_i\lambda''_j$ is the pairing in terms of the Cartan matrix $F_{ij}$ and  $\rho$ is the Weyl vector of $\mathfrak{su}(N)$ which is just $\rho=[1,\ldots,1]$.

In $G_k$ WZW models there are special primary operators known as simple currents \cite{Intriligator:1989zw}. These have the special property that their OPE with other local primaries in the theory has a unique singular term. For the denominator theory $SU(N)_{2N}$, the action of a simple current $\cJ$ is completely captured by the outer automorphism of highest weight representations which is naturally associated to $Z(SU(N))$: 
\be\label{Jaction}
{\cal J}\cdot [\lambda_1,\ldots,\lambda_{N-1}]= [2N-\sum\limits_{i=1}^{N-1} \lambda_i,\lambda_1,\cdots,\lambda_{N-2}]\ed
\ee
Furthermore, it is possible to organize the action of the simple current ${\cal J}$ on the highest weight representations of $\mathfrak{su}(N)_{2N}$ into two disjoint sets, known as the long and short orbits, defined as:
\be\label{orbits}
R_{L}=\left\lbrace \Lambda' \big| ~ {\cal J}^l\cdot \Lambda' \neq \Lambda' \ec \forall l\in {Z}(SU(N)) \right\rbrace\ec
~R_{S}=\left\lbrace \Lambda' \big| ~\exists \; d\,|\,N ~{\mathrm{s.t.}} ~{\cal J}^d\cdot \Lambda' = \Lambda'  \right\rbrace\ed
\ee
The structure of $R_L$ and $R_S$ depends on the particular value of $N$, see \cite{Isachenkov:2014zua}. However, when $N$ is prime, there is a unique short orbit corresponding to the $[2,\cdots,2]$ representation. 

Using both equations \eqref{Jaction} and \eqref{orbits} we can find a modular invariant partition function for the denominator $SU(N)_{2N}$ CFT appearing in $\cT_{SU(N),\rm{IR}}$ \cite{Beltaos:2010ka}. For $N$ a prime number, this invariant reads 
\be
Z_{SU(N)_{2N}}=\sum_{\Lambda'\in R_L}|\chi_{\Lambda'}+\ldots+\chi_{{\cal J}^{N-1}\cdot\Lambda'}|^2+N\sum_{\Lambda'\in R_S}|\chi_{\Lambda'}|^2\,.
\ee

In the full coset CFT $\cT_{SU(N),\rm{IR}}$ there is an elegant way to describe how the chiral algebra $\mathfrak{su}(N)_{2N}$ embeds into $\mathfrak{so}(2(N^2-1))_1$ using the so-called branching functions $b_{\Lambda, \Lambda'}$. These functions allow us to decompose the characters of $\mathfrak{so}(2(N^2-1))_1$ into characters of $\mathfrak{su}(N)_{2N}$:
\be
\chi_\Lambda(q) = \sum_{\Lambda'}b_{\Lambda, \Lambda'}(q)\chi_{\Lambda'}(q)\ec
\ee
where $q=\exp(2\pi i \tau)$ and $\tau$ is the torus modular parameter. With this, we can follow the prescription outlined in Appendix \ref{app: coset} to arrive at the final expression for the torus partion function of $\cT_{SU(N),\rm{IR}}$ which is given by: 
\be\label{eq: SO/SU partition function}
Z_{\cT_{SU(N),\rm{IR}}}=\sum_{\Lambda=\0,\vv,\s,\cc}\left(N\sum_{\Lambda'\in R_L}|b_{\Lambda,\Lambda'}|^2+\sum_{\Lambda'\in R_S}|\sum_{i=1}^{N} b_{\Lambda,\Lambda',i}|^2\right)\ec
\ee
where the split branching functions $b_{\Lambda,\Lambda',i}$ are defined as follows:
\be
b_{\Lambda,\Lambda',i} = \frac{1}{N}(b_{\Lambda,\Lambda'}+\widehat\chi_{\Lambda,\Lambda',i})\ec\quad \sum_{i=1}^N\widehat{\chi}_{\Lambda,\Lambda',i}=0\ed
\ee

As expected from section \ref{sec: generalities}, since $\cT_{SU(N),\rm{IR}}$ has a $\bZ\one_N$ 1-form symmetry, the vacuum exhibits a $N$-fold degeneracy. The topological local operators implementing this symmetry are the simple currents of $SU(N)_{2N}$. The presence of this symmetry will play an important role in the study of the model. To see why, let us further rewrite the partition function as follows. As we mentioned above, when $N$ is prime, there is a unique short orbit corresponding to the $[2,\cdots,2]$ representation. The associated $\widehat{\chi}_{{\Lambda,\Lambda',i}}$ characters are actually known explicitly \cite{Isachenkov:2014zua}, we listed them in Appendix \ref{app: coset}. Crucially, these characters satisfy the following property:
\be
\widehat{\chi}_{{\Lambda,[2,\cdots,2],i}} = \widehat{\chi}_{{\Lambda,[2,\cdots,2],j}}\qquad ,\qquad \forall i,j \not= 1\,.
\ee
Therefore, the coset CFT partition function  can be rewritten as
\bea\label{eq: generic univereses}
&Z_{\cT_{SU(N),\rm{IR}}} =  \sum_{\Lambda=\0,\vv,\s,\cc}\left(N\sum_{\Lambda'\in R_L}|b_{\Lambda,\Lambda'}|^2+|b_{\Lambda,[2,\cdots,2],1}+(N-1)b_{\Lambda,[2,\cdots,2],2}|^2\right) = Z_1 + (N-1)Z_2\ec
\eea
where:
\bea\label{eq: general Z1 and Z2}
&Z_1 = \sum_{\Lambda=\0,\vv,\s,\cc}\left(\sum_{\Lambda'\in R_L}|b_{\Lambda,\Lambda'}|^2+| b_{\Lambda,[2,\cdots,2],1}|^2+(N-1)|b_{\Lambda,[2,\cdots,2],2}|^2\right)\ec \\
&Z_2 = \sum_{\Lambda=\0,\vv,\s,\cc}\left(\sum_{\Lambda'\in R_L}|b_{\Lambda,\Lambda'}|^2+(N-2)|b_{\Lambda,[2,\cdots,2],2}|^2 + (b_{\Lambda,[2,\cdots,2],1}\overline{b_{\Lambda,[2,\cdots,2],2}} + \textrm{h.c.}) \right)\ed
\eea
Our findings show that, for $N$ prime, $N-1$ of the $N$ universes appearing in the coset CFT $\cT_{SU(N),\rm{IR}}$ are the same.\footnote{Notice that $Z_1$ is the diagonal modular invariant usually found in the literature after performing field identification and fixed point resolution \cite{Schellekens:1989uf,Schellekens:1990xy,Fuchs:1995tq}. On the other hand, $Z_2$ is a non-diagonal modular invariant which is generically different from $Z_1$.} Our goal in the following will be to analyze the physics in these universes for small values of $N$. Furthermore, by studying mass deformations, we will show that our results are also consistent with the IR dynamics of $SU(N)$ adjoint QCD$_2$ coupled to a {\it single} Majorana fermion which has been studied in \cite{Komargodski:2020mxz}.

\subsection{3d Chern-Simons Description}\label{sec: CS general}
In order to make further progress with our analysis of the IR coset description for adjoint QCD, we would like to understand its symmetry structure. A very important tool for this task is the correspondence between a WZW model and Chern-Simons (CS) theory on an interval \cite{Witten:1988hf} which is often referred to as the Chern-Simons/WZW correspondence (see e.g. \cite{Moore:1988qv,Elitzur:1989nr,Fuchs:2002cm}). 

\begin{figure}[t]
\centering
\begin{tikzpicture}
	\draw [fill = red!30!white] (0,0) -- (1,1) -- (1,3) -- (0,2) -- cycle;
	\node at (-0.5,1.5) {\small L} ;
	\fill [fill = yellow, opacity = 0.2] (0,0) -- (4,0) -- (4,2) -- (5,3) -- (1,3) -- (0,2) -- cycle;
	\draw [densely dotted, red!70!black] (2.5, 1.98) to (4.5, 1.98);
	\draw [densely dotted, red!70!black] (2.5, 0.98) to (4.5, 0.98);
	\draw [fill = blue!30!white, opacity = 0.8] (4,0) -- (5,1) -- (5,3) -- (4,2) -- cycle;
	\node at (5.5,1.5) {\small R};
	\draw [very thick, red!70!black] (2.8, 1.48) arc [start angle = 0, end angle = 360, y radius = 0.5, x radius = 0.3];	
	\draw [very thick, blue!30!white!60!red] (4.8, 1.48) arc [start angle = 0, end angle = 360, y radius = 0.5, x radius = 0.3];
	\node at (2.6, 2.5) {CS theory};
	
	\begin{scope}[shift={(8, 0.)}]
	\draw [fill = red!30!white] (0,0) -- (1,1) -- (1,3) -- (0,2) -- cycle;
	\node at (-0.5,1.5) {\small L} ;
	\fill [fill = yellow, opacity = 0.2] (0,0) -- (4,0) -- (4,2) -- (5,3) -- (1,3) -- (0,2) -- cycle;
	\draw [very thick, red!70!black] (0.5, 1.5) to [out = 20, in = 160] (2.5, 1.5) to [out = -20, in = -160] (4.5, 1.5);
	\draw [fill = blue!30!white, opacity = 0.8] (4,0) -- (5,1) -- (5,3) -- (4,2) -- cycle;
	\node at (5.5,1.5) {\small R};
	\filldraw [red!70!black] (0.5, 1.5) circle [radius = 0.08];
	\filldraw [red!70!black] (4.5, 1.5) circle [radius = 0.08];
	\node at (2.6, 2.5) {CS theory};

	\end{scope}
\end{tikzpicture}
\caption{\label{fig: WZW/CS}%
CS/WZW correspondence, left: Verlinde lines of the boundary CFT are bulk topological lines moved parallel to the boundary. Right: correlation functions of primary operators are bulk lines ending perpendicularly on the two boundaries.}
\end{figure}
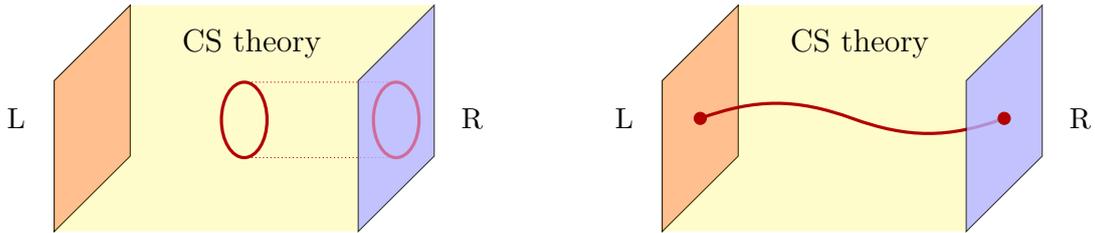

The basic idea is to consider a $G_k$ Chern-Simons theory on a $3$-dimensional manifold of the form $M_2 \times I$, where $I$ denotes an interval, with suitable (anti-)holomorphic boundary conditions. The boundary dual description for this system is the corresponding diagonal WZW model. Non-diagonal variants can be obtained by inserting topological surface defects parallel to the boundaries \cite{Fuchs:2002cm}.\footnote{This description is connected (by the folding trick along the surface) to the more standard case where a $G_k \times G_{-k}$ CS theory with topological boundary conditions on a single boundary is considered.} Among all possible symmetries of the WZW theory, a distinct subset is the one generated by lines that commute with both the left and right chiral algebras. These are called \emph{Verlinde lines} \cite{Verlinde:1988sn}. These lines are easily understood within the Chern-Simons/WZW correspondence, they are Wilson lines of the 3d CS theory along the two-dimensional spacetime $M_2$, see figure \ref{fig: WZW/CS}. 

In the context of WZW coset CFTs, the interplay between this correspondence and the phenomenon of decomposition hides some subtleties. See the recent discussion \cite{Cordova:2023jip}. In what follows we will provide some comments on how the correspondence works in these cases and then apply it to our model \eqref{eq: so coset}.

The bulk description for a WZW coset of the form $G_k/H_{k'}$ is Chern-Simons theory with gauge group $G_k \times H_{-k'}$ \cite{Moore:1989yh}. However, when the boundary theory has a 1-form symmetry $\Gamma\one$, the situation is subtler. In the context of Chern-Simons theory, this is reflected by certain lines being bosonic and constituting a gaugeable algebra $\Gamma$. These bosonic lines admit topological boundary conditions, implying that their endpoints are topological local operators of the boundary theory generating the $1$-form symmetry. As a result, this presentation describes the coset CFT with all its universes. Alternatively, one can consider the gauged theory $(G_k \times H_{-k'})/\Gamma\one$ with standard (anti-)holomorphic boundary conditions on the boundary.\footnote{More generally, $\Gamma\one$ is not necessarily a group-like symmetry. In these cases one needs to perform non-Abelian anyon condensation \cite{Cordova:2023jip}.}  In this alternative presentation there are no topological local operators and the theory has a unique ground state. This theory is therefore dual to the coset CFT projected into a single universe. As we are interested in the global properties of the IR coset, we will solely focus on the former description.

Let us now discuss the Verlinde lines of the coset theory. Topological Dirichlet boundary conditions for a subset of lines $\widehat{L} \in \Gamma\one$ imply that on the boundary we have the identification $\widehat{L}|_{\partial M_3} \sim \unit$. Consequently, any bulk line $L \in G_k \times H_{-k'}$ is subject to the identification\footnote{This is the 3d analog of the \emph{field identification} explained around \eqref{eq: field identification} in Appendix \ref{app: coset}.}
\be\label{eq: fusion identification}
L|_{\partial M_3}  \sim \left(L \otimes \widehat{L}\right)|_{\partial M_3} \,.
\ee
Within this set, there is a subset of lines that do not braid with $\widehat{L}$ indicating that they do not have a mixed anomaly with $\Gamma\one$ on the boundary. These lines correspond to global symmetries that act on a fixed vacuum, implying that they are not spontaneously broken. Conversely, lines that do braid with $\widehat{L}$ exhibit a mixed anomaly with $\Gamma\one$ on the boundary, indicating that they are spontaneously broken. We emphasize that the latter do not form a closed subset since by fusing such lines we can trivialize the anomaly. 

Bulk lines that are fixed under the fusion \eqref{eq: fusion identification} become non simple (namely they split into a sum of other lines) when pushed to the boundary. Bulk topological junctions $\mu \in \mathrm{Hom}(L \times \widehat{L}, L)$ become topological local operators $\mu \in \mathrm{Hom}(L,L)$ living on the line at the boundary, representing fusion between 1-form symmetry generators and the line $L$. The existence of non trivial local operators living on a line is the hallmark of its non-simplicity (see e.g. \cite{Bhardwaj:2017xup}).\footnote{This is the 3d analog of the \emph{fixed point resolution} described around \eqref{eq: fixed point slplitting} in Appendix \ref{app: coset}.}

In the CS/WZW correspondence, it is also possible to derive correlation functions of primary operators within the WZW chiral algebra. These are interpreted as a bulk line ending on the two boundaries, see Figure \ref{fig: WZW/CS}. The quantum dimensions of topological lines correspond to the holomorphic weights of the corresponding primary operators. Note that the bulk lines exhibiting a mixed 't Hooft anomaly with $\Gamma\one$ cannot end perpendicularly to the boundary and therefore they do not correspond to any primary operator, they are not Verlinde lines.

\begin{figure}[t]
\centering
\begin{tikzpicture}
	\draw [fill = red!30!white] (-2,0) -- (-1,1) -- (-1,3) -- (-2,2) -- cycle;
	\begin{scope}[shift={(0.75, 0.)}]
	\draw [fill = brown!30!white] (0,0) -- (1,1) -- (1,3) -- (0,2) -- cycle;
	\end{scope}
	\begin{scope}[shift={(2.75, 0.)}]
	\draw [fill = brown!30!white] (0,0) -- (1,1) -- (1,3) -- (0,2) -- cycle;
	\end{scope}
	\fill [fill = brown, opacity = 0.4] (0.75,0) -- (2.75,0) -- (3.75,1) -- (3.75,3) -- (1.75,3) -- (0.75,2) -- cycle;
	\node at (-3.2,1.5) {\small $\left(\frac{\mathfrak{so}(2N^2-2)_1}{\mathfrak{su}(N)_{2N}}\right)$} ;
	\fill [fill = yellow, opacity = 0.2] (-2,0) -- (6,0) -- (7,2) -- (7,3) -- (-1,3) -- (-2,2) -- cycle;
	\draw [fill = blue!30!white] (6,0) -- (7,1) -- (7,3) -- (6,2) -- cycle;
	\node at (8.2,1.5) {\small $\overline{\left(\frac{\mathfrak{so}(2N^2-2)_1}{\mathfrak{su}(N)_{2N}}\right)}$};
	\node at (-0.2,1.5) {\small $\mathfrak{su}(N)_{2N}$};
	\node at (-0.2,1.) {\small CS};
	\node at (5.,1.5) {\small $\mathfrak{su}(N)_{2N}$};
	\node at (5.,1.) {\small CS};
	\node at (2.25,1.5) {\small $\mathfrak{su}(N)$};
	\node at (2.25,1.) {\small YM};

\end{tikzpicture}
\caption{\label{fig: 2d RG flow}%
3d Construction of 2d QCD. In this setup, the 3d yellow regions host a topological Chern-Simons theory $SU(N)_{2N}$, whereas the brown region supports an $SU(N)$ Yang-Mills theory with no Chern-Simons level and finite coupling. The IR limit corresponds to collapsing the brown dynamical region.}
\end{figure}
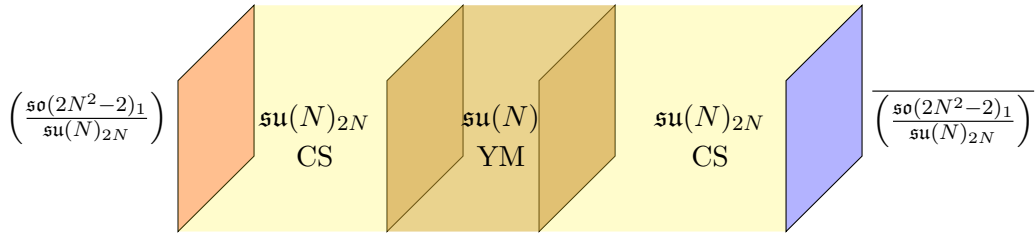

The 3d CS description discussed in this section also illuminates the symmetry structure of $\cT_{SU(N)}$ for finite $g_{\YM}$ \cite{Delmastro:2022prj,Komargodski:2020mxz}. This is possible since we can consider a 3d slab geometry where we place $2(N^2-1)$ free Majorana fermions fermions on both sides. See figure \ref{fig: 2d RG flow}. In the intermediate region of the slab we can gauge the global symmetry $SU(N)$ whose currents sit in the chiral algebra $\mathfrak{su}(N)_{2N}\subset\mathfrak{so}(2(N^2-1))_1$. In this context, the non-trivial RG flow happens in the intermediate region which is non-topological. However, the algebra $\mathfrak{so}(2(N^2-1))_1/\mathfrak{su}(N)_{2N}$ exists for any value of the coupling since its localized near the boundary of the slab. This property strengthens our symmetry-based perspective on the dynamics of adjoint QCD since it precludes the possibility that the chiral algebra and its rich symmetry structure are IR emergent.

\section{IR Analysis of \tps{\matht{\cT_{SU(2)}}}{SU(2)}}\label{sec: N=2}

Let us now proceed to apply the general analysis outlined in the previous sections to specific examples. The first model that we study is based on the gauge group $SU(2)$. Following section \ref{sec: generalities}, $\cT_{SU(2)}$ has an IR description in terms of the coset CFT:
\be\label{eq: N=2 coset}
\cT_{SU(2),\rm{IR}}=\frac{\Spin(6)_1}{SU(2)_4}\ec
\ee
with central charge \eqref{eq: SO/SU central charge} given by $c=1$. Therefore, the $2$ universes of $\cT_{SU(2),\rm{IR}}$ will be described by certain rational points on the $c=1$ conformal manifold, which we now aim to identify.

The partition function of \eqref{eq: N=2 coset} can be obtained following the general prescription of section \ref{sec: generalities}. We denote the representation of the denominator CFT $SU(2)_4$ by $\Lambda'=[\lambda]$, $0\leq \lambda\leq 4$, whose corresponding dimensions is $h_{\lambda}=\frac{1}{24}\lambda(\lambda+2)$. The simple current $\cJ$ is associated with the representation $\Lambda'=[4]$, with action specified by \eqref{Jaction}. Note that the only representations with vanishing monodromy under the action of ${\cal J}$ are the ones with trivial $N$-ality mod 2. The structure of long and short orbits is given by
\be
R_L=\{[[0],[4]]\}\ec\quad R_S=\{[2]\}\ec
\ee
where we also made explicit that $R_L$ consists of a single orbit comprising the identity and the simple current. From the general formula \eqref{eq: SO/SU partition function} we have that:
\be\label{su(2)cosetPF}
Z_{\cT_{SU(2),\rm{IR}}}=\sum_{\Lambda=\0,\vv,\s,\cc}\left(2|b_{\Lambda,[0]}|^2+|b_{\Lambda,[2],1} + b_{\Lambda,[2],2}|^2 \right)\ec
\ee
The above expression implies that there exists $12$ branching functions $b_{\Lambda,[0]},b_{\Lambda,[0],1},b_{\Lambda,[0],2}$ (these are known explicitly, see for instance \cite{Isachenkov:2014zua}) contributing to the final result. An important point is that these branching functions can be expressed as particular combinations of characters for a $\mathfrak{u}(1)_{12}$ chiral algebra. To see how, let us first recall the explicit expression for $\mathfrak{u}(1)_{12}$ characters 
\be\label{eq: U(1)_12 K}
K_\ell= \frac{1}{\eta(q)}\sum_{n\in{\mathbb Z}}q^{6\left(n+\frac{\ell}{12}\right)^2} y^{2(n+\frac{\ell}{12})}\ec\quad\ell\in\{-5,-4,\ldots, 5,6\}\ec
\ee
where we also included a $U(1)$ fugacity $y$. Using the explicit expression for the branching functions we can indeed verify that:
\be\label{branchdecomposition}
\begin{array}{rclcrcl}
b_{\0,[0]}&=&K_0 &\ec&\quad   b_{\0,[2],1} + b_{\0,[2],2}&=&K_4+K_{-4}\ec    \\
b_{\vv,[0]}&=&K_6 &\ec& \quad  b_{\vv,[2],1} + b_{\vv,[2],2}&=&K_2+K_{-2}\ec \\
b_{\s,[0]}&=&K_3 &\ec& \quad  b_{\s,[2],1}+b_{\s,[2],2}&=&K_1+K_5 \ec \\ 
b_{\cc,[0]}&=&K_{-3} &\ec& \quad  b_{\cc,[2],1} + b_{\cc,[2],2}&=&K_{-1}+K_{-5}\ed
\end{array}
\ee
In line with the general arguments presented in Section \ref{sec: generalities} (and further detailed in Appendix \ref{app: coset}), we observe that branching functions associated with short orbits decompose into sums of $\mathfrak{u}(1)_{12}$ characters. 

It is well known that on the $c=1$ conformal manifold parametrized by the radius $R$ there are two special points with enhanced $\mathfrak{u}(1)_{12}$ chiral algebra which can be respectively described by a free compact boson at radii $R=\sqrt{6}$ and $R=\sqrt{3/2}$ (as well as their T-dual values), in units where the self-dual radius occurs at $R_{\rm{sd}}=1$. 
We therefore propose that these two distinct gapless theories describe the physics in the two universes of $\cT_{SU(2),\rm{IR}}$, that is\footnote{This result was also independently obtained  in \cite{unpublished}. We thank Z. Komargodski and S. Seifnashri for related discussions.}
\be
Z_{\cT_{SU(2),\rm{IR}}} = Z^{c=1}_{R=\sqrt{6}} + Z^{c=1}_{R=\sqrt{3/2}}\ed
\ee
The above equation can be easily verified by plugging the branching function decomposition \eqref{branchdecomposition} into \eqref{su(2)cosetPF} to find
\begin{align}\label{eq: N=2 partition function final}
Z_{\cT_{SU(2),\rm{IR}}} &= 2\left(|K_0|^2+2|K_2|^2+2|K_3|^2+2|K_4|^2+|K_6|^2\right)+2|K_1+K_5|^2\\
&=Z^{c=1}_{R=\sqrt{6}} + Z^{c=1}_{R=\sqrt{3/2}}\ec\nonumber
\end{align}
where we assumed that that $y=1$ and hence $K_{-\ell}\simeq K_\ell$. This formula is in perfect agreement with the general expression given in \eqref{eq: generic univereses}.

The values $R=\sqrt{6}$ and $R=\sqrt{3/2}$ on the $c=1$ conformal manifold have also another special property. Namely, at these two points the chiral algebra $\mathfrak{u}(1)_{12}$ is well known to enhance to a full-fledged $\cN=(2,2)$ superconformal algebra (SCA) \cite{Boucher:1986bh,DiVecchia:1986fwg}.\footnote{Some useful properties of ${\cal N}=(2,2)$ superconformal algebras and their characters are collected in Appendix \ref{sec: app N=2}.} The IR theory at these two values is, in fact, spacetime supersymmetric. From the perspective of adjoint QCD, this can be seen as an example of RG flow resulting in emergent supersymmetry. For recent results on RG flows with emergent supersymmetry in two spacetime dimensions see e.g. \cite{Johnson-Freyd:2019wgb,Kikuchi:2022jbl,Delmastro:2022prj}. Although the presence of a superconformal algebra has not provided crucial insights in this section, its role will be significant in the discussion of section \ref{sec: N=3}.

As extensively studied in recent literature (see e.g. \cite{Frohlich:2006ch,Chang:2018iay,Thorngren:2021yso,Benini:2022hzx,Cordova:2023qei,Damia:2024xju,Bharadwaj:2024gpj} for a partial list of references), two-dimensional bosonic RCFTs exhibit a variety of topological defect lines that generate non-invertible 0-form global symmetries. These are often emergent symmetries in the IR theory. However, as emphasized at the end of section \ref{sec: generalities}, the Verlinde lines of \(\text{Spin}(6)_1/SU(2)_4\) are exact symmetries of the model. This is because they are implemented by topological lines in the free Majorana theory, which commute with the \(SU(N)\) gauging. In the next subsection, we will provide a complete characterization of the global symmetries associated with Verlinde lines by studying their realization in Chern-Simons theory.

Finally, let us comment on the non-invertible topological lines which are already manifest in the presentation \eqref{eq: N=2 partition function final}. It is well known that rational points at radii $R$ and $R/N$ are connected by gauging a ${\mathbb Z}_N$ subset of the topological Verlinde lines featured by these theories \cite{Chang:2018iay,Thorngren:2021yso,Benini:2022hzx}. From this perspective, the theory \eqref{eq: N=2 partition function final} possess two non-invertible global symmetries generated by condensation of a ${\mathbb Z}_2$ and $\bZ_3$ (together with T-duality) subset of the Verlinde lines, hence mapping the two universes to each other.\footnote{For previous discussions on spontaneous (non-invertible) symmetry breaking in gapless systems, see \cite{Bhardwaj:2023bbf,Bhardwaj:2024qrf}.}
Note that the $\bZ_2$ non-invertible symmetry is not manifest in the Chern-Simons formulation of the theory, {\it i.e.} it does not correspond to a Verlinde line of the coset \eqref{eq: N=2 coset}. This is associated to the bosonizaztion of the $(-1)^{C_L}$ symmetry of adjoint QCD. Further comments on this point are reserved for section \ref{sec: mass def}.  

\subsection{Symmetries from 3d Chern-Simons}
Following the discussion from section \ref{sec: CS general}, we would like to realize the coset theory $\cT_{SU(2), \textrm{IR}}$ using a Chern-Simons theory with gauge group
\be\label{cs6142}
\Spin(6)_1 \times SU(2)_{-4}\ec
\ee
placed on a slab geometry with suitable boundary conditions.

The non-trivial gauge invariant operators of a Chern-Simons theory are topological line operators $a[\gamma]$, also known as anyons, labeled by the integrable representations of the corresponding gauge group. In the case of interest, the topological lines are parameterized by $a=[\Lambda,\Lambda']$ with $\Lambda=\0,\vv,\s,\cc$ and $\Lambda' = 0,\cdots, 4$ corresponding to the integrable representations of the $\mathfrak{so}(6)_1$ and $\mathfrak{su}(2)_{-4}$ chiral algebra respectively.

\begin{figure}[t]\label{fig: braiding and spin}
\centering
\begin{tikzpicture}
	\draw [thick] (0.14, 0.5) node[above left] {\small$a[\gamma]$} arc [start angle = 180, end angle = 157, radius = 5];
	\filldraw [white] (0.21, 1.34) circle [radius = 0.12];
	\draw [thick, black, rotate around = {-10: (0.23, 2.12)}] (0.23, 2.12) arc [start angle = 110, end angle = 451, x radius = 1.1, y radius = 0.4] node[pos = 0.65, below, shift={(0,-0.1)}] {\small$b[\gamma']$};
       \node[right] at (2.3, 1.6) {\large$ = \quad \frac{S_{a,b}}{S_{0,a}} $};
\begin{scope}[shift={(4.5,0)}]
	\draw [thick] (0.14, 0.5) node[above right] {\small$a[\gamma]$} arc [start angle = 180, end angle = 157, radius = 5];
\end{scope}
\begin{scope}[shift={(9.5,0)}]
	\draw [thick] (0, 0.5) node[above left] {\small$a[\gamma]$} -- (0, 1.4);
	\draw [thick] (0, 1.4) arc [start angle = 180, end angle = -135, radius = 0.25];
	\draw [thick] (0, 1.65) -- (0, 2.3);
	\node [right] at (0.9, 1.4) {\large$ = \;\; \theta_{a}$};
	\draw [thick] (2.5, 0.5) node[above right] {\small$a[\gamma]$} -- (2.5, 2.3);
\end{scope}
\end{tikzpicture}
\caption{
Anyon configuration corresponding to the braiding and the topological spin $\theta_a$.}
\end{figure}%

Some non trivial information about the correlation functions of the boundary CFT is encoded in the $S$ and $T$ matrices of the bulk theory, which compute the braiding and the topological spins $\theta_a = \exp{\left(2\pi i h_a \right)}$  of the corresponding anyons (see figure \ref{fig: braiding and spin}).\footnote{More refined data of the theory are encoded in the so-called F matrix of the theory which, together with $S$ and $T$, defines a Modular Tensor Category (MTC) and completely determine the 3d CS theory (see \cite{Ng:2024jkm} for a very recent discussion). However, in what follows we do not need this data and we will only work with S and T matrices.} One can also introduce the notion of the quantum dimension $d_a$ of a line, defined as 
\be
d_a = \frac{S_{a,0}}{S_{0,0}}\,,
\ee
which computes the correlation function of single anyon defined on a closed contractible cycle.

Clearly, the S-matrix of the theory \eqref{cs6142} can be decomposed into the ones of the individual theories $\Spin(6)_1$ and $SU(2)_{-4}$:
\bea
S_{[\Lambda_1,\Lambda_1'],[\Lambda_2,\Lambda_2']} = S_{\Lambda_1,\Lambda_2}^{\Spin(6)_1}\left(S_{\Lambda_1',\Lambda_2'}^{SU(2)_{4}}\right)^*\ec
\eea
where
\bea\label{eq: S matrices su2 so6}
S^{SU(2)_{k}}_{\Lambda_1,\Lambda_2} = \frac{\sqrt{2}}{\sqrt{k+2}}\sin{\left(\frac{\pi(\Lambda_1+1)(\Lambda_2+1)}{k+2}\right)}&\ec \quad S^{\Spin(2n)_1} = \frac{1}{2}\mat{1&1&1&1\\1&1&-1&-1\\1&-1&i^{n}&-i^{n}\\1&-1&-i^{n}&i^{n}}\,.
\eea 
The S-matrix of the theory can also be used to obtain the fusion algebra of anyons. This is achieved by using the Verlinde formula:
\be\label{eq: Verlinde}
a \times b = \sum\limits_c N_{ab}^c \,c \quad,\quad N_{ab}^c = \sum\limits_d \frac{S_{ad}S_{bd}S^*_{cd}}{S_{0d}}\,.
\ee
Similarly, the topological spins factorize into
\be
\theta_{[\Lambda,\Lambda']} = \theta^{\Spin(6)_1}_{\Lambda}\theta^{SU(2)_4}_{\Lambda'} = \exp{\left(2\pi i (h_a^{\Spin(6)_1} - h_a^{SU(2)_4})\right)}\ec
\ee
where
\be\label{eq: spins}
h^{\Spin(2n)_1}_{\Lambda} = \begin{cases}
    & 0 \quad,\quad \text{for $\Lambda=\0$}\\
    & \frac{1}{2} \quad,\quad \text{for $\Lambda=\vv$}\\
    & \frac{n}{8} \quad,\quad \text{for $\Lambda=\s,\cc$}\,.
\end{cases} 
\quad,\quad h^{SU(2)_k}_{\Lambda'}= \frac{\Lambda'(\Lambda' +2)}{4(k+2)}\,.
\ee

From \eqref{eq: S matrices su2 so6}, \eqref{eq: Verlinde}  and \eqref{eq: spins} it is easy to show that the line $[\0,4]$ generate a $\bZ_2$ symmetry and it has spin $\theta_{[\0,4]}=1$. It follows that it can be gauged, i.e. the line $\cA = [\0,0] \oplus [\0,4]$ is a gaugeable algebra. As already emphasized in the general discussion of section \ref{sec: generalities}, the presence of this algebra is related to the existence of the 1-form $\bZ_2$ symmetry in the boundary theory: imposing Dirichlet boundary conditions for $[\0,4]$, its end-point becomes a topological local operator of the boundary theory, thus generating a $\bZ_2$ 1-form symmetry.

Let us now describe the boundary interpretation of the bulk anyons. Within the full set of lines $[\Lambda,\Lambda']$, the ones with even values of $\Lambda'$ do not braid with $[\0,4]$ while the ones with odd values of $\Lambda'$ admit a non-trivial braiding with $[\0,4]$. From the perspective of the boundary theory, this is the analog of the monodromy condition studied before. This implies that, while the former do not share a mixed anomaly with the boundary 1-form symmetry, the latter have a non-trivial 1-form symmetry charge and therefore non-trivial anomaly. Consequently, they serve as topological domain walls separating different universes of the boundary theory.
 
Because of the Dirichlet boundary condition, on the boundary we also have the identification
\be\label{eq: indentification su2}
[\Lambda,\Lambda'] \sim [\Lambda,\Lambda'] \times [\0,4] = [\Lambda,4-\Lambda']\,,
\ee
where in the last step we have used the Verlinde formula \eqref{eq: Verlinde} to compute fusion of lines.
Similarly to the 2d analysis, we have four fixed lines $[\Lambda,2]$ of quantum dimension 2. Following the general analysis of section \ref{sec: generalities}, they split into eight lines denoted by $[\Lambda,2]_{1,2}$ of quantum dimension 1 and defined such that $[\Lambda,2]_{1}+[\Lambda,2]_{2} = [\Lambda,2]$.\footnote{Following the identification \eqref{eq: indentification su2} and the splitting, the S-matrix and spins must be adjusted accordingly. These new data can be constructed by imposing consistency conditions on the unresolved set of lines, similar to the procedure described in \cite{Moore:1989yh}.}

Therefore we have:
\begin{itemize}
    \item A set of $12$ lines of quantum dimension $1$ that do not braid with $[\0,4]$ hence being unbroken symmetries of the model. They are given by
    \be\label{twelve}
    [\Lambda,0]\ec\quad [\Lambda,2]_1\ec\quad [\Lambda,2]_2
 \ec\quad \Lambda=\0,\vv,\s,\cc\,.
\ee
The lines $[\Lambda,0]$ generate the $\bZ_4$ symmetry of the $\Spin(6)_1$ CFT. The set of topological lines \eqref{twelve} implement a $\bZ_{12}$ symmetry generated by the Verlinde lines of the $\mathfrak{u}(1)_{12}$ chiral algebra present in the two universes. Their  topological spins are 
\be
(h_{[\Lambda,0]})_{\Lambda=\0,\cdots,\cc} = \left(0,\frac{1}{2},\frac{3}{8},\frac{3}{8}\right)\ec\quad (h_{[\Lambda,2]_{1,2}})_{\Lambda=\0,\cdots,\cc} = \left(\frac{2}{3},\frac{1}{6},\frac{1}{24},\frac{1}{24}\right)\ed
\ee
These are indeed the correct holomorphic weights (mod $1$) for primary operators of the $\mathfrak{u}(1)_{12}$ chiral algebra, as expected from the standard CS/WZW dictionary.
\item A set of $4$ lines which braid with $[\0,4]$, hence being spontaneously broken. They are of the form $[\Lambda,1]$ with $\Lambda=\0,\cdots,\cc$. Notice that these are non-invertible topological lines. They obey the following fusion rule:
\be
[\Lambda_1,1]\times [\Lambda_2,1] = [\Lambda_1\times \Lambda_2,0] + [\Lambda_1\times \Lambda_2,2]_1+[\Lambda_1\times \Lambda_2,2]_2\,.
\ee
By choosing one of them, say the $[\0,1]$ line, all the other lines are obtained by fusion with the $12$ unbroken lines. Therefore we can claim without loss of generality that the topological line $[\0,1]$ generates the spontaneously broken symmetry. The above fusion rules make it evident that these lines implement a global symmetry described by a Tambara-Yamagami $TY(\bZ_3)$ category. The group $\bZ_3$ is interpreted as the subgroup of the unbroken symmetry $\bZ_{12}$.
 \end{itemize} 
These results are in complete agreement with the analysis provided in the previous subsection. The property that the universes of $\cT_{SU(2),\rm{IR}}$ are described by two different theories is completely natural in the Chern-Simons picture where it is interpreted as a consequence of spontaneous non-invertible symmetry breaking.\footnote{For other studies of spontaneous non-invertible symmetry breaking see \cite{Chang:2018iay,Thorngren:2019iar,Damia:2023ses,Bhardwaj:2023idu}.}

\subsection{Mass Deformations}\label{sec: mass def}
Having obtained an explicit description for $\cT_{SU(2),\rm{IR}}$ allows us to explore how its dynamics is modified when we introduce massive deformations.

 Let us consider deforming the UV theory ${\cal T}_{\rm{UV}}$ by a mass operator ${\cal O}_{\rm{UV}}$. The operator $\cO_{\rm{UV}}$ is mapped, under the RG flow, to an operator in the IR theory ${\cal T}_{\rm{IR}}$ that we denote by $\cO_{\rm{IR}}$. The local operator $\cO_{\rm{UV}}$ is charged under the $U(1)_A\times U(1)_V$ symmetry, therefore in order to obtain an explicit description for $\cO_{\rm{IR}}$ we need to track its charges under the IR $U(1)_L\times U(1)_R$ symmetry.

As already explained in section \ref{sec: generalities}, the UV symmetry $U(1)_V \times U(1)_A$ has a mixed 't Hooft anomaly captured by the anomaly inflow action
\be
\cA_{\rm{UV}} = \frac{i2(N^2-1)}{2\pi} \int A_V d A_A\,,
\ee
where $A_{A,V}$ are respectively classical background gauge fields for $U(1)_{A,V}$. Clearly, the IR fixed point must reproduce this anomaly. 

The IR $\cN=(2,2)$ superconformal algebra fixes the value of the $U(1)_L \times U(1)_R$ mixed 't Hooft anomaly in terms of the CFT central charge \cite{Benini:2013cda}:
\be
c = 3k_R = -3k_L\ec
\ee
where $k_{L,R}$ are the coefficients appearing in the two point functions of flavor R-symmetry currents.\footnote{We are using the standard supersymmetric normalization where the R-charge  defined by \eqref{eq: hac Qac} takes fractional values.} In order to compare it with the UV global symmetry $U(1)_V\times U(1)_A$, we need to identify both charges and background gauge fields as follows:
\be\label{eq: change of basis}
3(Q_R \pm Q_L)=Q_{A,V}\ec  \qquad \frac{1}{6}(A_R \pm A_L) = A_{A,V}\ed
\ee
With the above identification we can express the IR mixed anomaly as
\be
\cA_{\rm{IR}} = \frac{i\widetilde{k}}{2\pi}\int A_Ad A_V\ec \qquad \widetilde{k}= 6 c = 2(N^2-1)\ec
\ee
ensuring the appropriate match between the UV and IR mixed 't Hooft anomalies.

The theory of adjoint QCD \eqref{eq: Adj QCD action} allows for two independent Dirac masses (together with their complex conjugates)
\be
\cO_{1} = {\rm Tr}(\Psi_L^{\dagger}\Psi_R) \ec\quad \cO_{2} = {\rm Tr}(\Psi_L\Psi_R)\,.
\ee
By expressing the Dirac fermion as $\Psi=\psi_1+i\psi_2$, we can write the masses for either of the adjoint Majorana components denoted by $\psi_{1,2}$ as follows:
\be
\cO_{\psi_1}\equiv\cO_1+\cO_2=2\psi_1^L\psi_1^R \ec\quad \cO_{\psi_2}\equiv\cO_1-\cO_2=2\psi_2^L\psi_2^R\,.
\ee
The Dirac massess $\cO_{1,2}$ have respectively charges $Q_A = 2,0$ and $Q_V = 0,2$ under $U(1)_{A,V}$. The mapping \eqref{eq: change of basis} that we derived using the above 't Hooft anomaly matching argument implies that $\cO_{1,2}$ correspond to IR operators with charges $Q_L = Q_R = \frac{1}{3}$ and $Q_R = -Q_L = \frac{1}{3}$ respectively. In particular, we have that:
\be\label{eq: IR deformation}
\begin{cases}
\cO_{1,\rm{IR}} = V_{2,0} \in |K_2|^2\ec R=\sqrt{6}\ec\\
\cO_{1,\rm{IR}} = \varnothing \ec R=\sqrt{3/2}\ec
\end{cases}\quad
\begin{cases}
\cO_{2,\rm{IR}} = \varnothing\ec R=\sqrt{6}\ec\\
\cO_{2,\rm{IR}} =  V_{1,0} \in K_2K_{-2}\ec R=\sqrt{3/2}\ec
\end{cases}
\ee
where $V_{n,w}$ denotes vertex operators with charges $(n,w)$ under $U(1)_n\times U(1)_w$ momentum and winding symmetries of the IR theory $\cT_{SU(2),\rm{IR}}$ described by a sum of two $c=1$ theories with radii $R=\sqrt{6}$ and $R=\sqrt{3/2}$. These operators can be expressed as
\be
V_{n,w}(z,\bar z)=\,:e^{i n \phi(z,\bar z)}e^{i w \tilde\phi(z,\bar z)}:\ec
\ee
where $\phi$ and $\tilde\phi$ are respectively a compact scalars at radius $R$ and its T-dual version.\footnote{At a given radius $R$, the map between $Q_{L,R}$ and $n,w$ reads
\be
Q_{L,R}=\frac{1}{R^2}\left(n\pm wR^2\right)
\ee
In particular, for $R=\sqrt{6}$, the above expression together with \eqref{eq: change of basis} leads to the charge assignment of \cite{Gopakumar:2012gd}.} 
The symbol $\varnothing$ in \eqref{eq: IR deformation} has been introduced to denote that $\cO_{\rm{IR}}$ is decoupled at long distances to leading order in $m/g_{\YM}$. Generically, there will be higher order contributions starting at order $(m/g_{\YM})^2$ that eventually gap the theory. Notice that the decoupling of $\cO_2$ in the universe described by the $R=\sqrt{6}$ compact boson agrees with the results of \cite{Gopakumar:2012gd}, though the dynamics concerning the flux tube sector at $R=\sqrt{3/2}$ have been overlooked so far.

If we consider instead the mass of a single Majorana fermion, we obtain
\be\label{majoranamap}
\cO_{\psi^1,\rm{IR}}= \begin{cases}
     V_{2,0}\ec\quad R=\sqrt{6}\ec\\
     V_{1,0}\ec\quad R=\sqrt{3/2}\ed
    \end{cases}
\ee
Different UV mass operators have distinct physical consequences in each universe. Notably, one of the two Dirac masses always decouples in each universe. This observation aligns with the fact that the two universes are related by a non-invertible symmetry transformation, under which both UV operators \(\cO_{1,2}\) are charged. These operators are exchanged under the action of the topological line implementing the $TY(\bZ_2)$ symmetry (see for instance \cite{Thorngren:2021yso}):
\be
TY(\bZ_2)\, : \,\, V_{2,0} \,\, {\rm at} \,\, R=\sqrt{6} \,\, \to \,\,  V_{1,0} \,\, {\rm at} \,\, R=\sqrt{3/2}\ed
\ee
Finally, let us also emphasize that the theory deformed by a single Majorana mass preserves a 0-form symmetry that participates in a mixed anomaly with the 1-form symmetry, thus implying deconfinement at long distances.

\subsection{\tps{$\cT_{SU(2)}$}{SU(2)} Massive Phases}\label{massivephases}
\begin{figure}[t]
$$
\scalebox{0.25}{

\includegraphics[]{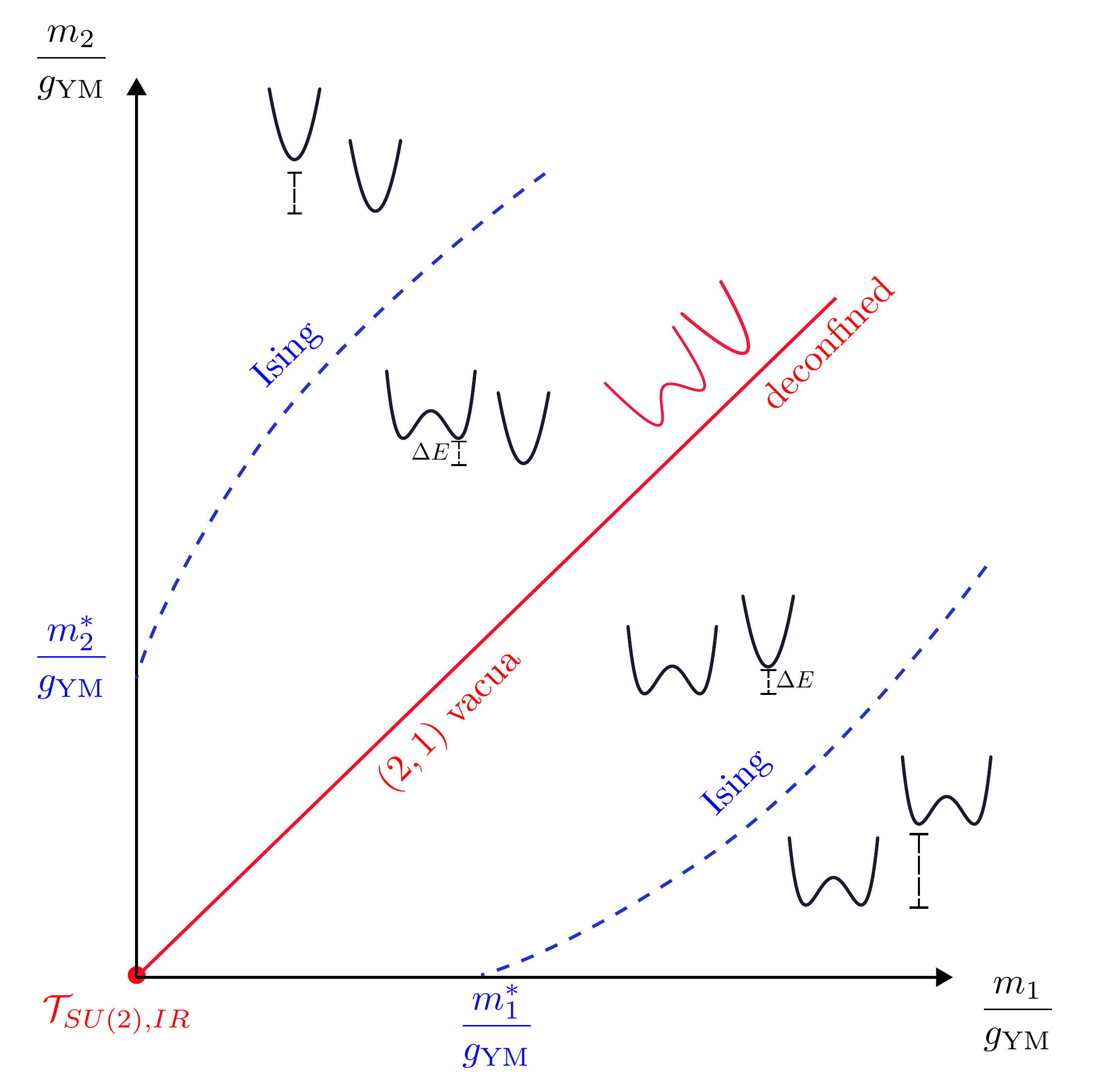}
}
\hspace{1cm}
$$
\caption{A portrait of the massive IR phases of $\cT_{SU(2)}$. At the origin, we have the massless CFT discussed in Section \ref{sec: N=2}. The red line represents a region where the theory is gapped and deconfined, while the dashed blue lines indicate phase transitions occurring in the flux tube sector. 
These phase transitions are second order and belong to the Ising universality class.  When depicting the vacua in specific regions of the diagram, we conventionally choose to place the universe with vanishing 1-form symmetry charge on the left. 
}
\label{fig: su2 phasediagram}
\end{figure}

In this section, our aim is to explore the massive IR phases of $\cT_{SU(2)}$ by dialing the two independent Dirac masses that were introduced earlier.
\paragraph{$\mathbf{m_1=m_2}$}Let us first consider perturbing the UV theory by a single Majorana fermion mass operator $\cO_{\psi_1}$ with coefficient $m_{\psi_1}$ (diagonal red line in figure \ref{fig: su2 phasediagram}).  From \eqref{majoranamap}, we already know that $\cO_{\psi_1}$ flows to the vertex operator $V_{2,0}$ in the universe described by the compact boson theory at radius $R= \sqrt{6}$. The operator $V_{2,0}$ preserves an invertible subgroup $\bZ_2 \times U(1)_w \subset U(1)_n \times U(1)_w$ which has an anomaly parametrized by the non-trivial element $p \in H^3(\bZ_2 \times U(1)_w ,U(1) )  \cong \bZ_2$.  Therefore, the $R=\sqrt{6}$ universe perturbed by $V_{2,0}$ cannot be trivially gapped and the minimal gapped theory matching the anomaly has two vacua.  A similar analysis for the universe $R=\sqrt{3/2}$ shows that the operator $V_{1,0}$ only preserves an anomaly-free subgroup $U(1)_w \subset U(1)_n \times U(1)_w$. We thus conclude that the IR theory is compatible with a trivially gapped theory with a unique vacuum. Summarizing, the perturbation $\cO_{\psi_1}$ leads to a flow ending in a gapped theory with three vacua: two of them in the first universe and a trivially gapped vacuum in the second one.

The operator $\cO_{\psi_1}$ is invariant under the non-invertible TY$(\bZ_2)$ symmetry connecting the two universes, implying that the IR theory must be self-dual under a $\bZ_2$ symmetry gauging. Assuming that the IR is gapped, a minimal TY$(\bZ_2)$-invariant theory has two vacua in one universe and a trivially gapped vacuum in the other, consistent with the above description.\footnote{Notice that the vacua of the gapped theory must fit into a non-negative integer matrix (NIM) representation of the non-invertible symmetry \cite{Komargodski:2020mxz,Choi:2023xjw}. This requirement already forbids a trivially gapped vacuum in the IR and is related to the anomaly of the non-invertible TY$(\bZ_2)$ symmetry (see \cite{Thorngren:2021yso,Zhang:2023wlu,Antinucci:2023ezl}).} The existence of such a (non-invertible) symmetry connecting two universes implies deconfinement. Our findings are consistent with the picture advocated in \cite{Komargodski:2020mxz} for $SU(2)$ QCD with a single adjoint Majorana fermion. This suggests that no phase transition occurs between the regime \(m_{\psi_1} \ll g_{\YM}\) (where the above analysis is valid) and \(m_{\psi_1} \gg g_{\YM}\) (where the analysis in \cite{Komargodski:2020mxz} applies).

\paragraph{$\mathbf{m_{1,2}\gg g_{\YM}}$} Let us now describe the physics obtained after perturbing with one of the two Dirac mass operators $\cO_{1,2}$ with corresponding coefficients $m_{1,2}$. For $m_1\gg g_{\YM}$, the adjoint Dirac fermion can be integrated out and the resulting theory is just $2d$ pure $SU(2)$ Yang-Mills theory which is known to have a unique trivially gapped
confining vacuum in the IR. Notice that this scenario usually involves a choice of local counterterms \cite{Vafa:1983tf}. As discussed in section \ref{sec: generalities}, $2d$ pure $SU(2)$ Yang-Mills theory is subject to decomposition. To project into any of the two universes one has to gauge the $1$-form symmetry $\bZ_2\one$ and obtain $SO(3)_{\pm}$ gauge theories respectively.\footnote{We are using the nomenclature of \cite{Aharony:2013hda}, where $SO(3)_{\pm}$ respectively denote the gauge theory with gauge group $SO(3)$ and discrete $\theta$-term $0,\pi$.} We perform a choice of local counterterms to identify the low-energy gapped vacuum with the vacuum of $SO(3)_+$ Yang-Mills theory, labelled by the singlet representation of $SU(2)$.

In the fermionic theory, the effective description for $m_2\gg g_{\YM}$ is obtained by acting with $\bZ_2^{C_L}$. Due to the mixed anomaly between  $\bZ_2^{C_L}$ and $\bZ\one_N$, the theory is stacked with an SPT which flips the 1-form symmetry charge of the two universes. Moreover, since $\bZ_2^{C_L}$ has a further mixed anomaly with $\bZ_2^{\rm F}$, its action changes the spin statistics of a given state. Consequently, the lowest energy state of the fermionic theory at $m_2\gg g_{\YM}$ has non-trivial 1-form symmetry charge and fermionic statistics. Therefore, it should be regarded as the ground state of $SO(3)_-$ YM theory stacked with the Arf counterterm (see Appendix \ref{app: bosonization}).

Upon gauging $(-1)^F$, the bosonic ground state splits into two degenerate vacua, whilst the fermionic one leads to a single one.
In turn, $\bZ_2^{C_L}$ gets mapped to the non-invertible $TY(\bZ_2)$ symmetry extensively discussed in previous sections.

\paragraph{$\mathbf{m_{1}\ll g_{\YM}}$} The limit $m_1\ll g_{\YM}$ can be accessed by deforming $\cT_{SU(2),\rm{IR}}$ with \eqref{eq: IR deformation}. In this case, both (non-invertible) symmetries TY$(\bZ_3)$ and TY$(\bZ_2)$ connecting the two universes are explicitly broken and the deformed theory is compatible with a confined phase at low energy. In the universe described by a compact boson at radius $R=\sqrt{6}$, the flow ends up in a gapped phase with two degenerate vacua. Concurrently, in the universe $R=\sqrt{3/2}$, the deformation decouples at first order in the mass parameter $m_1$ while it receives higher order contributions from less relevant terms. 
Thus, the Dirac mass operator $\cO_1$ makes $\cT_{SU(2),\rm{IR}}$ gapped with two vacua in the universe with lowest energy. This is the same confined phase that we observed in the regime $m_{1}\gg g_{\YM}$, suggesting that no phase transition occurs on the horizontal $m_1$ axis in the zero flux tube sector.

\paragraph{$\mathbf{m_{2}\ll g_{\YM}}$}A similar analysis can be repeated by turning on the Dirac mass operator $\cO_2$ with coefficient $m_2 \ll g_{\YM}$ but, compared to the previous paragraph, the roles played by the universes are now exchanged. Looking at \eqref{eq: IR deformation} we can deduce that, upon deforming with such an operator, $\cT_{SU(2),\rm{IR}}$ flows to a single trivially gapped vacuum. Note that, in the opposite limit $m_2\gg g_{\YM}$, the low energy description exhibits a single gapped vacuum associated with (bosonized) pure $SO(3)_{-}$ Yang-Mills theory stacked with the Arf counterterm. We therefore conclude that, on the vertical $m_2$ axis depicted in Figure \ref{fig: su2 phasediagram}, no phase trasition occur in the universe with lowest energy.

\paragraph{$\mathbf{m_{1}>m_2}$}Next, we turn our attention to the lower right section of the diagram, bounded by the red diagonal line where \( m_1 = m_2 \) and the horizontal axis where \( m_2 = 0 \) (see Figure \ref{fig: su2 phasediagram}). In the regime where \( m_{\psi_1} \gg g_{\text{YM}} \), we can introduce a positive mass \( m_{\psi_2} \) for the remaining adjoint Majorana fermion. In this part of the diagram, we can leverage the analysis of the massive phase of \( SU(2) \) adjoint QCD with a single adjoint Majorana fermion as conducted in \cite{Komargodski:2020mxz}. Dialing \( m_{\psi_2} > 0 \) lifts the degeneracy of the gapped phase along the \( m_1 = m_2 \) line, leading to a universe with a doubly degenerate ground state. This behavior is consistent with what is observed in the regimes where \( m_2 \ll m_1 \ll g_{\rm{YM}} \) and \( m_2 \ll g_{\rm{YM}} \ll m_1 \). Therefore, we propose that there should be no discontinuities in the description of the massive phase within the lower wedge of the diagram in Figure \ref{fig: su2 phasediagram}. 

In the flux tube sector, when $m_{2}\ll g_{\YM}\ll m_1$, the theory is gapped with one vacuum, while it has two degenerate vacua when $m_{1,2}\gg g_{\YM}$. Therefore this universe exhibits a phase transition of the Ising type. This transition will extend upon varying $m_2$ (dashed blue line in Figure \ref{fig: su2 phasediagram}) and eventually cross the $m_2=0$ axis at some special value $m_1^*/g_{\text{YM}}$.\footnote{
The bending of the blue line in diagram \ref{fig: su2 phasediagram} can be explained by noting that the leading large \( m_{\psi_1} \) correction to the gluon propagator is negative. This effect can be reliably calculated using perturbation theory, as fermion loop diagrams are suppressed by \( g_{\YM}/m_{\psi_1} \) in this regime. The leading contribution can be extracted from the general result provided in \cite{Peskin:1995ev}. Interestingly, the final result is analogous to that in massive QED, differing only by a positive proportionality constant.
} 
This phase transition corresponds to the emergent supersymmetry on the confining string.\footnote{The emergence of supersymmetry in massive adjoint QCD has been found long ago at infinite volume within light-cone quantization \cite{Kutasov:1993gq}. Recently, it has been argued that it persists at finite volume as well \cite{Dempsey:2024ofo}.} It is well established that $SU(N)$ gauge theory coupled to a single adjoint Majorana fermion of mass $m_\psi$ develops $\cN=(1,1)$ supersymmetry at $m_{\psi}=\pm m_{\rm{susy}}$, with $m_{\rm{susy}}= g_{\rm YM}\sqrt{N/2\pi}$. When considering a system with two Majorana fermions, with masses $m_{\psi_1}$ and $m_{\psi_2}$, a similar phenomenon should occur at \cite{Popov:2022vud,Dempsey:2024ofo}:
\be
m_{\psi_2}=m_{\rm susy}(m_{\psi_1})\ec \quad m_{\rm susy}(m_{\psi_1}\to \infty)\to g_{\rm YM}\sqrt{\frac{N}{2\pi}}\ed
\ee
The property that supersymmetry is linearly realized in the ground state for $m_{\psi_2}=+m_{\rm susy}(m_{\psi_1})$ is in perfect agreement with our expectation that no phase transitions should occur (in the lowest energy universe) in the $m_2<m_1$ wedge of the phase diagram. Conversely, in the remaining universe which is lifted upon mass deformation, the spectrum develops an exact Bose-Fermi degeneracy with vanishing Witten index \cite{Komargodski:2020mxz,Cherman:2019hbq,Dempsey:2024ofo}. Supersymmetry is, therefore, spontaneously broken, resulting in a massless Majorana goldstino. In the bosonized system, this phenomenon is manifest as a second order phase transition of the Ising type taking place on the confining string.\footnote{The observation that the $c=\frac12$ minimal model (either a free Majorana fermion or the Ising model) realizes supersymmetry non-linearly dates back to \cite{Kastor:1988ef,Friedan:1984rv}.}

\paragraph{$\mathbf{m_{2}>m_1}$} Finally, we analyze the upper left region of the diagram in Figure \ref{fig: su2 phasediagram}. This wedge of the diagram can be easily deduced using the $\bZ_2^{C_L}$ symmetry (non-invertible TY$(\bZ_2)$ symmetry in the bosonized theory) which maps $m_1 \leftrightarrow m_2$. States with bosonic (fermionic) statistic, under $m_{1,2}\rightarrow m_{2,1}$, are mapped to states with the opposite spin statistic and a different 1-form symmetry charge. Thus, the phase diagram shown in figure \ref{fig: su2 phasediagram} can be readily derived.

\subsection{Comments on the String Tension}\label{sec:tension}

An important observable to analyze in a gapped and confining phase is the effective tension of a confining string characterizing the flux tube sector. For adjoint QCD with gauge group $SU(N)$ such strings are often called $k$-strings, computing their effective tension is physically equivalent to understanding the behavior of a Wilson loop in a representation whose Young tableau has $k$ boxes. Even though adjoint QCD with gauge group $SU(2)$ has only one such string, we would like to compute its tension to give some comments on its expected behavior as a function of the two Dirac masses $m_{1,2}$.

As discussed in the previous section, for such a tension to be non-vanishing we must consider regions off the diagonal $m_1=m_2$ line in the  diagram (see Figure \ref{fig: su2 phasediagram}). We will thus focus on $m_2\leq m_1$, since the opposite region $m_1\leq m_2$ can be obtained by flipping the sign of one of the Majorana fermions. 

In the regime where both bare mass parameters satisfy $m_{1,2}\ll g_{\YM}$, we consider the IR fixed point $\cT_{SU(2), \rm{IR}}$ deformed by the operators \eqref{eq: IR deformation}. Since $\cT_{SU(2), \rm{IR}}$ is described by (a sum of) free compact bosons, the perturbed theory corresponds to two (decoupled) Sine-Gordon models. As a result, this particular RG flow can be solved using integrability techniques, enabling us to calculate the string tension analytically.\footnote{We thank Z. Komargodski for suggesting this approach.} Let us denote by \(\phi_i\) the compact scalar, where the subscript \(i = 1, 2\) labels the respective universes with radii $R_1 = \sqrt{6}$ and $R_2= \sqrt{3/2}$. Their partition function is
\be
Z = \sum_{i=1}^{2}\int [\cD\phi_i]\exp[-S_i]\ec
\ee
where:
\bea\label{eq: sin gordon}
S_1=\int &\frac{1}{4\pi \beta^2} (\partial\phi_1)^2 -2\mu_1\cos{2\phi_1}\ec \quad 
S_2=\int &\frac{1}{16\pi\beta^2} (\partial\phi_2)^2 -2\mu_2\cos{\phi_2}\ed
\eea
Since the scaling dimension of the IR mass operators is $\Delta = \frac{1}{3}$, the couplings $\mu_i$ can be expressed as $\mu_i=m_i\Lambda^{2/3}$ where $\Lambda$ is proportional to $g_{\YM}$. The above presentation is the standard description of Sine-Gordon theory in terms of an exactly marginal coupling $\beta$ \cite{Coleman:1974bu,Zamolodchikov:1978xm, Dashen:1975hd, Korepin:1975zu}. Due to the non-invertible symmetries mapping the universes at the massless point, both theories in \eqref{eq: sin gordon} feature the same value of $\beta$, namely
\be
\beta^2=\frac{1}{R_1^2}=\frac{1}{4R_2^2}=\frac16\ed
\ee
Remarkably, using integrability \cite{Zamolodchikov:1995xk, Lukyanov:1996jj} it is possible to compute the vacuum energy density as function of the couplings $\mu_i$ and $\beta$: 
\be\label{vacenergy}
\cE_i[\mu_i]=-\mu_i^{\frac{1}{1-\beta^2}} \kappa(\beta)\ec 
\ee
where
\be
\kappa(\beta) \equiv \pi^{\xi}\tan\left(\frac{\pi \xi}{2}\right)\left[\frac{\Gamma\left(\frac{\xi}{2}\right)}{\Gamma\left(\frac{1+\xi}{2}\right)}\right]^2\left[\frac{ \Gamma\left(1-\beta^2\right)}{\Gamma\left(\beta^2\right)}\right]^\frac{1}{1-\beta^2}\ec \quad \xi\equiv \frac{\beta^2}{1-\beta^2}\ed
\ee
The $k$-string tension is in general determined by:
\be
T_k = \cE_k-\cE_0\ec
\ee
where $\cE_k$ denotes the vacuum energy density of the universe with 1-form symmetry charge equal to $k$.
Using \eqref{vacenergy}, we obtain the following expression for the only non-trivial string tension in $\cT_{SU(2)}$:
\be\label{eq: SU(2) tension}
T_1=\kappa_0~ \Lambda^{4/5}\left(m_2^{6/5}-m_1^{6/5}\right)\ec  
\ee
where $\kappa_0\equiv \kappa(1/\sqrt{6}) \simeq 2.4572$. Therefore, the tension of the confining string $T_1$ scales with a fractional power of the mass parameters. A similar behavior was also observed in the charge-$q$ multiflavor Schwinger model with $N_f\geq 2$ \cite{Armoni:2018bga}.\footnote{When $N_f=2$ one can show that the result reported in \cite{Armoni:2018bga} can be derived in complete analogy to the reasoning used to determine \eqref{eq: SU(2) tension}.}

We emphasize that \eqref{eq: SU(2) tension} is strictly valid for $m_{1,2}\ll g_{\YM}$, as operators different from \eqref{eq: IR deformation} do arise at higher orders. Consequently, when exploring different regions of the phase diagram, the dependence of the string tension on the relevant parameters may change drastically. In fact, the string tension can also be explicitly determined in certain regions of the phase diagram \ref{fig: su2 phasediagram} (and also for higher values of $N$) which are \emph{outside} the validity of the methods described so far. This is what happens when we separately give masses to the adjoint Majoranas $m_{\psi_1}\gg g_{\YM}$ and $m_{\psi_2} \ll g_{\YM}$. Since the theory with $m_{\psi_1}\gg g_{\YM}$ is gapped, the analysis of \cite{Komargodski:2020mxz} applies and it has been found that, for small values of $2\leq N\leq 5$, the tension reads
\be\label{eq: sine scaling}
T_k\sim g_{\YM} m_{\psi_2} \sin\frac{\pi k}{N}\sim g_{\YM}(m_2-m_1)\sin\left(\frac{\pi k}{N}\right)\ec
\ee
this behavior is commonly referred in the literature as the {\it sine} scaling \cite{Douglas:1995nw}. Note also that the dependence on the mass parameters is now linear. Moreover, when either one of the two independent Dirac masses is large, i.e. $m_{1,2}\gg g_{\YM}$, it is also known that the effective $k$-string tension must reproduce the \emph{Casimir} scaling behavior of pure $2d$ YM theory:
\be\label{casimirscaling}
T_k \sim g^2_{\YM} k(N-k)\ed
\ee
See \cite{Strassler:1998id,Armoni:2003ji,Armoni:2003nz,Armoni:2011dw} and references therein for extensive literature on this subject.

Explicitly determining the function \( T_k \) and understanding its evolution across the diagram \ref{fig: su2 phasediagram} to reproduce the scaling behaviors in \eqref{eq: SU(2) tension}, \eqref{eq: sine scaling} and \eqref{casimirscaling} is a challenging problem that we aim to address in future work. A crucial step towards this goal is extending the analysis to larger values of \( N \), where adjoint QCD exhibits multiple non-trivial \( k \)-strings. To this end, in Section \ref{sec: N=3}, we explore adjoint QCD with the gauge group \( SU(3) \). 

\section{IR Analysis of \tps{\matht{\cT_{SU(3)}}}{SU(3)}}\label{sec: N=3}
A more challenging example that we analyze in this work appears in the IR limit of $SU(3)$ adjoint QCD with one adjoint Dirac fermion. This model is described by the coset CFT:
\be\label{eq: N=3 coset}
\cT_{SU(3),\rm{IR}}=\frac{\Spin(16)_1}{SU(3)_6}\,,
\ee
with central charge $c=8/3$. From the general analysis of section \ref{sec: generalities}, we expect that the partition function decomposes as
\be\label{eq:N=3 partition sum}
Z_{\cT_{SU(3), \rm{IR}}} = Z^{c=8/3}_1 + 2Z^{c=8/3}_2\ed
\ee
Our goal is therefore to determine the partition function in each of the two universes. A priori, this is a complicated problem since both theories are strongly coupled and therefore computing their exact partition function seems hopeless. 

An important point, that we already alluded to at the end of section \ref{sec: N=2}, is that the RG flow to \eqref{eq: N=3 coset} is another example of emergent supersymmetry in two-dimensions \cite{Gopakumar:2012gd}. In particular, the coset \eqref{eq: N=3 coset} admits a $\cN=(2,2)$ superconformal algebra and its central charge value $c=8/3$ coincides with the $k=16$ element in the discrete series of minimal models $\cM_k$ with $\cN=(2,2)$ superconformal algebra and central charge $c=3k/(k+2), k\in \mathbb{N}$, based on the coset
\be\label{susyminimal}
\cT_{\cM_k} = \frac{SU(2)_k\times U(1)_{2}}{U(1)_{k+2}}\ec
\ee
see Appendix \ref{sec: app N=2} for further details. All modular invariant partition functions for theories admitting a $\mathcal{N}=(2,2)$ superconformal algebra and central charge $c=8/3$ have been systematically classified in the work \cite{Gray:2008je}.\footnote{The reader may already be familiar with the ``ADE classification" of $\cN=(2,2)$ minimal models \cite{Gepner:1987qi,Vafa:1988uu}. The work of \cite{Gannon:1996hp,Gray:2008je} shows that in general there are many more modular invariants for the $\cN=(2,2)$ superconformal algebra. However, if we further require the
CFT to be invariant under spectral flow by half a unit (see for instance Appendix A of \cite{Belin:2020nmp} and references therein), then the possible modular invariants are classified by the ADE series \cite{Vafa:1988uu,Cecotti:1992rm}. (In string theory,
it is common to refer to this condition as preserving spacetime supersymmetry.) As we will see, our arguments \emph{do not} rely on
spectral flow and we will only consider the classification  of invariants obtained in \cite{Gray:2008je}.} Our task is thus greatly simplified and reduced to identifying both $Z^{c=8/3}_1$ and $Z^{c=8/3}_{2}$ among the list of modular invariant partition functions. 

The highest weight representations for the coset \eqref{eq: N=3 coset} are labeled by $(\Lambda,\Lambda')$ where as before $\Lambda=\0,\vv,\s,\cc$ and $\Lambda'=[\lambda_1,\lambda_2]$ with $\lambda_1+\lambda_2\leq 6$. The simple current is ${\cal J}={\cal O}_{\0,[6,0]}$ and it acts on the $SU(3)_6$ label following \eqref{Jaction}. Making use of \eqref{eq: SU(N) dimensions}, one easily finds that the vanishing monodromy condition becomes
\be
\lambda_1-\lambda_2 = 0 \mod 3\ed
\ee
It also follows that the structure of long and short orbits read 
\be\begin{array}{lcl}
R_L&=&\left\lbrace (\Lambda,\Lambda')\ec~\Lambda' \in\left\lbrace [0,0],[6,0],[0,6] \right\rbrace , \left\lbrace [3,0],[3,3],[0,3] \right\rbrace,
\left\lbrace [1,1],[4,1],[1,4] \right\rbrace
\right\rbrace\ec\\
R_S&=&\left\lbrace (\Lambda,\Lambda') \, \, , \,\, \Lambda' =[2,2] \right\rbrace\ed
\end{array}\ee
As opposed to the $\cT_{SU(2), \rm{IR}}$ theory, explicit expressions for the branching functions of $\cT_{SU(3), \rm{IR}}$ are not available in the literature. One can however write formal expressions for modular invariants in terms of ${\cal N}=(2,2)$ superconformal algebra characters. Some aspects of the theory $\cT_{SU(3), \rm{IR}}$ were also studied using this point of view in  \cite{Gopakumar:2012gd,Isachenkov:2014zua}. In order to determine the partition $Z_{\cT_{SU(3),\rm{IR}}}$ we identified the following list of branching functions:
\bea\label{eq: N=3 branching functions}
&b_{\0,[0,0]} =
\frac{1}{2}\left( \rm{ch}^{(\textrm{NS})}_{0} + \widetilde{ch}^{(\textrm{NS})}_{0}\right)\,,\hspace{1cm} b_{\s,[0,0]} = \frac{1}{2}\left( \rm{ch}^{(\textrm{R})}_{1} + \widetilde{ch}^{(\textrm{R})}_{1} \right)\ec \\
&b_{\0,[1,1]} = \frac{1}{2}\left( \rm{ch}^{(\textrm{NS})}_{\frac{1}{6}} + \widetilde{ch}^{(\textrm{NS})}_{\frac{1}{6}}\right)\,,\hspace{1cm} b_{\s,[1,1]} = \frac{1}{2}\left( \rm{ch}^{(\textrm{R})}_{\frac{2}{3}}+ \widetilde{ch}^{(\textrm{R})}_{\frac{2}{3}} \right)\ec\\
&b_{\0,[3,0]} = \frac{1}{2}\left( \rm{ch}^{(\textrm{NS})}_{\frac{1}{3}} + \widetilde{ch}^{(\textrm{NS})}_{\frac{1}{3}}\right)\,,\hspace{1.01cm} b_{\s,[3,0]} = \frac{1}{2}\left( \rm{ch}^{(\textrm{R})}_{\frac{1}{3}} + \widetilde{ch}^{(\textrm{R})}_{\frac{1}{3}} \right)\ec\\
&b_{\0,[2,2]_1} = \frac{1}{2}\left( \rm{ch}^{(\textrm{NS})}_{\frac{1}{9}} + \widetilde{ch}^{(\textrm{NS})}_{\frac{1}{9}}\right)\,,\hspace{0.87cm} b_{\s,[2,2]_1} = \frac{1}{2}\left( \rm{ch}^{(\textrm{R})}_{\frac{1}{9}}+ \widetilde{ch}^{(\textrm{R})}_{\frac{1}{9}} \right)\ec\\
&b_{\0,[2,2]_2} = \frac{1}{2}\left( \rm{ch}^{(\textrm{NS})}_{\frac{11}{18}} + \widetilde{ch}^{(\textrm{NS})}_{\frac{11}{18}}\right)\,,\hspace{0.87cm} b_{\s,[2,2]_2} = \frac{1}{2}\left( \rm{ch}^{(\textrm{R})}_{\frac{1}{9}'}+ \widetilde{ch}^{(\textrm{R})}_{\frac{1}{9}'}  \right)\ec\\
&b_{\vv,[0,0]} =
\frac{1}{2}\left( \rm{ch}^{(\textrm{NS})}_{0} - \widetilde{ch}^{(\textrm{NS})}_{0}\right)\,,\hspace{1cm} b_{\cc,[0,0]} = \frac{1}{2}\left( \rm{ch}^{(\textrm{R})}_{1} - \widetilde{ch}^{(\textrm{R})}_{1} \right)\ec \\
&b_{\vv,[1,1]} = \frac{1}{2}\left( \rm{ch}^{(\textrm{NS})}_{\frac{1}{6}} - \widetilde{ch}^{(\textrm{NS})}_{\frac{1}{6}}\right)\,,\hspace{1.015cm} b_{\cc,[1,1]} = \frac{1}{2}\left( \rm{ch}^{(\textrm{R})}_{\frac{2}{3}}- \widetilde{ch}^{(\textrm{R})}_{\frac{2}{3}} \right)\ec\\
&b_{\vv,[3,0]} = \frac{1}{2}\left( \rm{ch}^{(\textrm{NS})}_{\frac{1}{3}} - \widetilde{ch}^{(\textrm{NS})}_{\frac{1}{3}}\right)\,,\hspace{1.cm} b_{\cc,[3,0]} = \frac{1}{2}\left( \rm{ch}^{(\textrm{R})}_{\frac{1}{3}} - \widetilde{ch}^{(\textrm{R})}_{\frac{1}{3}} \right)\ec\\
&b_{\vv,[2,2]_1} = \frac{1}{2}\left( \rm{ch}^{(\textrm{NS})}_{\frac{1}{9}} - \widetilde{ch}^{(\textrm{NS})}_{\frac{1}{9}}\right)\,,\hspace{0.87cm} b_{\cc,[2,2]_1} = \frac{1}{2}\left( \rm{ch}^{(\textrm{R})}_{\frac{1}{9}}- \widetilde{ch}^{(\textrm{R})}_{\frac{1}{9}} \right)\ec\\
&b_{\vv,[2,2]_2} = \frac{1}{2}\left( \rm{ch}^{(\textrm{NS})}_{\frac{11}{18}} - \widetilde{ch}^{(\textrm{NS})}_{\frac{11}{18}}\right)\,,\hspace{0.87cm} b_{\cc,[2,2]_2} = \frac{1}{2}\left( \rm{ch}^{(\textrm{R})}_{\frac{1}{9}'}- \widetilde{ch}^{(\textrm{R})}_{\frac{1}{9}'}  \right)\ec\\
\eea
where $\textrm{ch}_h^{(\textrm{NS})}$, $\textrm{ch}_h^{(\textrm{R})}$ denote the (extended fermionic) characters whose lowest dimensional ${\cal N}=(2,2)$ primary has dimension $h$ and $\widetilde{\textrm{ch}}^{(\textrm{NS})}_h$,$\widetilde{\textrm{ch}}^{(\textrm{R})}_h$ are the same characters with an additional insertion of $(-1)^{\textrm{F}}$. Note that these branching functions were not completely spelled out in the references \cite{Gopakumar:2012gd,Isachenkov:2014zua}. See Appendix \ref{sec: app N=2} for further details.

From the above discussion it follows that the partitions functions for both universes $Z^{c=8/3}_1$ and $Z^{c=8/3}_2$ must be realized in terms modular invariants for the $\cN=(2,2)$ minimal model $\cM_{16}$.  This model is realized by the coset \eqref{susyminimal} whose modular invariants are defined by a matrix $\widetilde{M}$ that can be shown to factorize as
\be
\widetilde M=M_{SU(2)}\times M\ec
\ee
where the matrix $M_{SU(2)}$ corresponds to a modular invariant for the $SU(2)_k$ CFT subject to the ADE classification of \cite{Cappelli:1987xt}.

In \cite{Gopakumar:2012gd}, a modular invariant realized by one of the universes of the theory \eqref{eq: N=3 coset} has been identified with the one denoted as $\widetilde M^{4,2}$ \cite{Gray:2008je}.\footnote{Relevant aspects of the classification \cite{Gray:2008je} are reviewed in Appendix \ref{sec: app N=2}.} The partition function reads
\bea\label{eq: Z M42}
Z_{\widetilde M^{4,2}}  = \frac{1}{2}\Biggl[& \sum\limits_{h=0,\frac{1}{9},\frac{1}{6},\frac{1}{3}}|\rm{ch}^{(\textrm{NS})}_h|^2 +2|\rm{ch}^{(\textrm{NS})}_{\frac{11}{18}}|^2  + \left(|\rm{ch}^{(\textrm{NS})}_h|^2 \rightarrow |\widetilde {\rm{ch}}^{(\textrm{NS})}_h|^2\right)\\
&+\sum\limits_{h= \frac{1}{9},\frac{1}{3},\frac{2}{3},1}|\rm{ch}^{(\textrm{R})}_h|^2 + 2 |\rm{ch}^{(\textrm{R})}_{\frac{1}{9}'}|^2
+ \left(|\rm{ch}^{(\textrm{R})}_h|^2 \rightarrow |\widetilde{ \rm{ch}}^{(\textrm{R})}_h|^2\right)\Biggr]\,.
\eea

By comparison with the general expression \eqref{eq: general Z1 and Z2} and making use of the complete list of branching functions that we proposed in \eqref{eq: N=3 branching functions}  one concludes that 
\be
Z^{c=8/3}_1 = Z_{\widetilde{M}^{4,2}}\ed
\ee
This result can be easily understood from the elementary fact that when the partition function \eqref{eq: Z M42} is expressed in term of the branching functions \eqref{eq: N=3 branching functions} no off-diagonal combinations of (extended) 
$\cW$-algebra characters can appear (see Appendix \ref{sec: app N=2}).

Finally, we are also able to completely determine the partition function in the remaining universes.  From \eqref{eq: general Z1 and Z2}, following similar steps as above, we can in fact single out the modular invariant denoted by $\tilde E^{16}_2$ in \cite{Gray:2008je}: 
\be
Z^{c=8/3}_2 = Z_{\widetilde E^{16}_2}\ec
\ee
where
\bea\label{eq: Z E162}
Z_{\widetilde E^{16}_2} = \frac{1}{2}\Biggl[& \sum\limits_{h=0,\frac{1}{6},\frac{1}{3},\frac{11}{18}}|\rm{ch}^{(\textrm{NS})}_h|^2 +\left(\rm{ch}^{(\textrm{NS})}_{\frac{1}{9}}\overline{\rm{ch}^{(\textrm{NS})}_{\frac{11}{18}}} + h.c. \right) + \left(\rm{ch}^{(\textrm{NS})}_h \rightarrow \widetilde{\rm{ch}}^{(\textrm{NS})}_h\right)\\
&+\sum\limits_{h= \frac{1}{3},\frac{2}{3},\frac{1}{9}',1}|\rm{ch}^{(\textrm{R})}_h|^2 + \left(\rm{ch}^{(\textrm{R})}_{\frac{1}{9}}\overline{\rm{ch}^{(\textrm{R})}_{\frac{1}{9}'}} + h.c. \right)
+ \left(\rm{ch}^{(\textrm{R})}_h \rightarrow \widetilde{ \rm{ch}}^{(\textrm{R})}_h\right)
\Biggr]\ed
\eea
Therefore, the final expression for the partition function of the coset theory $\cT_{SU(3),\textrm{IR}}$ is given by:
\be\label{eq: N=3 partition sum final}
Z_{\cT_{SU(3), \textrm{IR}}}=Z_{\widetilde M^{4,2}}+ 2Z_{\widetilde E^{16}_2}\ed
\ee
Some comments on this result are in order:
\begin{itemize}
 
\item An important consistency check is that  twisted Ramond characters are missing from the partition function $Z_{\cT_{SU(3), \textrm{IR}}}$. In order for this to be consistent with \eqref{eq: N=3 partition sum final} one needs
\be
 \widetilde{\rm ch}^\textrm{R}_{\frac19}+2\widetilde{\rm ch}^\textrm{R}_{\frac{1}{9}'}=0\ec\quad
 \widetilde{\rm ch}^\textrm{R}_{h=\frac13,\frac23,1}=0 \ed
\ee
These are the necessary conditions for $b_{\s,\Lambda'}=b_{\cc,\Lambda'}$ in \eqref{eq: N=3 branching functions}.

\item  $\cT_{SU(3)}$ is a spin theory 
and we expect its IR description to depend on the spin structure. However,  the partition functions \eqref{eq: Z M42} and \eqref{eq: Z E162} obtained after bosonization 
are actually summed over the four torus spin structures (see Appendix \ref{app: bosonization}). It is easy to obtain the spin-dependent partition functions by applying fermionization, these consist of characters in a fixed sector of the $(-1)^{\textrm{F}}$ symmetry on both the spatial and temporal cycle. For instance, the fermionic partition function in the $(\textrm{NS},\textrm{NS})$ sector is
\be
Z_{\cT_{SU(3),\textrm{IR}}}^{(\textrm{NS},\textrm{NS})} = Z_{1}^{(\textrm{NS},\textrm{NS})} + 2Z_{2}^{(\textrm{NS},\textrm{NS})}\ec
\ee
where
\be
\begin{split}
Z_{1}^{(\textrm{NS},\textrm{NS})} &= \sum\limits_{h=0,\frac{1}{9},\frac{1}{6},\frac{1}{3}}|\textrm{ch}^{\textrm{NS}}_h|^2 +2|\rm{ch}^{\textrm{NS}}_{\frac{11}{18}}|^2\ec \\
Z_{2}^{(\textrm{NS},\textrm{NS})} &= \sum\limits_{h=0,\frac{1}{6},\frac{1}{3},\frac{11}{18}}|\textrm{ch}^{\textrm{NS}}_h|^2 +\left(\rm{ch}^{\textrm{NS}}_{\frac{1}{9}}\overline{\rm{ch}^{\textrm{NS}}_{\frac{11}{18}}} + h.c. \right)\ec
\end{split}
\ee
and similarly for all remaining choices of spin structures.

\end{itemize}

To conclude this section we would like to offer some comments on topological line operators. In our analysis of the coset $\cT_{SU(2), \textrm{IR}}$ from section \ref{sec: N=2} it was easy to determine the existence of a finite global symmetry whose gauging could map one universe to the other. Instead, for $\cT_{SU(3), \textrm{IR}}$ it is not at all evident  whether there exist non-invertible symmetries connecting the different universes. These symmetries must arise from a self-duality of the coset WZW theory \eqref{eq: N=3 coset} under gauging of a certain subalgebra of its topological lines. More precisely, in the next susbsection we will show that there is indeed a topological line implementing a self-duality upon gauging a non-invertible subalgebra of the set of Verlinde lines of $\cT_{SU(3), \textrm{IR}}$. The modular invariants of \eqref{eq: N=3 partition sum final} are therefore mapped to each other by those generalized gaugings.

However, it is important to note that the Chern-Simons analysis will only reveal the topological lines that commute with the chiral algebra $\Spin(16)_1/SU(3)_6$. In general, as observed for the $\cT_{SU(2), \rm{IR}}$ theory exhibiting a TY$(\bZ_2)$ symmetry, the theory may possess additional global symmetries that are not manifest in the Chern-Simons/WZW approach.

\subsection{Symmetries from 3d Chern-Simons}\label{CSSU3}
Let us now comment on the 3d Chern-Simons description of $\cT_{SU(3),\rm{IR}}$. The corresponding 3d TQFT is
\be
\Spin(16)_1 \times SU(3)_{-6}\ec
\ee
whose anyons can be parametrized by $(\Lambda,\Lambda')$ where $\Lambda=(\0,\vv,\s,\cc)$ correspond to $\Spin(16)_1$ integrable representations while $\Lambda'=[\lambda_1,\lambda_2]$ with $\lambda_1+\lambda_2 \leq 6$ refers to $SU(3)_{-6}$ integrable representations. Similarly to $\cT_{SU(2),\rm{IR}}$, the data defining the 3d TFT factorize into the ones of $\Spin(16)_1$ and $SU(3)_{-6}$ respectively. In particular we have
\be
S_{(\Lambda_1,\Lambda_1'),(\Lambda_2,\Lambda_2')} = S_{\Lambda_1,\Lambda_2}^{\Spin(16)_1}S_{\Lambda'_1,\Lambda'_2}^{*SU(3)_k} \quad ,\quad \theta_a = \exp{\left( 2\pi i (h_a^{\Spin(16)_1}-h_a^{SU(3)_k})\right)}\ec
\ee
where the generic data of $\Spin(2n)_1$ already appear in section \ref{sec: N=2}, while
\bea
&S_{\Lambda'_1,\Lambda'_2}^{SU(3)_k}  = \frac{-i}{(k+3)\sqrt{3}} \sum\limits_{w\in \text{Weyl}(SU(3))} \det(w) \,\exp{\left( -2\pi i \frac{w(\Lambda'_1+\rho)\cdot (\Lambda_2'+\rho)}{k+3}\right)}\ec\\
& h^{SU(3)_6}_{\lambda}=\frac{\Lambda \cdot (\Lambda+2\rho)}{2(k+3)}\,,
\eea
where $\rho=[1,1]$ is the Weyl vector of $SU(3)$.

One can show that the anyons
\be\label{eq: anyons Z3}
(\0; [0,0])\,,\,(\0;[6,0])\,,\,(\0;[0,6])
\ee
generate a $\bZ_3$ symmetry and have spin zero. They generate a gaugeable algebra and (after imposing Dirchlet boundary conditions) their topological endpoints implement a boundary $\bZ\one_3$ 1-form symmetry, as expected.

Since we imposed Dirichlet boundary conditions on the anyons \eqref{eq: anyons Z3}, the following boundary identification holds:
\be
(\Lambda;[\lambda_1,\lambda_2]) \sim (\Lambda;[\lambda_1,\lambda_2])\times(\0;[6,0]) \sim  (\Lambda;[\lambda_1,\lambda_2])\times(\0;[0,6])\,.
\ee
Evidently, the $\Spin(16)_1$ label is transparent to this identification while the $SU(3)_{-6}$ representations form orbits 
  \bea
&[0,0]\sim [6,0] \sim [0,6 ]\quad,\quad [0,3] \sim [3,0] \sim [3,3]\\
&[1,1] \sim [4,1] \sim[1,4 ]\quad,\quad [2,2] \,\,\,\text{ is a fixed point,}
 \eea
that can be derived using \eqref{eq: Verlinde}. Lines kept fixed by the identification split into a sum of three simple lines, i.e.
 \be
( \Lambda; [2,2]) = ( \Lambda, [2,2])_1+( \Lambda; [2,2])_2+( \Lambda; [2,2])_3\,.
 \ee
 After this analysis we can discuss the structure of boundary line operators coming from the CS description.
\begin{itemize}
\item We find $24$ boundary lines which do not braid with the one-form symmetry and therefore they are unbroken symmetries acting on each universe. A set of representatives is
\be\label{eq: N=3 Verlinde}
    (\Lambda;[0,0]) \quad,\quad
    (\Lambda;[0,3])    \quad,\quad
    (\Lambda;[1,1])   \quad,\quad
    (\Lambda;[2,2])_{1,2,3}   \,,
\ee
with $\Lambda = \0,\vv,\s,\cc$. These lines are generically non-invertible while the subset spanned by $(\Lambda;[0,0])$ generates the only invertible $\bZ_2 \times \bZ_2$ sub-algebra associated to the center of $\Spin(16)$.  The full fusion algebra can be easily computed from \eqref{eq: Verlinde}.

We can also compute the spin of these lines and therefore the holomorphic dimension of the corresponding character. We find
\bea
& (h_{(\Lambda;0,0)})_{\Lambda=\0,\cdot,\cc }= \left(0,\frac{1}{2},0,0\right) \quad,\quad (h_{(\Lambda;0,3)  })_{\Lambda=\0,\cdot,\cc } = \left(\frac{1}{3},\frac{5}{6},\frac{1}{3},\frac{1}{3}\right)\\
&(h_{(\Lambda;1,1)})_{\Lambda=\0,\cdot,\cc}  = \left(\frac{2}{3},\frac{1}{6},\frac{2}{3},\frac{2}{3}\right) \quad,\quad (h_{(\Lambda;2,2)_i})_{\Lambda=\0,\cdot,\cc}  = \left(\frac{1}{9},\frac{11}{18},\frac{1}{9},\frac{1}{9}\right) \quad\forall i \,,
\eea
in agreement with the boundary analysis in \eqref{eq: N=3 branching functions}. Therefore these lines are in one to one correspondence with the primary operators of the CFTs describing the three universes in \eqref{eq:N=3 partition sum}.

\item There are $24$ boundary lines that braid with the one-form symmetry. A set of representatives is
\bea\label{eq: N=3 non-invertible lines}
&(\Lambda; [0, 1] )\;\,\qquad\qquad,\qquad \;\, (\Lambda; [0, 2] ) \;\,\qquad\qquad,\qquad \;\,  (\Lambda; [1, 2] )\\
&(\Lambda;[ 0, 5] )\equiv(\Lambda^*; [0, 1] )^{\dagger} \quad,\quad (\Lambda; [0, 4] )\equiv(\Lambda^*; [0, 2] )^{\dagger}     \quad , \quad (\Lambda; [1, 3] )\equiv(\Lambda^*; [1, 2] )^{\dagger}   
\eea
with $\Lambda = \0,\vv,\s,\cc$ and $\Lambda^*$ is the conjugate representation of $\Lambda$. We denote the orientation reversal of the topological line $(\Lambda,\Lambda')$ as $(\Lambda,\Lambda')^\dagger$. All of them obey non invertible fusion rules, signaling the fact that the three universes are not all described by the same CFT.\footnote{By looking at the fusion algebra, we can show that all these lines implement a gauging of a non invertible symmetry. However, as in the case of $\cT_{SU(2), \rm{IR}}$, the Chern-Simons picture can be blind to more hidden self-dualities of the model that can make this gauging easier.} 
\end{itemize}

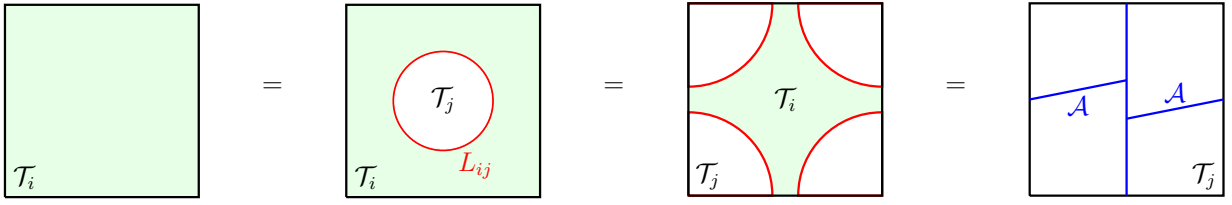
\begin{figure}[t]
$$
\scalebox{0.85}{
   \raisebox{-4 em}{ \begin{tikzpicture}
		
		\filldraw[white!90!green] (0,0)--(0,3)--(3,3)--(3,0)--(0,0);
		 \node[black] at (0.3,0.3) {$\cT_{i}$};
\draw[line width=1] (0,0)--(0,3)--(3,3)--(3,0)--(0,0);
   \end{tikzpicture}}
   \qquad =\qquad
   \raisebox{-4. em}{  \begin{tikzpicture}
		
		\filldraw[white!90!green] (0,0)--(0,3)--(3,3)--(3,0)--(0,0);
		 \node[black] at (0.3,0.3) {$\cT_{i}$};
		  \draw[line width = 2.,red] (1.5,1.5) circle [radius=0.75];
		   \filldraw[white] (1.5,1.5) circle [radius=0.75];
		   \node[black] at (1.5,1.5) {$\cT_{j}$};
		   \node[red] at (2,0.5) {$L_{ij}$};
\draw[line width=1] (0,0)--(0,3)--(3,3)--(3,0)--(0,0);

		   \end{tikzpicture}}
		   
    \qquad =\qquad      \raisebox{-4. em}{  \begin{tikzpicture}

                \filldraw[line width=1,white!90!green] (0,0)--(0,3)--(3,3)--(3,0)--(0,0);
     		\draw[color=red,line width=1,fill=white] (1.7,3.) arc (180: 270: 1.3)--(3,3)-- cycle;
		  \draw[color=red,line width=1,fill=white] (0,1.7) arc (270: 360: 1.3)--(0,3)-- cycle;
		  \draw[color=red,line width=1,fill=white] (1.3,0) arc (0: 90: 1.3)--(0,0)-- cycle;
		   \draw[color=red,line width=1,fill=white] (3,1.3) arc (90: 180: 1.3)--(3,0)-- cycle;
    
    		\draw[line width=1] (0,0)--(0,3)--(3,3)--(3,0)--(0,0);

		    \node[black] at (1.5,1.5) {$\cT_{i}$};
		     \node[black] at (0.3,0.3) {$\cT_{j}$};

			        \end{tikzpicture}}

 \qquad =\qquad 
     \raisebox{-4 em}{  \begin{tikzpicture}
		\draw[line width=1] (0,0)--(0,3)--(3,3)--(3,0)--(0,0);
		\draw[line width=1,blue] (1.5,0)--(1.5,3);
	        \draw[line width=1,blue] (3,1.5)--(1.5,1.5-0.3) ;
	        \draw[line width=1,blue] (0,1.5)--(1.5,1.5+0.3);
	         \node[blue] at (0.75,1.4) {$\cA$};
	          \node[blue] at (1.5+0.75,1.6) {$\cA$};
	            \node[black] at (2.7,0.3) {$\cT_{j}$};
	          \end{tikzpicture}}

}
\hspace{1cm}
$$
\caption{Topological manipulations which realize the universes $\cT_j$ as a gauging of the universe $\cT_i$. Here $L_{ij}$ is a generic topological line with a mixed anomaly with the 1-form symmetry, thus serving as a domain wall between universes $i$ and $j$. It is such that $L_{ij}\times L_{ij}^{\dagger} = \cA$ where $L_{ij}^{\dagger}$ is its orientation reversal and $\cA$ is a non-simple topological line. Factors involving quantum dimensions of the lines involved have been omitted in the figure.
\label{fig: gauging procedure}}
\end{figure}
The latter set of lines, representing spontaneously broken symmetries of $\cT_{SU(3)}$, serve as domain walls between distinct universes. Due to the $\mathbb{Z}_3$ 1-form symmetry, reversing the orientation of one line does not produce the same symmetry action. When these lines are fused with their orientation reversals, they produce non-simple lines $\cA_i$ made out of the set \eqref{eq: N=3 Verlinde}, i.e. Verlinde lines preserved in all three universes. In particular, by using \eqref{eq: Verlinde}, we get
\bea\label{eq: N=3 noninv}
& (\Lambda, [0,1])\times  (\Lambda^{*},[0,5]) = (\0,[0,0]) + (\0,[1,1]) \equiv \cA_1\\
&(\Lambda,[0,2])\times  (\Lambda^{*},[0,4]) = (\0,[0,0]) + (\0,[1,1]) + \sum_i (\0,[ 2, 2])_i\equiv \cA_2\\
&(\Lambda, [1,2])\times  (\Lambda^{*},[1,3]) = (\0,[0,0]) + 3(\0,[0,3]) + 4(\0,[1,1]) + 2\sum_i (\0,[2, 2])_i\equiv \cA_3
\eea
This fusion algebra can be used to understand which operation one can perform to pass from one universe to the other. Following \cite{Diatlyk:2023fwf}, we can interpret  the lines in \eqref{eq: N=3 non-invertible lines} as domain walls separating theories related by the gauging of $\cA_i$.  This can be shown by performing several topological manipulation on the torus partition function as in figure \ref{fig: gauging procedure}. In particular, these arguments imply that non-simple lines $\cA_i$, obtained by fusion a topological defect with its orientation reversal, form a gauge-able algebra.\footnote{Notice that just giving the object $\cA_i$ is not enough to specify an algebra since we need to specify morphisms $m:\cA \times \cA \rightarrow \cA$. However as shown in \cite{Diatlyk:2023fwf} , these morphisms are completely determined by the fusion algebra \eqref{eq: N=3 noninv}. In the following we will not get into the determination of this extra datum.}

We observe that there are (at least) three different algebras connecting the universes. This implies the existence of several self-dualities under which a given universe remains invariant. Notice that a similar conclusion was reached in section \ref{sec: N=2}, where we identified two different gaugings (i.e., a $\bZ_2$ or a $\bZ_3$ gauging) connecting the two universes.

\subsection{Comments on Mass and Quartic Deformations}\label{sec: N=3 mass def}
Let us now discuss the role of relevant deformations in $\cT_{SU(3)}$. In addition to the two independent mass operators $\cO_{1,2}$ introduced in section \ref{sec: mass def} here we will also consider the effects of quartic fermion interactions.

Recall that the Dirac mass term $\cO_1$ and $\cO_2$ have quantum numbers $(Q_A=\pm 2, Q_V=0)$ and $(Q_A=0, Q_V=\pm 2)$ under $U(1)_A\times U(1)_V$. These are identified with 
the following primaries of the theory $\cT_{SU(3), \rm{IR}}$ belonging to the NS sector:\footnote{In principle there is a more relevant operator in the R sector which however does not map to any quadratic operator in the fermionic theory.}
\be\label{eq: IR mass ops}
{\cal O}_{6,\pm 6}^{\rm{NS}}\ec\quad h_{6,\pm 6}=\frac16 \ec\quad Q_{6,\pm6}=\pm \frac{1}{3} \ec
\ee
where $h_{6, \pm 6}$ and $Q_{6,\pm 6}$ denote respectively the conformal dimension and $U(1)_R$ charge of the primaries. See Appendix \ref{sec: app N=2} for more details on the $\cN=(2,2)$ superconformal algebra conventions that we are adopting.

The observation that all mass terms have degenerate dimensions was previously noted in \cite{Gopakumar:2012gd}. This phenomenon arises because the operators in \eqref{eq: IR mass ops} belong to an extended $\cW$-symmetry multiplet whose (extended) character is ${\rm ch}^{\rm{NS}}_{1/6}$. Consequently, all four possible pairings are present in the partition function of the three universes, as seen in \eqref{eq: Z M42} and \eqref{eq: Z E162}. Therefore, mass deformations map to non-trivial IR operators in all universes. This contrasts sharply with the situation at $N=2$ discussed in section \ref{sec: mass def}.

Since all the masses correspond to the same extended character, the Verlinde lines of the Chern-Simons theory discussed in Section \ref{CSSU3} are unable to distinguish between them since they commute with the $\cW$-algebra. Consequently, these symmetries do not allow us to differentiate between Dirac and Majorana mass operators. In line with this, a straightforward calculation shows that the primary operator associated with ${\rm ch}^{\rm{NS}}_{1/6}$ is charged under the non-invertible symmetries that share a mixed anomaly with the $\bZ\one_3$ 1-form symmetry, consistent with a confining scenario. However, as discussed in Section \ref{sec: N=2}, we anticipate the presence of an additional global symmetry that distinguishes between the masses of Majorana and Dirac fermions. A natural candidate is the triality symmetry constructed by gauging a $\bZ_2\times \bZ_2$ symmetry with discrete torsion, known to appear in adjoint QCD$_2$ with a single Majorana \cite{Komargodski:2020mxz}. This is a manifest global symmetry of the UV theory of free fermions which is preserved under $SU(3)$ gauging. Given that our UV theory consists of two decoupled copies of the same model of free fermions, an identical symmetry should act diagonally on the two Majorana fermions and commute with the diagonal $SU(3)$ gauging. It would be interesting, however, to explicitly verify the presence of such a triality symmetry in $\cT_{SU(3), \rm{IR}}$ and analyze its phase diagram upon massive deformation, we leave this investigation for future work.

As in the discussion from Section \ref{sec:tension}, assuming that $\cT_{SU(3)}$ confines for $m_1 \neq m_2$ (and deconfines for $m_1=m_2$), we can still deduce the behavior of the string tension as a function of $m_{1,2}$ and $g_{\text{YM}}$ in the regime where $m_{1,2} \ll g_{\text{YM}}$. In this regime, the string tension can be derived from the vacuum energy density by deforming the IR CFT $\mathcal{T}_{SU(3),\textrm{IR}}$ with the primary operator corresponding to the two independent Dirac masses. Since these masses have a conformal dimension of $\Delta = 1/3$ in the IR, dimensional analysis shows that 
\be
\mu_{1,2} = m_{1,2} \Lambda^{2/3}\,,
\ee
leading to a string tension scaling as
\be
T_k \propto \Lambda^{4/5} \left(m_1^{6/5} - m_2^{6/5}\right)\,,
\ee
exactly as observed for $\cT_{SU(2)}$.\footnote{Interestingly, the IR scaling dimension of the mass operators remains $\Delta = 1/3$, independent of $N$ \cite{Gopakumar:2012gd}.} Unfortunately, the unknown proportionality constant prevents us from determining the precise behavior of the tension as a function of $N$ and $k$.

Finally, let us analyze the effect of quartic fermion interactions. The theory $\cT_{SU(3)}$ has quartic operators of the form $\Tr(\Psi_L^{\dagger}\Psi_L\Psi_R^{\dagger}\Psi_R)$ (with all possible index contractions) whose quantum numbers under $U(1)_A\times U(1)_V$ are $Q_A = Q_V = 0$.  The most relevant operator of this form has IR scaling dimension $\Delta = \frac{2}{9}$.\footnote{This operator exists for any values of $N\geq 3$ and has IR scaling dimension $\Delta = \frac{2(N-2)}{3N}$. For $N=2$, there is a unique double-trace quartic operator which is invariant under all symmetries of the theory.} The work \cite{Gopakumar:2012gd} raised concerns that such deformations might potentially destabilize the IR fixed point. However, a straightforward outcome of our analysis is that the fate of these deformations can be determined using a symmetry-based argument.  The quartic operator appears in the expansion of $|\rm \rm{ch}^{\rm{NS}}_{1/9}|^2$ which is present in the universes $\widetilde{M}^{4,2}$ but not in $\widetilde{E}_{2}^{16}$. This implies that the operator is charged under the non-invertible symmetries of the theory and therefore not generated along the RG flow.  This provides another clear example of how non-invertible symmetries can violate the conventional notion of naturalness (for further examples, see \cite{Chang:2018iay,Thorngren:2021yso,Komargodski:2020mxz}).\footnote{For $N \geq 3$, we expect that adding the quartic operator to the Lagrangian will restore confinement. Conversely, for $N = 2$, adding the double-trace operator will leave the theory in a deconfined phase. For a related discussion in $SU(N)$ adjoint QCD$_2$ with $N_f=1$, see \cite{Cherman:2024onj}.}

\section*{Acknowledgments}

We are grateful to A. Antinucci, R. Argurio, A. Armoni, D. Delmastro, D. Hofman, Z. Komargodski, G. Rizi, S. Seifnashri and Y. Wang for discussions and comments. J.A.D.~and L.T.~are Postdoctoral Researchers of the F.R.S.-FNRS (Belgium).~The research of J.A.D.~, G.G.~and L.T.~is funded through an ARC advanced project, and further supported by IISN-Belgium (convention 4.4503.15).

\appendix

\section{Details on Bosonization}\label{app: bosonization}
In this appendix, we review 2d bosonization, focusing specifically on fermionic CFTs and theories of free Majorana fermions. Every fermionic theory possesses a $\mathbb{Z}_2$ symmetry known as $(-1)^{\textrm{F}}$, which assigns a value of $-1$ to fermionic states. On a spin manifold, fermionic states gain a plus or minus sign under a $2\pi$ rotation around non-contractible cycles, depending on the choice of spin structures. In particular, on a torus, antiperiodic fermions belong to what is commonly referred to as the Neveu-Schwarz (NS) sector, whereas periodic fermions belong to the Ramond (R) sector. In the context of CFTs, the characters corresponding to these sectors are denoted by $\text{ch}^{\textrm{NS}}$ and $\text{ch}^{\textrm{R}}$. Importantly, the subsector represented by $\mathbb{P}\left(\text{ch}^{\textrm{NS}/\textrm{R}}\right)$, where $\mathbb{P} = \frac12(1 + (-1)^{\textrm{F}})$, encompasses the bosonic states. These states are commonly referred to as $(\textrm{NS}/\textrm{R}, \textrm{R})$ states, in contrast to fermionic states, which are often labeled as $(\textrm{NS}/\textrm{R}, \textrm{NS})$ states.

Gauging $(-1)^{\textrm{F}}$ projects out fermionic states from the NS sector and includes the invariant states from the twisted (R) sector. This procedure is equivalent to summing over the spin structures and is commonly referred to as the GSO projection. The resulting bosonized (gauged) partition function can be expressed as
\be
Z^+_{\textrm{B}} = \frac{1}{2} \left( Z_F^{(\textrm{NS,NS})}+Z_F^{(\textrm{NS,R})}+Z_F^{(\textrm{R,NS})}+Z_F^{(\textrm{R,R})}\right)\ec
\ee
where $Z_F^{(X,Y)}$ is the fermionic torus partition function with states in the $(X,Y)$ sector. Note that the same partition function can also be interpreted as a sum over spin structures, as the insertion of $(-1)^{\textrm{F}}$ defects changes the spin structure of the manifold.\footnote{More precisely, the set of spin structures defines a torsor over $H^1(X,\bZ_2)$, parameterizing the set of background gauge fields for $(-1)^{\textrm{F}}$.}

We can further extend the gauging of $(-1)^{\textrm{F}}$ by incorporating a discrete torsion, parameterized by the Arf invariant. The Arf invariant acts as a topological phase, denoted as $(-1)^{\text{Arf}[s]}$, assigning a phase of $(-1)$ to odd spin structures $s$, i.e. to (R,R) states.
This produces the bosonic partition function
\be
Z^-_{\textrm{B}} =  \frac{1}{2} \left( Z_F^{(\textrm{NS,NS})}+Z_F^{(\textrm{NS,R})}+Z_F^{(\textrm{R,NS})}-Z_F^{(\textrm{R,R})}\right)\,.
\ee
Let us apply this general construction to $2N$ free Majorana fermions. The torus partition function of the fermionic theory factorizes as 
\be
Z_{F}^{(X,Y)} = |d_{X,Y}|^2\,,
\ee
where $d_{\textrm{X,Y}}$ is a single fermionc character that depends on the choice of the spin structure parametrized by X,Y $=$ NS,R. The non-Abelian bosonization map implies
\be
d_{\textrm{NS,Y}} = \chi_0 \pm \chi_v\ec \qquad d_{\textrm{R,Y}} = \chi_s \pm \chi_c \ec
\ee
where $\chi_i$ are the $\mathfrak{so}(2N)_1$ characters. Therefore, the two bosonized theories are
\be
Z^\pm_{\textrm{B}} = \frac{1}{2}\left(|d_{\textrm{NS,NS}}|^2+|d_{\textrm{NS,R}}|^2+|d_{\textrm{R,NS}}|^2\pm|d_{\textrm{R,R}}|^2\right)\,.
\ee
We then find
\bea
&Z^+_{\textrm{B}} = |\chi_0|^2+|\chi_v|^2+|\chi_s|^2+|\chi_c|^2\ec\\
&Z^-_{\textrm{B}} = |\chi_0|^2+|\chi_v|^2+\chi_s\overline{\chi_c}+\chi_c\overline{\chi_s}\ed
\eea
We recognize $Z^+_{\textrm{B}}$ as the diagonal $\Spin(2N)_1$ WZW model while $Z^-_{\textrm{B}}$ is one of the non-diagonal models. Notice however that, due to the $\bZ_2$  outer automorphism of the $\mathfrak{so}(2N)$ algebra, in the absence of fugacities $\chi_s = \chi_c$ implying that $Z^+_{\textrm{B}}(\tau)=Z^-_{\textrm{B}}(\tau)$ as torus partition functions. This implies the existence of a non-invertible duality symmetry in the bosonic theory, which we identify with the invertible $(-1)^{C_L}$ symmetry of the fermionic theory.

\section{GKO Construction, Field Identification and Fixed Points}\label{app: coset}

This section outlines the salient features of $G/H$ coset WZW theories. All the fundamental aspects of this construction can be found in \cite{DiFrancesco:1997nk} (see also \cite{EberhardtWZW} for a recent review). For the subtleties involving identification currents, we refer the reader to the original works \cite{Schellekens:1989uf, Fuchs:1995tq, Schellekens:1989am, Schellekens:1990xy}.

Given a simply connected group $G$ and an integer $k$, there exists a WZW CFT denoted by $G_k$. The Kac-Moody algebra $\mathfrak{g}_k$ is generated by the current $J^a$, with $a$ an adjoint index in the Lie algebra $\mathfrak{g}$, and OPE
\be \label{eq: current algebra}
J^a(z)J^b(w)\sim \frac12 \frac{k}{(z-w)^2}+\frac{f^g_{abc}J^c(w)}{z-w}\ed
\ee
The WZW model for $G_k$ is a rational CFT, with its Hilbert space decomposing into a finite set of representations of the current algebra generated by ${J^a}$. Each representation is labeled by an integrable highest weight $\Lambda$, which is an array of fundamental weights $[\lambda_1, \ldots, \lambda_r]$, where $r = \text{rank}(\mathfrak{g})$, subject to $\sum_{i} \lambda_i \leq k$. The state-operator correspondence assigns a chiral primary ${\cal O}_\Lambda$ to each highest weight state $|\Lambda\rangle$ in the module ${\cal H}_\Lambda$. The rest of the module is generated by the enveloping algebra $U(\mathfrak{g}_k)$ formed by the negative modes of the currents.

Through the Sugawara construction, a Virasoro algebra with central charge naturally embeds within the chiral algebra $\mathfrak{g}_k$, given by:
\be
c(\mathfrak{g}_k)=\frac{k \, {\rm dim}(\mathfrak{g})}{k+h^g} \,,
\ee
where $h^g$ is the dual Coxeter number associated to the Lie algebra $\mathfrak{g}$.

Consider a normal subgroup $H$ of $G$. The group embedding $H\subset G$ induces an embedding of their associated chiral algebras $\mathfrak{h}_{k'}\subset \mathfrak{g}_{k}$ with $k'=I(G,H)k$ and $I(G,H)$ the embedding index. For the class of theories \eqref{eq: so coset} considered in this work, $I(G,H)=2N$. In particular, the Virasoro generators of the coset theory $G_k/H_{k'}$ must belong to the commutant of $U(\mathfrak{h}_{k'})$ within $U(\mathfrak{g}_{k})$. A natural choice is $L_n=L_n^g-L_n^h$, with $L_n^g$ the Virasoro generators constructed from the currents \eqref{eq: current algebra} and analogously for $L_n^h$. The resulting central charge reads
\be\label{eq:general coset central charge}
c=c(\mathfrak{g}_k)-c(\mathfrak{h}_{k'})\,.
\ee 
In this context, a conformal embedding occurs when $c(\mathfrak{g}_k)=c(\mathfrak{h}_{k'})$ and the associated coset WZW theory becomes topological. This is the class of theories arising in the infrared of adjoint QCD$_2$ with $N_f=1$ \cite{Komargodski:2020mxz}.  

The embedding of the chiral algebras induces a natural decomposition of their respective representation spaces. Specifically, given ${\cal H}_\Lambda$ and ${\cal H}_{\Lambda'}$ the modules of $\mathfrak{g}_k$ and $\mathfrak{h}_{k'}$ respectively, one has ${\cal H}_\Lambda=\sum_{\Lambda'} {\cal H}_{\Lambda,\Lambda'}\otimes {\cal H}_{\Lambda'}$. Therefore, highest weight states, hence chiral primary operators, in the coset theory are labeled by pairs $(\Lambda,\Lambda')$ of highest weight representations of $\mathfrak{g}_k$ and $\mathfrak{h}_{k'}$. When considering the theory on a torus with complex structure $\tau$ this leads to the following character decomposition
\be \label{eq:character decomp}
\chi_{\Lambda}(\tau)=\sum_{\lambda'}b_{\Lambda,\Lambda'}(\tau)\chi_{\Lambda'}(\tau)\,,
\ee
where $b_{\Lambda,\Lambda'}(\tau)$ are called branching functions. The latter inherit their modular properties from the characters of the $G_k$ and $H_{k'}$ theories, making them the natural candidates for the conformal blocks of the coset $G_k/H_{k'}$ theory. In particular, the defining equation \eqref{eq:character decomp} instructs that the modular $S$-matrix acting on $b_{\Lambda,\Lambda'}$ reads
\be\label{eq:coset S}
S_{\Lambda_1,\Lambda'_1;\Lambda_2,\Lambda'_2}=S^g_{\Lambda_1,\Lambda_2}(S^h_{\Lambda'_1,\Lambda'_2})^*\,.
\ee
The branching functions can be assembled into a modular invariant in order to construct partition functions for the coset theory. Given a pair of modular invariant matrices $M_{(\Lambda_1,\Lambda_2)}$ and $M_{(\Lambda'_1,\Lambda'_2)}$ for $G_k$ and $H_{k'}$ respectively, a modular invariant under the action of \eqref{eq:coset S} is given by the tensor product
\be\label{eq: GKO matrix}
M_{\Lambda_1,\Lambda'_1;\Lambda_2,\Lambda'_2}\equiv M_{\Lambda_1,\Lambda_2}M_{\Lambda'_1,\Lambda'_2}
\ee
and the partition function is obtained as
\be\label{eq: GKO partition function}
Z^{coset}=\sum_{(\Lambda_1,\Lambda'_1),(\Lambda_2,\Lambda'_2)} M_{\Lambda_1,\Lambda_2;\Lambda'_1,\Lambda'_2}b_{\Lambda_1,\Lambda'_1}\overline{b_{\Lambda_2,\Lambda'_2}}
\ee 
The procedure just outlined goes under the name of GKO construction \cite{Goddard:1984vk}. For the case of conformal embeddings, the branching functions are just positive integer numbers and the GKO partition function is a constant, consistently with the coset theory being a TQFT.

However, modular invariance is not a sufficient condition to define a consistent RCFT. In addition, the S-matrix must be unitary and lead to non-negative fusion coefficients by means of the Verlinde formula. Violations to these conditions arise due to additional selection rules satisfied by the branching functions. In general, these selection rules are originated by the presence of {\it indentifiation currents} and systematic procedures to resolve these degeneracies have been developed in the literature, going under the name of {\it field identification} and   {\it fixed point resolution} \cite{Schellekens:1989uf}. As it turns out, these subtleties are in one to one correspondence with the presence of 1-form symmetries, hence being of key relevance for our purposes. 

\paragraph{Selection rules from simple currents.} A distinguished subset of primaries are the {\it simple currents} \cite{Intriligator:1989zw}. By definition, a simple current whose fusion with any other primary is of the form
\be\label{eq: simple current def}
{\cal J}(z){\cal O}(w)\sim \frac{{\cal O}'(w)}{(z-w)^{t_{\cal O}}}+ \, {\rm regular}\ec\quad t_{\cal O}=h_{\cal O}-h_{{\cal O}'}+h_J\ed
\ee  
The value of $t_{\cal O}$ mod 1 is called the {\it monodromy} of a given operator with respect to the simple current ${\cal J}$, and we will henceforth use the same notation $t_{\cal O}$ to refer to the monodromy. In addition, the monodromy preserves the fusion algebra, {\it i.e.} if ${\cal O}_k\subset {\cal O}_i\times {\cal O}_j$ then $t_{{\cal O}_k}=t_{{\cal O}_i}+t_{{\cal O}_j}$.  
For a given WZW theory $G_k$, the monodromy values are in one-to-one correspondence with elements of the center $Z(G)$. Therefore, representations of the Kac-Moody algebra $\mathfrak{g}_k$ organize themselves into conjugacy classes graded by $Z(G)$. 

Since $G_k$ is a rational CFT, the action \eqref{eq: simple current def} of simple currents is necessarily of finite order, there exists a positive integer $N$ such that ${\cal J}^N\sim 1$. For the WZW theory $G_k$, simple currents are nevertheless not associated with generators of any global symmetry. 

This situation may change drastically when considering a coset WZW theory $G_k/H_{k'}$. If a simple current ${\cal J}$ is present in such a theory, it will be determined by certain combinations of simple currents in $G_k$ and $H_{k'}$. More precisely, its action over the coset primaries ${\cal O}$ reads
\be
{\cal J}(z){\cal O}_{\Lambda,\Lambda'}(w)\sim \frac{{\cal O}_{{\cal J}_G\cdot\Lambda,{\cal J}_H\cdot\Lambda'}}{(z-w)^{t_{\Lambda,\Lambda'}}} + {\rm regular}\ed
\ee

As it turns out \cite{Schellekens:1990xy}, a simple current in a coset theory satisfies $h_{\cal J}=0$ and it is usually referred as the {\it identification current}. Regarding $G_k/H_{k'}$ as a gauge theory with gauge group $H$, the simple current ${\cal J}$ becomes a topological operator generating a 1-form symmetry $\Gamma^{(1)}\subset Z(H)$. For the class of theories considered in this article, ${\cal J}$ has trivial component in $G_{k}={\rm Spin}(2(N^2-1))_1$ and generates a 1-form symmetry $\bZ_N=Z(SU(N))$, consistently with the symmetry of adjoint QCD$_2$ with gauge group $SU(N)$.

The presence of the identification current places stringent constraints on the structure of the coset WZW theory \cite{Schellekens:1989uf}, embodied by the following selection rules:

\begin{itemize}

\item Operators with non-trivial monodromy are excluded from the spectrum of the $G_{k}/H_{k'}$ theory,
\be\label{eq: selection rule 1}
t_{\Lambda,\Lambda'}\neq 0 \,\, \Rightarrow \,\, b_{\Lambda,\Lambda'}=0\ed
\ee   

\item Branching functions corresponding to representations within the same orbit under the action of the identification current ${\cal J}$ describe the same module in the coset theory,
\be\label{eq: selection rule 2}
\forall (\Lambda,\Lambda') \quad | \quad t_{\Lambda,\Lambda'}=0 \quad \Rightarrow \quad b_{\Lambda,\Lambda'}=b_{{\cal J}^l\cdot(\Lambda,\Lambda')} \ec\quad S_{\Lambda_1,\Lambda'_1;\Lambda_2,\Lambda'_2}=S_{{\cal J}^l\cdot(\Lambda_1,\Lambda'_1);{\cal J}^{l'}\cdot(\Lambda_2,\Lambda'_2)}\ec
\ee 
for any $0\leq l,l'<N$, with $N$ the order of ${\cal J}$. 

\end{itemize}

From the perspective of ${\cal J}$ as a generator of a 1-form symmetry, these conditions are well-founded: local operators cannot carry charge under a 1-form symmetry. In fact, as we will explicitly demonstrate in examples where the gauge group is $SU(N)$, the vanishing monodromy condition is equivalent to having trivial $N$-ality.

\paragraph{Orbits of simple currents and fixed points.} The selection rule \eqref{eq: selection rule 2} organizes the representations $(\Lambda,\Lambda')$ satisfying \eqref{eq: selection rule 1} into two disjoint sets, namely the long and short orbits with respect to the action of the simple current ${\cal J}$. More precisely
\bea
R_{L}&=\left\lbrace (\Lambda,\Lambda') \quad \big| \quad {\cal J}^l\cdot (\Lambda,\Lambda') \neq (\Lambda,\Lambda') \, , \, \forall l\in {\mathbb Z}_N \right\rbrace\ec\\
R_{S}&=\left\lbrace (\Lambda,\Lambda') \quad \big| \quad \exists \; d\,|\,N \quad  {\rm s.t.} \quad {\cal J}^d\cdot (\Lambda,\Lambda') = (\Lambda,\Lambda')  \right\rbrace \ed
\eea
A prototypical example of a long orbit is the one involving the trivial representation, hence containing the simple current itself.
Throughout this work, we will focus on cases for which $N$ is prime, hence $d=1$ in the definition of $R_S$ above. Operators associated to representations $(\Lambda,\Lambda')\in R_S$ are therefore called {\it fixed points}.

As a consequence of the above selection rules, and in order to assemble the GKO invariant \eqref{eq: GKO matrix}, it becomes a natural choice to make use of the simple current invariants of $G_{k}$ and $H_{k'}$ respectively.\footnote{\label{footnote}It is important to note that this is not a necessary condition for the consistency of the procedure. Furthermore, for certain chiral algebras, simple current invariants may not exist (e.g., when the simple current has half-integer spin) \cite{Schellekens:1990xy}. In fact, there is no fundamental obstacle to deriving the GKO invariant from diagonal modular invariants. The benefit of using simple current invariants, when they do exist, is that they naturally identify the orbits of representations under the action of the simple current. As mentioned in this section, this approach also prescribes an additional rescaling of the characters to obtain a properly normalized partition function \cite{Schellekens:1989uf}. In practice, this rescaling involves dividing the entire sum by $N$.}For the case of $SU(N)_k$, this singles out the so-called $D$-type invariants, as mentioned in \ref{sec: generalities}. Consequently, the modular invariant partition function \eqref{eq: GKO partition function} takes the following generic form \cite{Schellekens:1989am}
\be\label{eq: GKO current invariant}
Z^{\textrm{GKO}}=\sum_{(\Lambda,\Lambda')\in [R_L]}|b_{\Lambda,\Lambda'}+\ldots+b_{{\cal J}^{N-1}\cdot(\Lambda,\Lambda')}|^2+N\sum_{(\Lambda,\Lambda')\in R_S}|b_{\Lambda,\Lambda'}|^2 \ec
\ee
where the factor of $N$ in front of the fixed point contribution is a distinctive property of simple current modular invariants (assuming $N$ is prime). The notation $[R_L]$ means that we are summing over a single representative for each long orbit in order to avoid overcounting.

The degeneracy of $S$-matrix elements imposed by \eqref{eq: selection rule 2} makes the $S$-matrix non-unitary. One way to amend this is to identify all the corresponding highest weight vectors with a single representative within each orbit in the coset theory, leading to a procedure dubbed {\it field identification} 
\be\label{eq: field identification}
b_{\Lambda,\Lambda'}\sim b_{{\cal J}\cdot(\Lambda,\Lambda')}\sim\ldots\sim b_{{\cal J}^{N-1}\cdot(\Lambda,\Lambda')}\ed
\ee
Correspondingly, the $S$-matrix should be projected to the reduced space of representations obtained under the above identification, toghether with the following rescaling of its matrix elements
\be\label{eq: rescaled S-matrix}
S_{\Lambda_1,\Lambda'_1;\Lambda_1,\Lambda'_1}\to \sqrt{N_{(\Lambda_1,\Lambda'_1)}N_{(\Lambda_2,\Lambda'_2)}} S_{\Lambda_1,\Lambda'_1;\Lambda_1,\Lambda'_1}\quad , \quad N_{(\Lambda,\Lambda')}=\left\lbrace\begin{array}{ccc}
N & , & (\Lambda,\Lambda')\in [R_L]\\
1 & , & (\Lambda,\Lambda')\in R_S
\end{array}\right.
\ee
It can be shown that the new $S$-matrix defined above is now unitary \cite{Schellekens:1989uf}. 

Note that, due to the field identification, the contribution of the long orbits is proportional to $N^2$. From now on, we will define the partition function of the coset theory by factoring out the common factor of $N$, namely
\be\label{eq: coset partition function}
Z^{\textrm{Coset}}=N\sum_{(\Lambda,\Lambda')\in [R_L]}|b_{\Lambda,\Lambda'}|^2+\sum_{(\Lambda,\Lambda')\in R_S}|b_{\Lambda,\Lambda'}|^2 \ed
\ee

\paragraph{Fixed point resolution.} Even if the partition function in \eqref{eq: coset partition function} is modular invariant and the rescaled $S$-matrix \eqref{eq: rescaled S-matrix} is unitary, the coset CFT still has some pathological features that prevent its identification with a conventional CFT. 
The most stringent one concerns the fact that fusion coeffients involving fixed point fields ({\it i.e.} $(\Lambda,\Lambda')\in R_S$) are not integral. This can be explicitly checked in WZW models with fixed points by plugging \eqref{eq: rescaled S-matrix} into the Verlinde formula. The conventional procedure of {\it fixed point resolution} \cite{Schellekens:1989uf} instructs to split the branching function into a sum of $N$ independent characters 
\be\label{eq: fixed point slplitting}
(\Lambda,\Lambda')\in R_S \,\, \Rightarrow \,\,  b_{\Lambda,\Lambda'}=\sum_{i=1}^N b_{\Lambda,\Lambda',i}\ec \quad {\rm with} \quad  b_{\Lambda,\Lambda',i}\equiv\frac{1}{N}\left(b_{\Lambda,\Lambda'}+\widehat\chi_{\Lambda,\Lambda',i}\right) \ec
\ee
provided that
\be\label{eq: shift constraint}
\sum_{i=1}^N\chi_{\Lambda,\Lambda',i}=0\ed
\ee
For the class of theories studied in this work, namely $G_k={\rm Spin}(2(N^2-1))_1$ and $H_{k'}=SU(N)_{2N}$, the shift characters $\{\widehat\chi_{\Lambda,\Lambda',i}\}$ turn out to be constants. For $N\geq 3$ one has \cite{Isachenkov:2014zua}:\footnote{For $N=2$ they are $
\widehat{\chi}_{{\cc,[2,\cdots,2],i}}=\pm(-1)^{i+1}$
while all the others are zero.
} 
\bea
& \widehat{\chi}_{{id,[2,\cdots,2],1}}= N-1 \,,\hspace{2cm} \widehat{\chi}_{{id,[2,\cdots,2],j}}= -1 \quad (\forall j \not=1)\ec\\ 
&\widehat{\chi}_{{v,[2,\cdots,2],1}}= 0 \,,\hspace{3cm}\widehat{\chi}_{{v,[2,\cdots,2],j}}= 0 \quad (\forall j \not=1)\ec\\
&\widehat{\chi}_{{s,[2,\cdots,2],1}}= -\frac{(N-1)^2}{2} \,,\hspace{1.25cm} \widehat{\chi}_{{s,[2,\cdots,2],j}}= \frac{N-1}{2} \quad (\forall j \not=1)\ec\\
&\widehat{\chi}_{{c,[2,\cdots,2],1}}= -\frac{N^2-1}{2} \,,\hspace{1.6cm}
\widehat{\chi}_{{c,[2,\cdots,2],j}}= -\frac{N+1}{2} \quad (\forall j \not=1)\,.
\eea

Quite importantly, the identity character has multiplicity $N$ and the partition function is not yet that of an ordinary CFT but a sum of the form
\be\label{eq: coset partition sum}
Z^{\textrm{Coset}}=Z_{\textrm{CFT}_1}+\ldots + Z_{\textrm{CFT}_N}\ed
\ee
As reviewed in section \ref{sec: generalities}, this decomposition is a natural consequence of the presence of a 1-form symmetry. Furthermore, as we will explicitly show in the examples considered throughout this work, different universes are generically connected by certain topological manipulations, implemented by non-invertible topological Verlinde lines of the coset WZW theory. This observation in turn implies that the partition functions $Z_{\textrm{CFT}_i}$ are realized by different modular invariants of the same chiral algebra. Let us emphasize that there are also cases for which all universes are described by exactly the same theory, not just the same chiral algebra. This occurs for instance when states belonging to different universes are solely connected by the action of invertible symmetries.    

Before concluding this section, in order to make contact with previous literature, let us depict the conventional approach to remove the multiplicity of the identity character and reduce \eqref{eq: coset partition function} to a single CFT, usually identified with $Z_{\textrm{CFT}_1}$ in \eqref{eq: coset partition sum}. Indeed, as a consequence of the constraint \eqref{eq: shift constraint}, one immediately verifies that for $(\Lambda,\Lambda')\in R_S$ 
\be
|b_{\Lambda,\Lambda'}|^2=N\sum_{i=1}^N|b_{\Lambda,\Lambda',i}|^2-\frac{1}{N}\sum_{i=1}^N|\widehat\chi_{\Lambda,\Lambda',i}|^2\ed
\ee  
Therefore, defining a residual partition function by:\footnote{A curious fact is that the set $\{\widehat\chi_{\Lambda,\Lambda',i}\}$ defines by itself a CFT, though not necessarily unitary \cite{Schellekens:1990xy}.}
\be
Z^{\textrm{Res}}\equiv \frac{1}{N}\sum_{(\Lambda,\Lambda')\in R_S}\sum_{i=1}^N|\widehat\chi_{\Lambda,\Lambda',i}|^2\ec
\ee
one concludes that
\be
\frac{1}{N}\left(Z^{\textrm{Coset}}+Z^{\textrm{Res}}\right)=\sum_{(\Lambda,\Lambda')\in [R_L]}|b_{\Lambda,\Lambda'}|^2+\sum_{(\Lambda,\Lambda')\in R_S}\sum_{i=1}^N |b_{\Lambda,\Lambda',i}|^2\ec
\ee
which, by construction, corresponds to the modular invariant partition function of a unitary CFT.

\section{\tps{\matht{\cN=(2,2)}}{N=(2,2)} Superconformal Symmetry}\label{sec: app N=2}

In this section we review some useful aspects of the ${\cal N}=(2,2)$ superconformal algebra (SCA). This algebra is generated by the stress tensor $T(z)$, the dimension 3/2 supercurrent $G(z)$, and the dimension 1 current $J(z)$ implementing the $U(1)_R$ R-symmetry. It admits several representations such as the standard antiperiodic boundary condition for the supercurrent which defines the Neveu-Schwarz (NS) sector and the periodic one leading to the Ramond (R) sector. We refer to the literature for more details \cite{Boucher:1986bh,Lin:2016gcl}.

For $c\leq 3$, all unitary representations of the ${\cal N}=(2,2)$ SCA belong to a discrete series of minimal models with central charge
\be
c=\frac{3k}{k+2}\ec \quad k\in{\mathbb N}\ec
\ee
with their bosonic subalgebra represented by the coset 
\be\label{eq: minimal model}
\frac{SU(2)_k\times U(1)_{2}}{U(1)_{k+2}}\ed
\ee
Since these minimal models are particular instances of rational CFT's, they admit a finite set of representations. For a given value of $k$, these modules are labeled by a pair of integers 
\be\label{eq: Pk}
(r,s)\in P_k \equiv \left\lbrace 0\leq r\leq k \,\, , \,\, |s-[r+s]|\leq a \right\rbrace\ec
\ee
where $[x]\in\{0,1\}$ denotes the value of $x$ mod 2. With this notation, representations of the NS (R) superconformal algebras are characterized by $[r+s]=0$ ($[r+s]=1$).  

Primaries for these modules have the following conformal dimension and $U(1)_R$ charge
\be
h_{rs}=\frac{r(r+2)-s^2}{4(k+2)}+\frac{[r+s]}{8}\ec \quad Q_{rs}= \frac{s}{k+2}-\frac{[r+s]}{2}\label{eq: hac Qac}\ec
\ee
and the rest of the module ${\cal H}_{rs}$ is generated by acting with the negative modes of $T(z)$, $G(z)$ and $J(z)$ as usual.
The torus characters associated to these representations are defined as
\be
{\rm ch}_{rs}(\tau,z)={\rm Tr}_{{\cal H}_{rs}}q^{L_0-\frac{c}{24}} y^{J_0}\ec \quad q\equiv e^{2\pi i \tau} \,\, , \,\, y\equiv e^{2\pi i z}\ec
\ee
with $L_0$ and $J_0$ the zero modes of $T(z)$ and $J(z)$ respectively. In addition, twisted characters are constructed by the insertion of the chiral $(-1)^{\textrm{F}}$
\be
\widetilde{\rm ch}_{rs}(\tau,z)={\rm Tr}_{{\cal H}_{rs}}(-1)^{\textrm{F}} q^{L_0-\frac{c}{24}} y^{J_0}=e^{-i\pi Q_{rs}}{\rm ch}_{rs}\left(\tau,z+\frac12\right)\ec
\ee
where we represent the action of $(-1)^{\textrm{F}}$ within a module ${\cal H}_{rs}$ as $(-1)^{J_0-Q_{rs}}$. This prescription is well defined since the $U(1)_R$ charge gets shifted by integers upon the successive action of the higher modes. For sake of notational simplicity, we will henceforth omit the $(\tau,z)$ dependence of the characters. Moreover, throughout this work, the fugacity is set to $y=1$ unless otherwise specified.

Since fermionic theories require specifying a given spin structure, we need to work with the bosonic subalgebra to construct modular invariant combinations of the above characters. The characters of the bosonic subalgebra are obtained as
\bea
\chi_{rs}&\equiv \frac12{\rm Tr}_{{\cal H}_{rs}} \left(1+(-1)^{\textrm{F}}\right)q^{L_0-\frac{c}{24}} =\frac12\left({\rm ch}_{rs}+\widetilde{\rm ch}_{rs}\right)\ec\\
\widetilde\chi_{rs}&\equiv \frac12{\rm Tr}_{{\cal H}_{rs}} \left(1-(-1)^{\textrm{F}}\right)q^{L_0-\frac{c}{24}}=\frac12\left({\rm ch}_{rs}-\widetilde{\rm ch}_{rs}\right)\ed
\eea

In order to define a modular invariant non-negative integer matrix, one needs to enlarge \eqref{eq: Pk} to include the $\widetilde\chi_{rs}$ characters. This is done by including the image $(r,s)$ under the action of the order 2 simple current ${\cal J}\,:\, (r,s)\to (k-r,s+\bar k)$ with $\bar k\equiv k+2$, namely:\footnote{This is the simple current corresponding to the model \eqref{eq: minimal model}. It should not be confused with the identification currents pertaining to the coset WZW theories studied in the main text, which are related to the presence of a 1-form symmetry. Note also that the dimensions listed in \eqref{eq: hac Qac} may violate the unitarity bound ($h\geq 0$) when evaluated on the images of the indices under the action of ${\cal J}$. The proper interpretation of this notation is that the state of lowest dimension surviving the projection in $\widetilde\chi_{rs}$ has dimension $\tilde h_{rs}=h_{k-r,s+\bar k}$ mod 1. } 
\be
\chi_{k-r,s+\bar k}\equiv \widetilde\chi_{rs}\ed
\ee

Therefore, modular invariant partition functions are written as combinations of the characters on this enlarged set
\be
Z=\sum_{(r,s),(r',s')\in Q_k}\widetilde M_{rs;r's'}\chi_{rs}\overline\chi_{r's'}\ec
\ee
where
\be
Q_k\equiv \left\lbrace (r,s)\in P_k \, \cup \, (k-r,s+\bar k) \right\rbrace\ed
\ee

This class of modular invariants have been classified for all values of $k$, see for instance \cite{Gray:2008je}. The relevant cases for the purposes of this work are $k=1$ and $k=16$.

\subsection{Modular Invariants for \tps{{$k=1$}}{k=1}}

We begin by listing the allowed values of the indices $(r,s)$ in the enlarged set $Q_1$
\bea
NS \, &: \, \, \left\lbrace  (0,0) , (1,1) , (1,-1)  \right\rbrace \cup \left\lbrace  (1,3) , (0,4) , (0,2)  \right\rbrace\ec \label{eq: extended set k=1} \\
R \, &: \, \, \left\lbrace  (0,1) , (1,2) , (1,0)  \right\rbrace \cup \left\lbrace  (1,4) , (0,5) , (0,3)  \right\rbrace \ed
\eea

Modular invariants for odd values of $k$ are characterized by three integers $(v,z,n)$. Taking $k=1$ fixes $v=3$ and one is left with four possibilities characterized by $z\in\{1,2\}$ and $n\in\{0,1\}$. The modular matrix reads
\be\label{eq: M k=1}
\widetilde M_{r s;r's'} =1 \quad {\rm if} \quad \left\lbrace
\begin{array}{c}
r'={\cal J}^{(r+s)n}r \\
r'=r+(r+s)n \,\, {\rm mod} \, 2\\
r'=2z s\,\, {\rm mod} \, 3
\end{array}\ed\right.
\ee

Since for $k=1$ the central charge is $c=1$, the four modular invariants should be realized by the ${\cal N}=(2,2)$ RCFT's along the $c=1$ conformal manifold. Adopting the conventions for which the self-dual radius is $R=1$, these four theories occur at $R=\sqrt{6}$, $R=\sqrt{3/2}$ and their $T$-dual radii \cite{Waterson:1986ru}. All them four enjoy an enhanced $U(1)_{12}$ chiral algebra. This connection can be sharpened by explicit construction of the characters $\chi_{rs}$ in terms of characters of the $U(1)_{12}$ chiral algebra, namely
\be\label{eq: chi to K}
\chi_{rs}=K_{2r-3[r+s]}\ec \quad K_\ell= \eta^{-1}\sum_{n\in{\mathbb Z}}q^{6\left(n+\frac{\ell}{12}\right)^2}\ec \quad \ell\in {\mathbb Z}_{12}\ec
\ee
where $\eta$ denotes the standard Dedekind eta-function.

Making use of \eqref{eq: M k=1} together with \eqref{eq: chi to K} one may explicitly check that the four possibilities correspond to the four values of the radius listed above. As an illustrative example, consider the modular invariant with $(v,z,n)=(3,2,0)$. Due to $n=0$, one immediately has $r'=r$. Moreover, the condition for $s$ becomes $s'=s \,\, {\rm mod} \, 2 \, \, \wedge \,\, s'=s \,\, {\rm mod} \, 3$ which, for elements in the set \eqref{eq: extended set k=1} imposes $s'=s$. Then we get a diagonal modular invariant
\be Z_{(3,2,0)}=\sum_{(rs)\in Q_1}|\chi_{rs}|^2=\sum_{\ell\in{\mathbb Z}_{12}}|K_\ell|^2\ec
\ee
corresponding to $R=\sqrt{6}$.

\subsection{Modular Invariants for \tps{{$k=16$}}{k=16}}

The WZW theory \eqref{eq: N=3 coset} possess an enhanced ${\cal W}$-symmetry due to the existence of additional chiral conserved higher-spin currents \cite{Gopakumar:2012gd}. Consequently, the universes are described by modular invariants for which the ${\cal N}=(2,2)$ characters are naturally grouped in orbits under the action of these currents. It is customary to denote these extended characters by ${\rm ch}_{h}$, with a single label $h$ denoting the lowest scaling dimension appearing in the multiplet. To illustrate this structure, consider the multiplet of the identity in the NS sector
\be
\begin{array}{ccccccccccccc}{\rm ch}^{(\textrm{NS})}_{0}&=&{\rm ch}^{(\textrm{NS})}_{00}&+&{\rm ch}^{(\textrm{NS})}_{16,12}&+&{\rm ch}^{(\textrm{NS})}_{16,-12}&+&{\rm ch}^{(\textrm{NS})}_{16,6}&+&{\rm ch}^{(\textrm{NS})}_{16,-6}&+&{\rm ch}^{(\textrm{NS})}_{16,0}\ed\\
& & h=0 & & h=2 & & h=2 & & h=\frac72 & & h=\frac72 & & h=4
\end{array}
\ee
When turning to the twisted extended characters, one should keep in mind that additional relative signs arise between minimal model characters whose dimension differ by a half-integer. For instance, for the extended character involving the identity one has
\be
\widetilde{\rm ch}^{(\textrm{NS})}_{0}=\widetilde{\rm ch}^{(\textrm{NS})}_{00}+\widetilde{\rm ch}^{(\textrm{NS})}_{16,12}+\widetilde{\rm ch}^{(\textrm{NS})}_{16,-12}-\widetilde{\rm ch}^{(\textrm{NS})}_{16,6}-\widetilde{\rm ch}^{(\textrm{NS})}_{16,-6}+\widetilde{\rm ch}^{(\textrm{NS})}_{16,0}\ed
\ee

Now we list the values of $(r,s)$ for $Q_{16}$. In light of the previous discussion, we group these values as they appear in the extended characters relevant for the description of the theory \eqref{eq: N=3 coset}. In the second line within the brackets below we include the values obtained under $(r,s)\to (k-r,s+\bar k)$
\bea\label{eq: N=3 extended characters}
{\rm ch}^{(\textrm{NS})}_{\frac16} \, : \,\,&\left\lbrace\begin{array}{cccccc}
(6,0)&(10,0)&(6,6)&(6,-6)&(10,6)&(10,-6) \\
(10,18)&(6,18)&(10,24)&(10,12)&(6,24)&(6,12)\end{array}\right\rbrace \\
{\rm ch}^{(\textrm{NS})}_{\frac13} \, : \,\,&\left\lbrace\begin{array}{cccccc}
(4,0)&(12,0)&(12,6)&(12,-6)&(12,12)&(12,-12) \\
(12,18)&(4,18)&(4,24)&(4,12)&(4,30)&(4,6)\end{array}\right\rbrace \\
{\rm ch}^{(\textrm{NS})}_{\frac19} \, : \,\,&\left\lbrace\begin{array}{cccccc}
(2,0)&(14,0)&(14,6)&(14,-6)&(14,12)&(14,-12) \\
(14,18)&(2,18)&(2,24)&(2,12)&(2,30)&(2,6)\end{array}\right\rbrace \\
{\rm ch}^{(\textrm{NS})}_{\frac{11}{18}} \, : \,\,&\left\lbrace\begin{array}{ccc}
(8,0)&(8,6)&(8,-6) \\
(8,18)&(8,24)&(8,12)\end{array}\right\rbrace\\
{\rm ch}^{(\textrm{R})}_{1} \, : \,\,&\left\lbrace\begin{array}{cccccc}
(16,-15)&(16,15)&(16,-9)&(16,9)&(16,-3)&(16,3) \\
(0,3)&(0,33)&(0,9)&(0,27)&(0,15)&(0,21)\end{array}\right\rbrace \\
{\rm ch}^{(\textrm{R})}_{\frac13} \, : \,\,&\left\lbrace\begin{array}{cccccc}
(12,9)&(12,-9)&(12,3)&(12,-3)&(4,3)&(4,-3) \\
(4,27)&(4,9)&(4,21)&(4,15)&(12,21)&(12,15)\end{array}\right\rbrace \\
{\rm ch}^{(\textrm{R})}_{\frac23} \, : \,\,&\left\lbrace\begin{array}{cccccc}
(10,9)&(10,-9)&(10,3)&(10,-3)&(6,3)&(6,-3) \\
(6,27)&(6,9)&(6,21)&(6,15)&(10,21)&(10,15)\end{array}\right\rbrace \\
{\rm ch}^{(\textrm{R})}_{\frac19} \, : \,\,&\left\lbrace\begin{array}{cccccc}
(2,3)&(14,3)&(14,-3)&(14,9)&(14,-9)&(14,15) \\
(14,21)&(2,21)&(2,15)&(2,27)&(2,9)&(2,33)\end{array}\right\rbrace \\
{\rm ch}^{(\textrm{R})}_{\frac{1}{9'}} \, : \,\,&\left\lbrace\begin{array}{ccc}
(8,3)&(8,-3)&(8,9) \\
(8,21)&(8,15)&(8,27)\end{array}\right\rbrace 
\eea

The classification of modular invariants for the $k=16$  minimal model can be found in \cite{Gray:2008je}. As briefly mentioned in the main text, the representation of these models in terms of the coset \eqref{eq: minimal model} induces the following factorization for the modular matrix 
\be
\widetilde M_{rs;r's'}=M^{SU(2)}_{r,r'}M_{s,s'}\ec
\ee
where the $M^{SU(2)}$ factor corresponds to a modular invariant for the $SU(2)_k$ WZW theory and, as such, falls into the standard A-D-E classification \cite{Cappelli:1987xt}.

Due to the extended ${\cal W}$-symmetry of the coset \eqref{eq: N=3 coset}, only the invariants respecting the multiplet structure \eqref{eq: N=3 extended characters} should be considered. For $k=16$, this already singles out the $M^{SU(2)}$ corresponding to the D-type and the exceptional invariants of $SU(2)_{16}$
\be\label{eq: SU(2) D-type}
M^{SU(2)}_D\rightarrow\sum_{\substack{r={\rm even}\\ r\neq 8
}}|\chi_r+\chi_{16-r}|^2+2|\chi_8|^2\ec
\ee
\be\label{eq: SU(2) E-type}
M^{SU(2)}_E\rightarrow|\chi_0+\chi_{16}|^2+|\chi_4+\chi_{12}|^2+|\chi_6+\chi_{10}|^2+|\chi_8|^2+\left(\chi_2+\chi_{14}\right)\overline\chi_8 + \textrm{c.c.}
\ee
It is easy to check that the above invariants respect the pairing of the $r$-indices implied by the extended characters \eqref{eq: N=3 extended characters}.

The remaining part of the modular matrix, namely the matrix $M$ acting on the $s$-index, is characterized by three integers $(v,z,x)$ with $v=3$, $z=1$, $x\in\{1,3\}$ or $v=9$, $z\in\{1,8\}$, $x\in\{1,3\}$, and reads
\be\label{eq: M k=16}
\widetilde M_{\frac{9}{v} s , \frac{9}{v} s'} =1 \quad {\rm if} \quad \left\lbrace
\begin{array}{c}
s'=z s \,\, {\rm mod} \, \frac{v^2}{9}\\
s'=x s\,\, {\rm mod} \, 4
\end{array}\ed\right.
\ee
By inspection, one finds that the appropriate pairing of indices that follows from the multiplet structure \eqref{eq: N=3 extended characters} fixes $v=3$, $z=1$, $x=1$. Combining this with either \eqref{eq: SU(2) D-type} or \eqref{eq: SU(2) E-type} leads to the
modular invariant partition functions in \eqref{eq: Z M42} and \eqref{eq: Z E162} respectively.

\bibliographystyle{ytphys}
\baselineskip=0.85\baselineskip
\bibliography{refs}

\providecommand{\href}[2]{#2}\begingroup\raggedright\begin{thebibliography}{100}

\bibitem{Aharony:2013hda}
O.~Aharony, N.~Seiberg, and Y.~Tachikawa, ``{Reading between the lines of
  four-dimensional gauge theories},''
  \href{http://dx.doi.org/10.1007/JHEP08(2013)115}{{\em JHEP} {\bfseries 08}
  (2013) 115}, \href{http://arxiv.org/abs/1305.0318}{{\ttfamily arXiv:1305.0318
  [hep-th]}}.

\bibitem{Gaiotto:2014kfa}
D.~Gaiotto, A.~Kapustin, N.~Seiberg, and B.~Willett, ``{Generalized Global
  Symmetries},'' \href{http://dx.doi.org/10.1007/JHEP02(2015)172}{{\em JHEP}
  {\bfseries 02} (2015) 172}, \href{http://arxiv.org/abs/1412.5148}{{\ttfamily
  arXiv:1412.5148 [hep-th]}}.

\bibitem{Wilson:1974sk}
K.~G. Wilson, ``{Confinement of Quarks},''
  \href{http://dx.doi.org/10.1103/PhysRevD.10.2445}{{\em Phys. Rev. D}
  {\bfseries 10} (1974) 2445--2459}.

\bibitem{Seiberg:1994rs}
N.~Seiberg and E.~Witten, ``{Electric - magnetic duality, monopole
  condensation, and confinement in N=2 supersymmetric Yang-Mills theory},''
  \href{http://dx.doi.org/10.1016/0550-3213(94)90124-4}{{\em Nucl. Phys. B}
  {\bfseries 426} (1994) 19--52},
  \href{http://arxiv.org/abs/hep-th/9407087}{{\ttfamily arXiv:hep-th/9407087}}.
  [Erratum: Nucl.Phys.B 430, 485--486 (1994)].

\bibitem{tHooft:1974pnl}
G.~'t~Hooft, ``{A Two-Dimensional Model for Mesons},''
  \href{http://dx.doi.org/10.1016/0550-3213(74)90088-1}{{\em Nucl. Phys. B}
  {\bfseries 75} (1974) 461--470}.

\bibitem{Migdal:1975zg}
A.~A. Migdal, ``{Recursion Equations in Gauge Theories},'' {\em Sov. Phys.
  JETP} {\bfseries 42} (1975) 413.

\bibitem{Gross:1993hu}
D.~J. Gross and W.~Taylor, ``{Two-dimensional QCD is a string theory},''
  \href{http://dx.doi.org/10.1016/0550-3213(93)90403-C}{{\em Nucl. Phys. B}
  {\bfseries 400} (1993) 181--208},
  \href{http://arxiv.org/abs/hep-th/9301068}{{\ttfamily arXiv:hep-th/9301068}}.

\bibitem{Gross:1993yt}
D.~J. Gross and W.~Taylor, ``{Twists and Wilson loops in the string theory of
  two-dimensional QCD},''
  \href{http://dx.doi.org/10.1016/0550-3213(93)90042-N}{{\em Nucl. Phys. B}
  {\bfseries 403} (1993) 395--452},
  \href{http://arxiv.org/abs/hep-th/9303046}{{\ttfamily arXiv:hep-th/9303046}}.

\bibitem{Aharony:2023tam}
O.~Aharony, S.~Kundu, and T.~Sheaffer, ``{A string theory for two dimensional
  Yang-Mills theory. Part I},''
  \href{http://dx.doi.org/10.1007/JHEP07(2024)063}{{\em JHEP} {\bfseries 07}
  (2024) 063}, \href{http://arxiv.org/abs/2312.12266}{{\ttfamily
  arXiv:2312.12266 [hep-th]}}.

\bibitem{Dalley:1992yy}
S.~Dalley and I.~R. Klebanov, ``{String spectrum of (1+1)-dimensional large N
  QCD with adjoint matter},''
  \href{http://dx.doi.org/10.1103/PhysRevD.47.2517}{{\em Phys. Rev. D}
  {\bfseries 47} (1993) 2517--2527},
  \href{http://arxiv.org/abs/hep-th/9209049}{{\ttfamily arXiv:hep-th/9209049}}.

\bibitem{Bhanot:1993xp}
G.~Bhanot, K.~Demeterfi, and I.~R. Klebanov, ``{(1+1)-dimensional large N QCD
  coupled to adjoint fermions},''
  \href{http://dx.doi.org/10.1103/PhysRevD.48.4980}{{\em Phys. Rev. D}
  {\bfseries 48} (1993) 4980--4990},
  \href{http://arxiv.org/abs/hep-th/9307111}{{\ttfamily arXiv:hep-th/9307111}}.

\bibitem{Katz:2013qua}
E.~Katz, G.~Marques~Tavares, and Y.~Xu, ``{Solving 2D QCD with an adjoint
  fermion analytically},''
  \href{http://dx.doi.org/10.1007/JHEP05(2014)143}{{\em JHEP} {\bfseries 05}
  (2014) 143}, \href{http://arxiv.org/abs/1308.4980}{{\ttfamily arXiv:1308.4980
  [hep-th]}}.

\bibitem{Gross:1995bp}
D.~J. Gross, I.~R. Klebanov, A.~V. Matytsin, and A.~V. Smilga, ``{Screening
  versus confinement in (1+1)-dimensions},''
  \href{http://dx.doi.org/10.1016/0550-3213(95)00655-9}{{\em Nucl. Phys. B}
  {\bfseries 461} (1996) 109--130},
  \href{http://arxiv.org/abs/hep-th/9511104}{{\ttfamily arXiv:hep-th/9511104}}.

\bibitem{Cherman:2019hbq}
A.~Cherman, T.~Jacobson, Y.~Tanizaki, and M.~\"Unsal, ``{Anomalies, a mod 2
  index, and dynamics of 2d adjoint QCD},''
  \href{http://dx.doi.org/10.21468/SciPostPhys.8.5.072}{{\em SciPost Phys.}
  {\bfseries 8} no.~5, (2020) 072},
  \href{http://arxiv.org/abs/1908.09858}{{\ttfamily arXiv:1908.09858
  [hep-th]}}.

\bibitem{Komargodski:2020mxz}
Z.~Komargodski, K.~Ohmori, K.~Roumpedakis, and S.~Seifnashri, ``{Symmetries and
  strings of adjoint QCD$_{2}$},''
  \href{http://dx.doi.org/10.1007/JHEP03(2021)103}{{\em JHEP} {\bfseries 03}
  (2021) 103}, \href{http://arxiv.org/abs/2008.07567}{{\ttfamily
  arXiv:2008.07567 [hep-th]}}.

\bibitem{Dempsey:2021xpf}
R.~Dempsey, I.~R. Klebanov, and S.~S. Pufu, ``{Exact symmetries and threshold
  states in two-dimensional models for QCD},''
  \href{http://dx.doi.org/10.1007/JHEP10(2021)096}{{\em JHEP} {\bfseries 10}
  (2021) 096}, \href{http://arxiv.org/abs/2101.05432}{{\ttfamily
  arXiv:2101.05432 [hep-th]}}.

\bibitem{Dempsey:2022uie}
R.~Dempsey, I.~R. Klebanov, L.~L. Lin, and S.~S. Pufu, ``{Adjoint Majorana
  QCD$_{2}$ at finite N},''
  \href{http://dx.doi.org/10.1007/JHEP04(2023)107}{{\em JHEP} {\bfseries 04}
  (2023) 107}, \href{http://arxiv.org/abs/2210.10895}{{\ttfamily
  arXiv:2210.10895 [hep-th]}}.

\bibitem{Witten:1983ar}
E.~Witten, ``{Nonabelian Bosonization in Two-Dimensions},''
  \href{http://dx.doi.org/10.1007/BF01215276}{{\em Commun. Math. Phys.}
  {\bfseries 92} (1984) 455--472}.

\bibitem{Gopakumar:2012gd}
R.~Gopakumar, A.~Hashimoto, I.~R. Klebanov, S.~Sachdev, and K.~Schoutens,
  ``{Strange Metals in One Spatial Dimension},''
  \href{http://dx.doi.org/10.1103/PhysRevD.86.066003}{{\em Phys. Rev. D}
  {\bfseries 86} (2012) 066003},
  \href{http://arxiv.org/abs/1206.4719}{{\ttfamily arXiv:1206.4719 [hep-th]}}.

\bibitem{Isachenkov:2014zua}
M.~Isachenkov, I.~Kirsch, and V.~Schomerus, ``{Chiral Primaries in Strange
  Metals},'' \href{http://dx.doi.org/10.1016/j.nuclphysb.2014.06.004}{{\em
  Nucl. Phys. B} {\bfseries 885} (2014) 679--712},
  \href{http://arxiv.org/abs/1403.6857}{{\ttfamily arXiv:1403.6857 [hep-th]}}.

\bibitem{Ambrosino:2023dik}
F.~Ambrosino and S.~Komatsu, ``{2d QCD and Integrability, Part I: 't Hooft
  model},'' \href{http://arxiv.org/abs/2312.15598}{{\ttfamily arXiv:2312.15598
  [hep-th]}}.

\bibitem{Ambrosino:2024prz}
F.~Ambrosino and S.~Komatsu, ``{2d QCD and Integrability, Part II: Generalized
  QCD},'' \href{http://arxiv.org/abs/2406.11078}{{\ttfamily arXiv:2406.11078
  [hep-th]}}.

\bibitem{Dempsey:2023fvm}
R.~Dempsey, I.~R. Klebanov, S.~S. Pufu, and B.~T. S\o{}gaard, ``{Lattice
  Hamiltonian for adjoint QCD$_{2}$},''
  \href{http://dx.doi.org/10.1007/JHEP08(2024)009}{{\em JHEP} {\bfseries 08}
  (2024) 009}, \href{http://arxiv.org/abs/2311.09334}{{\ttfamily
  arXiv:2311.09334 [hep-th]}}.

\bibitem{Bergner:2024ttq}
G.~Bergner, S.~Piemonte, and M.~\"Unsal, ``{Investigating two-dimensional
  adjoint QCD on the lattice},''
  \href{http://dx.doi.org/10.1007/JHEP07(2024)048}{{\em JHEP} {\bfseries 07}
  (2024) 048}, \href{http://arxiv.org/abs/2404.03801}{{\ttfamily
  arXiv:2404.03801 [hep-lat]}}.

\bibitem{Dempsey:2024ofo}
R.~Dempsey, I.~R. Klebanov, S.~S. Pufu, and B.~T. S\o{}gaard, ``{Small Circle
  Expansion for Adjoint QCD$_2$ with Periodic Boundary Conditions},''
  \href{http://arxiv.org/abs/2406.17079}{{\ttfamily arXiv:2406.17079
  [hep-th]}}.

\bibitem{Kutasov:1993gq}
D.~Kutasov, ``{Two-dimensional QCD coupled to adjoint matter and string
  theory},'' \href{http://dx.doi.org/10.1016/0550-3213(94)90420-0}{{\em Nucl.
  Phys. B} {\bfseries 414} (1994) 33--52},
  \href{http://arxiv.org/abs/hep-th/9306013}{{\ttfamily arXiv:hep-th/9306013}}.

\bibitem{Ji:2019ugf}
W.~Ji, S.-H. Shao, and X.-G. Wen, ``{Topological Transition on the Conformal
  Manifold},'' \href{http://dx.doi.org/10.1103/PhysRevResearch.2.033317}{{\em
  Phys. Rev. Res.} {\bfseries 2} no.~3, (2020) 033317},
  \href{http://arxiv.org/abs/1909.01425}{{\ttfamily arXiv:1909.01425
  [cond-mat.str-el]}}.

\bibitem{Kutasov:1994xq}
D.~Kutasov and A.~Schwimmer, ``{Universality in two-dimensional gauge
  theory},'' \href{http://dx.doi.org/10.1016/0550-3213(95)00106-3}{{\em Nucl.
  Phys. B} {\bfseries 442} (1995) 447--460},
  \href{http://arxiv.org/abs/hep-th/9501024}{{\ttfamily arXiv:hep-th/9501024}}.

\bibitem{Goddard:1984vk}
P.~Goddard, A.~Kent, and D.~I. Olive, ``{Virasoro Algebras and Coset Space
  Models},'' \href{http://dx.doi.org/10.1016/0370-2693(85)91145-1}{{\em Phys.
  Lett. B} {\bfseries 152} (1985) 88--92}.

\bibitem{Delmastro:2021otj}
D.~Delmastro, J.~Gomis, and M.~Yu, ``{Infrared phases of 2d QCD},''
  \href{http://dx.doi.org/10.1007/JHEP02(2023)157}{{\em JHEP} {\bfseries 02}
  (2023) 157}, \href{http://arxiv.org/abs/2108.02202}{{\ttfamily
  arXiv:2108.02202 [hep-th]}}.

\bibitem{Hellerman:2006zs}
S.~Hellerman, A.~Henriques, T.~Pantev, E.~Sharpe, and M.~Ando, ``{Cluster
  decomposition, T-duality, and gerby CFT's},''
  \href{http://dx.doi.org/10.4310/ATMP.2007.v11.n5.a2}{{\em Adv. Theor. Math.
  Phys.} {\bfseries 11} no.~5, (2007) 751--818},
  \href{http://arxiv.org/abs/hep-th/0606034}{{\ttfamily arXiv:hep-th/0606034}}.

\bibitem{Aminov:2019hwg}
G.~Aminov, ``{Spontaneous symmetry breaking in pure 2D Yang-Mills theory},''
  \href{http://dx.doi.org/10.1103/PhysRevD.101.105017}{{\em Phys. Rev. D}
  {\bfseries 101} no.~10, (2020) 105017},
  \href{http://arxiv.org/abs/1911.03494}{{\ttfamily arXiv:1911.03494
  [hep-th]}}.

\bibitem{Sharpe:2019ddn}
E.~Sharpe, ``{Undoing decomposition},''
  \href{http://dx.doi.org/10.1142/S0217751X19502336}{{\em Int. J. Mod. Phys. A}
  {\bfseries 34} no.~35, (2020) 1950233},
  \href{http://arxiv.org/abs/1911.05080}{{\ttfamily arXiv:1911.05080
  [hep-th]}}.

\bibitem{Tanizaki:2019rbk}
Y.~Tanizaki and M.~\"Unsal, ``{Modified instanton sum in QCD and
  higher-groups},'' \href{http://dx.doi.org/10.1007/JHEP03(2020)123}{{\em JHEP}
  {\bfseries 03} (2020) 123}, \href{http://arxiv.org/abs/1912.01033}{{\ttfamily
  arXiv:1912.01033 [hep-th]}}.

\bibitem{Nguyen:2021naa}
M.~Nguyen, Y.~Tanizaki, and M.~\"Unsal, ``{Noninvertible 1-form symmetry and
  Casimir scaling in 2D Yang-Mills theory},''
  \href{http://dx.doi.org/10.1103/PhysRevD.104.065003}{{\em Phys. Rev. D}
  {\bfseries 104} no.~6, (2021) 065003},
  \href{http://arxiv.org/abs/2104.01824}{{\ttfamily arXiv:2104.01824
  [hep-th]}}.

\bibitem{Sharpe:2022ene}
E.~Sharpe, ``{An introduction to decomposition},''
  \href{http://arxiv.org/abs/2204.09117}{{\ttfamily arXiv:2204.09117
  [hep-th]}}.

\bibitem{Intriligator:1989zw}
K.~A. Intriligator, ``{Bonus Symmetry in Conformal Field Theory},''
  \href{http://dx.doi.org/10.1016/0550-3213(90)90001-T}{{\em Nucl. Phys. B}
  {\bfseries 332} (1990) 541--565}.

\bibitem{Beltaos:2010ka}
E.~Beltaos and T.~Gannon, ``{The $W_{N}$ minimal model classification},''
  \href{http://dx.doi.org/10.1007/s00220-012-1473-4}{{\em Commun. Math. Phys.}
  {\bfseries 312} (2012) 337--360},
  \href{http://arxiv.org/abs/1004.1205}{{\ttfamily arXiv:1004.1205 [hep-th]}}.

\bibitem{Schellekens:1989uf}
A.~N. Schellekens and S.~Yankielowicz, ``{Field Identification Fixed Points in
  the Coset Construction},''
  \href{http://dx.doi.org/10.1016/0550-3213(90)90657-Y}{{\em Nucl. Phys. B}
  {\bfseries 334} (1990) 67--102}.

\bibitem{Schellekens:1990xy}
A.~N. Schellekens and S.~Yankielowicz, ``{Simple Currents, Modular Invariants
  and Fixed Points},'' \href{http://dx.doi.org/10.1142/S0217751X90001367}{{\em
  Int. J. Mod. Phys. A} {\bfseries 5} (1990) 2903--2952}.

\bibitem{Fuchs:1995tq}
J.~Fuchs, B.~Schellekens, and C.~Schweigert, ``{The resolution of field
  identification fixed points in diagonal coset theories},''
  \href{http://dx.doi.org/10.1016/0550-3213(95)00623-0}{{\em Nucl. Phys. B}
  {\bfseries 461} (1996) 371--406},
  \href{http://arxiv.org/abs/hep-th/9509105}{{\ttfamily arXiv:hep-th/9509105}}.

\bibitem{Witten:1988hf}
E.~Witten, ``{Quantum Field Theory and the Jones Polynomial},''
  \href{http://dx.doi.org/10.1007/BF01217730}{{\em Commun. Math. Phys.}
  {\bfseries 121} (1989) 351--399}.

\bibitem{Moore:1988qv}
G.~W. Moore and N.~Seiberg, ``{Classical and Quantum Conformal Field Theory},''
  \href{http://dx.doi.org/10.1007/BF01238857}{{\em Commun. Math. Phys.}
  {\bfseries 123} (1989) 177}.

\bibitem{Elitzur:1989nr}
S.~Elitzur, G.~W. Moore, A.~Schwimmer, and N.~Seiberg, ``{Remarks on the
  Canonical Quantization of the Chern-Simons-Witten Theory},''
  \href{http://dx.doi.org/10.1016/0550-3213(89)90436-7}{{\em Nucl. Phys. B}
  {\bfseries 326} (1989) 108--134}.

\bibitem{Fuchs:2002cm}
J.~Fuchs, I.~Runkel, and C.~Schweigert, ``{TFT construction of RCFT correlators
  1. Partition functions},''
  \href{http://dx.doi.org/10.1016/S0550-3213(02)00744-7}{{\em Nucl. Phys. B}
  {\bfseries 646} (2002) 353--497},
  \href{http://arxiv.org/abs/hep-th/0204148}{{\ttfamily arXiv:hep-th/0204148}}.

\bibitem{Verlinde:1988sn}
E.~P. Verlinde, ``{Fusion Rules and Modular Transformations in 2D Conformal
  Field Theory},'' \href{http://dx.doi.org/10.1016/0550-3213(88)90603-7}{{\em
  Nucl. Phys. B} {\bfseries 300} (1988) 360--376}.

\bibitem{Cordova:2023jip}
C.~Cordova and D.~Garc\'\i{}a-Sep\'ulveda, ``{Non-Invertible Anyon Condensation
  and Level-Rank Dualities},''
  \href{http://arxiv.org/abs/2312.16317}{{\ttfamily arXiv:2312.16317
  [hep-th]}}.

\bibitem{Moore:1989yh}
G.~W. Moore and N.~Seiberg, ``{Taming the Conformal Zoo},''
  \href{http://dx.doi.org/10.1016/0370-2693(89)90897-6}{{\em Phys. Lett. B}
  {\bfseries 220} (1989) 422--430}.

\bibitem{Bhardwaj:2017xup}
L.~Bhardwaj and Y.~Tachikawa, ``{On finite symmetries and their gauging in two
  dimensions},'' \href{http://dx.doi.org/10.1007/JHEP03(2018)189}{{\em JHEP}
  {\bfseries 03} (2018) 189}, \href{http://arxiv.org/abs/1704.02330}{{\ttfamily
  arXiv:1704.02330 [hep-th]}}.

\bibitem{Delmastro:2022prj}
D.~Delmastro and J.~Gomis, ``{RG flows in 2d QCD},''
  \href{http://dx.doi.org/10.1007/JHEP09(2023)158}{{\em JHEP} {\bfseries 09}
  (2023) 158}, \href{http://arxiv.org/abs/2211.09036}{{\ttfamily
  arXiv:2211.09036 [hep-th]}}.

\bibitem{unpublished}
Z.~Komargodski, M.~Martone, and S.~Seifnashri, ``{Unpublished},''.

\bibitem{Boucher:1986bh}
W.~Boucher, D.~Friedan, and A.~Kent, ``{Determinant Formulae and Unitarity for
  the N=2 Superconformal Algebras in Two-Dimensions or Exact Results on String
  Compactification},''
  \href{http://dx.doi.org/10.1016/0370-2693(86)90260-1}{{\em Phys. Lett. B}
  {\bfseries 172} (1986) 316}.

\bibitem{DiVecchia:1986fwg}
P.~Di~Vecchia, J.~L. Petersen, M.~Yu, and H.~B. Zheng, ``{Explicit Construction
  of Unitary Representations of the N=2 Superconformal Algebra},''
  \href{http://dx.doi.org/10.1016/0370-2693(86)91099-3}{{\em Phys. Lett. B}
  {\bfseries 174} (1986) 280--284}.

\bibitem{Johnson-Freyd:2019wgb}
T.~Johnson-Freyd, ``{Supersymmetry and the Suzuki chain},''
  \href{http://arxiv.org/abs/1908.11012}{{\ttfamily arXiv:1908.11012
  [math.QA]}}.

\bibitem{Kikuchi:2022jbl}
K.~Kikuchi, ``{Emergent SUSY in two dimensions},''
  \href{http://arxiv.org/abs/2204.03247}{{\ttfamily arXiv:2204.03247
  [hep-th]}}.

\bibitem{Frohlich:2006ch}
J.~Frohlich, J.~Fuchs, I.~Runkel, and C.~Schweigert, ``{Duality and defects in
  rational conformal field theory},''
  \href{http://dx.doi.org/10.1016/j.nuclphysb.2006.11.017}{{\em Nucl. Phys. B}
  {\bfseries 763} (2007) 354--430},
  \href{http://arxiv.org/abs/hep-th/0607247}{{\ttfamily arXiv:hep-th/0607247}}.

\bibitem{Chang:2018iay}
C.-M. Chang, Y.-H. Lin, S.-H. Shao, Y.~Wang, and X.~Yin, ``{Topological Defect
  Lines and Renormalization Group Flows in Two Dimensions},''
  \href{http://dx.doi.org/10.1007/JHEP01(2019)026}{{\em JHEP} {\bfseries 01}
  (2019) 026}, \href{http://arxiv.org/abs/1802.04445}{{\ttfamily
  arXiv:1802.04445 [hep-th]}}.

\bibitem{Thorngren:2021yso}
R.~Thorngren and Y.~Wang, ``{Fusion Category Symmetry II: Categoriosities at
  $c$ = 1 and Beyond},'' \href{http://arxiv.org/abs/2106.12577}{{\ttfamily
  arXiv:2106.12577 [hep-th]}}.

\bibitem{Benini:2022hzx}
F.~Benini, C.~Copetti, and L.~Di~Pietro, ``{Factorization and global symmetries
  in holography},'' \href{http://dx.doi.org/10.21468/SciPostPhys.14.2.019}{{\em
  SciPost Phys.} {\bfseries 14} no.~2, (2023) 019},
  \href{http://arxiv.org/abs/2203.09537}{{\ttfamily arXiv:2203.09537
  [hep-th]}}.

\bibitem{Cordova:2023qei}
C.~Cordova and G.~Rizi, ``{Non-Invertible Symmetry in Calabi-Yau Conformal
  Field Theories},'' \href{http://arxiv.org/abs/2312.17308}{{\ttfamily
  arXiv:2312.17308 [hep-th]}}.

\bibitem{Damia:2024xju}
J.~A. Damia, G.~Galati, O.~Hulik, and S.~Mancani, ``{Exploring duality
  symmetries, multicriticality and RG flows at c = 2},''
  \href{http://dx.doi.org/10.1007/JHEP04(2024)028}{{\em JHEP} {\bfseries 04}
  (2024) 028}, \href{http://arxiv.org/abs/2401.04166}{{\ttfamily
  arXiv:2401.04166 [hep-th]}}.

\bibitem{Bharadwaj:2024gpj}
S.~Bharadwaj, P.~Niro, and K.~Roumpedakis, ``{Non-invertible defects on the
  worldsheet},'' \href{http://arxiv.org/abs/2408.14556}{{\ttfamily
  arXiv:2408.14556 [hep-th]}}.

\bibitem{Bhardwaj:2023bbf}
L.~Bhardwaj, L.~E. Bottini, D.~Pajer, and S.~Schafer-Nameki, ``{The Club
  Sandwich: Gapless Phases and Phase Transitions with Non-Invertible
  Symmetries},'' \href{http://arxiv.org/abs/2312.17322}{{\ttfamily
  arXiv:2312.17322 [hep-th]}}.

\bibitem{Bhardwaj:2024qrf}
L.~Bhardwaj, D.~Pajer, S.~Schafer-Nameki, and A.~Warman, ``{Hasse Diagrams for
  Gapless SPT and SSB Phases with Non-Invertible Symmetries},''
  \href{http://arxiv.org/abs/2403.00905}{{\ttfamily arXiv:2403.00905
  [cond-mat.str-el]}}.

\bibitem{Ng:2024jkm}
S.-H. Ng, E.~C. Rowell, and X.-G. Wen, ``{Recovering R-symbols from modular
  data},'' \href{http://arxiv.org/abs/2408.02748}{{\ttfamily arXiv:2408.02748
  [math.QA]}}.

\bibitem{Thorngren:2019iar}
R.~Thorngren and Y.~Wang, ``{Fusion category symmetry. Part I. Anomaly in-flow
  and gapped phases},'' \href{http://dx.doi.org/10.1007/JHEP04(2024)132}{{\em
  JHEP} {\bfseries 04} (2024) 132},
  \href{http://arxiv.org/abs/1912.02817}{{\ttfamily arXiv:1912.02817
  [hep-th]}}.

\bibitem{Damia:2023ses}
J.~A. Damia, R.~Argurio, F.~Benini, S.~Benvenuti, C.~Copetti, and L.~Tizzano,
  ``{Non-invertible symmetries along 4d RG flows},''
  \href{http://dx.doi.org/10.1007/JHEP02(2024)084}{{\em JHEP} {\bfseries 02}
  (2024) 084}, \href{http://arxiv.org/abs/2305.17084}{{\ttfamily
  arXiv:2305.17084 [hep-th]}}.

\bibitem{Bhardwaj:2023idu}
L.~Bhardwaj, L.~E. Bottini, D.~Pajer, and S.~Sch\"afer-Nameki, ``{Gapped Phases
  with Non-Invertible Symmetries: (1+1)d},''
  \href{http://arxiv.org/abs/2310.03784}{{\ttfamily arXiv:2310.03784
  [hep-th]}}.

\bibitem{Benini:2013cda}
F.~Benini and N.~Bobev, ``{Two-dimensional SCFTs from wrapped branes and
  c-extremization},'' \href{http://dx.doi.org/10.1007/JHEP06(2013)005}{{\em
  JHEP} {\bfseries 06} (2013) 005},
  \href{http://arxiv.org/abs/1302.4451}{{\ttfamily arXiv:1302.4451 [hep-th]}}.

\bibitem{Choi:2023xjw}
Y.~Choi, B.~C. Rayhaun, Y.~Sanghavi, and S.-H. Shao, ``{Remarks on boundaries,
  anomalies, and noninvertible symmetries},''
  \href{http://dx.doi.org/10.1103/PhysRevD.108.125005}{{\em Phys. Rev. D}
  {\bfseries 108} no.~12, (2023) 125005},
  \href{http://arxiv.org/abs/2305.09713}{{\ttfamily arXiv:2305.09713
  [hep-th]}}.

\bibitem{Zhang:2023wlu}
C.~Zhang and C.~C\'ordova, ``{Anomalies of (1+1)-dimensional categorical
  symmetries},'' \href{http://dx.doi.org/10.1103/PhysRevB.110.035155}{{\em
  Phys. Rev. B} {\bfseries 110} no.~3, (2024) 035155},
  \href{http://arxiv.org/abs/2304.01262}{{\ttfamily arXiv:2304.01262
  [cond-mat.str-el]}}.

\bibitem{Antinucci:2023ezl}
A.~Antinucci, F.~Benini, C.~Copetti, G.~Galati, and G.~Rizi, ``{Anomalies of
  non-invertible self-duality symmetries: fractionalization and gauging},''
  \href{http://arxiv.org/abs/2308.11707}{{\ttfamily arXiv:2308.11707
  [hep-th]}}.

\bibitem{Vafa:1983tf}
C.~Vafa and E.~Witten, ``{Restrictions on Symmetry Breaking in Vector-Like
  Gauge Theories},'' \href{http://dx.doi.org/10.1016/0550-3213(84)90230-X}{{\em
  Nucl. Phys. B} {\bfseries 234} (1984) 173--188}.

\bibitem{Peskin:1995ev}
M.~E. Peskin and D.~V. Schroeder,
  \href{http://dx.doi.org/10.1201/9780429503559}{{\em {An Introduction to
  quantum field theory}}}.
\newblock Addison-Wesley, Reading, USA, 1995.

\bibitem{Popov:2022vud}
F.~K. Popov, ``{Supersymmetry in QCD2 coupled to fermions},''
  \href{http://dx.doi.org/10.1103/PhysRevD.105.074005}{{\em Phys. Rev. D}
  {\bfseries 105} no.~7, (2022) 074005},
  \href{http://arxiv.org/abs/2202.04017}{{\ttfamily arXiv:2202.04017
  [hep-th]}}.

\bibitem{Kastor:1988ef}
D.~A. Kastor, E.~J. Martinec, and S.~H. Shenker, ``{RG Flow in N=1 Discrete
  Series},'' \href{http://dx.doi.org/10.1016/0550-3213(89)90060-6}{{\em Nucl.
  Phys. B} {\bfseries 316} (1989) 590--608}.

\bibitem{Friedan:1984rv}
D.~Friedan, Z.-a. Qiu, and S.~H. Shenker, ``{Superconformal Invariance in
  Two-Dimensions and the Tricritical Ising Model},''
  \href{http://dx.doi.org/10.1016/0370-2693(85)90819-6}{{\em Phys. Lett. B}
  {\bfseries 151} (1985) 37--43}.

\bibitem{Coleman:1974bu}
S.~R. Coleman, ``{The Quantum Sine-Gordon Equation as the Massive Thirring
  Model},'' \href{http://dx.doi.org/10.1103/PhysRevD.11.2088}{{\em Phys. Rev.
  D} {\bfseries 11} (1975) 2088}.

\bibitem{Zamolodchikov:1978xm}
A.~B. Zamolodchikov and A.~B. Zamolodchikov, ``{Factorized s Matrices in
  Two-Dimensions as the Exact Solutions of Certain Relativistic Quantum Field
  Models},'' \href{http://dx.doi.org/10.1016/0003-4916(79)90391-9}{{\em Annals
  Phys.} {\bfseries 120} (1979) 253--291}.

\bibitem{Dashen:1975hd}
R.~F. Dashen, B.~Hasslacher, and A.~Neveu, ``{The Particle Spectrum in Model
  Field Theories from Semiclassical Functional Integral Techniques},''
  \href{http://dx.doi.org/10.1103/PhysRevD.11.3424}{{\em Phys. Rev. D}
  {\bfseries 11} (1975) 3424}.

\bibitem{Korepin:1975zu}
V.~E. Korepin and L.~D. Faddeev, ``{Quantization of Solitons},''
  \href{http://dx.doi.org/10.1007/BF01028946}{{\em Theor. Math. Phys.}
  {\bfseries 25} (1975) 1039--1049}.

\bibitem{Zamolodchikov:1995xk}
A.~B. Zamolodchikov, ``{Mass scale in the sine-Gordon model and its
  reductions},'' \href{http://dx.doi.org/10.1142/S0217751X9500053X}{{\em Int.
  J. Mod. Phys. A} {\bfseries 10} (1995) 1125--1150}.

\bibitem{Lukyanov:1996jj}
S.~L. Lukyanov and A.~B. Zamolodchikov, ``{Exact expectation values of local
  fields in quantum sine-Gordon model},''
  \href{http://dx.doi.org/10.1016/S0550-3213(97)00123-5}{{\em Nucl. Phys. B}
  {\bfseries 493} (1997) 571--587},
  \href{http://arxiv.org/abs/hep-th/9611238}{{\ttfamily arXiv:hep-th/9611238}}.

\bibitem{Armoni:2018bga}
A.~Armoni and S.~Sugimoto, ``{Vacuum structure of charge k two-dimensional QED
  and dynamics of an anti D-string near an O1-plane},''
  \href{http://dx.doi.org/10.1007/JHEP03(2019)175}{{\em JHEP} {\bfseries 03}
  (2019) 175}, \href{http://arxiv.org/abs/1812.10064}{{\ttfamily
  arXiv:1812.10064 [hep-th]}}.

\bibitem{Douglas:1995nw}
M.~R. Douglas and S.~H. Shenker, ``{Dynamics of SU(N) supersymmetric gauge
  theory},'' \href{http://dx.doi.org/10.1016/0550-3213(95)00258-T}{{\em Nucl.
  Phys. B} {\bfseries 447} (1995) 271--296},
  \href{http://arxiv.org/abs/hep-th/9503163}{{\ttfamily arXiv:hep-th/9503163}}.

\bibitem{Strassler:1998id}
M.~J. Strassler, ``{QCD, supersymmetric QCD, lattice QCD and string theory:
  Synthesis on the horizon?},''
  \href{http://dx.doi.org/10.1016/S0920-5632(99)85012-9}{{\em Nucl. Phys. B
  Proc. Suppl.} {\bfseries 73} (1999) 120--132},
  \href{http://arxiv.org/abs/hep-lat/9810059}{{\ttfamily
  arXiv:hep-lat/9810059}}.

\bibitem{Armoni:2003ji}
A.~Armoni and M.~Shifman, ``{On k string tensions and domain walls in N=1
  gluodynamics},'' \href{http://dx.doi.org/10.1016/S0550-3213(03)00409-7}{{\em
  Nucl. Phys. B} {\bfseries 664} (2003) 233--246},
  \href{http://arxiv.org/abs/hep-th/0304127}{{\ttfamily arXiv:hep-th/0304127}}.

\bibitem{Armoni:2003nz}
A.~Armoni and M.~Shifman, ``{Remarks on stable and quasistable k strings at
  large N},'' \href{http://dx.doi.org/10.1016/j.nuclphysb.2003.08.021}{{\em
  Nucl. Phys. B} {\bfseries 671} (2003) 67--94},
  \href{http://arxiv.org/abs/hep-th/0307020}{{\ttfamily arXiv:hep-th/0307020}}.

\bibitem{Armoni:2011dw}
A.~Armoni, D.~Dorigoni, and G.~Veneziano, ``{k-String Tension from Eguchi-Kawai
  Reduction},'' \href{http://dx.doi.org/10.1007/JHEP10(2011)086}{{\em JHEP}
  {\bfseries 10} (2011) 086}, \href{http://arxiv.org/abs/1108.6196}{{\ttfamily
  arXiv:1108.6196 [hep-th]}}.

\bibitem{Gray:2008je}
O.~Gray, ``{On the complete classification of the unitary N=2 minimal
  superconformal field theories},''
  \href{http://dx.doi.org/10.1007/s00220-012-1478-z}{{\em Commun. Math. Phys.}
  {\bfseries 312} (2012) 611--654},
  \href{http://arxiv.org/abs/0812.1318}{{\ttfamily arXiv:0812.1318 [hep-th]}}.

\bibitem{Gepner:1987qi}
D.~Gepner, ``{Space-Time Supersymmetry in Compactified String Theory and
  Superconformal Models},''
  \href{http://dx.doi.org/10.1016/0550-3213(88)90397-5}{{\em Nucl. Phys. B}
  {\bfseries 296} (1988) 757}.

\bibitem{Vafa:1988uu}
C.~Vafa and N.~P. Warner, ``{Catastrophes and the Classification of Conformal
  Theories},'' \href{http://dx.doi.org/10.1016/0370-2693(89)90473-5}{{\em Phys.
  Lett. B} {\bfseries 218} (1989) 51--58}.

\bibitem{Gannon:1996hp}
T.~Gannon, ``{U(1)-m modular invariants, N=2 minimal models, and the quantum
  Hall effect},'' \href{http://dx.doi.org/10.1016/S0550-3213(97)00032-1}{{\em
  Nucl. Phys. B} {\bfseries 491} (1997) 659--688},
  \href{http://arxiv.org/abs/hep-th/9608063}{{\ttfamily arXiv:hep-th/9608063}}.

\bibitem{Belin:2020nmp}
A.~Belin, N.~Benjamin, A.~Castro, S.~M. Harrison, and C.~A. Keller,
  ``{$\mathcal{N}=2$ Minimal Models: A Holographic Needle in a Symmetric
  Orbifold Haystack},''
  \href{http://dx.doi.org/10.21468/SciPostPhys.8.6.084}{{\em SciPost Phys.}
  {\bfseries 8} no.~6, (2020) 084},
  \href{http://arxiv.org/abs/2002.07819}{{\ttfamily arXiv:2002.07819
  [hep-th]}}.

\bibitem{Cecotti:1992rm}
S.~Cecotti and C.~Vafa, ``{On classification of N=2 supersymmetric theories},''
  \href{http://dx.doi.org/10.1007/BF02096804}{{\em Commun. Math. Phys.}
  {\bfseries 158} (1993) 569--644},
  \href{http://arxiv.org/abs/hep-th/9211097}{{\ttfamily arXiv:hep-th/9211097}}.

\bibitem{Cappelli:1987xt}
A.~Cappelli, C.~Itzykson, and J.~B. Zuber, ``{The ADE Classification of Minimal
  and A1(1) Conformal Invariant Theories},''
  \href{http://dx.doi.org/10.1007/BF01221394}{{\em Commun. Math. Phys.}
  {\bfseries 113} (1987) 1}.

\bibitem{Diatlyk:2023fwf}
O.~Diatlyk, C.~Luo, Y.~Wang, and Q.~Weller, ``{Gauging non-invertible
  symmetries: topological interfaces and generalized orbifold groupoid in 2d
  QFT},'' \href{http://dx.doi.org/10.1007/JHEP03(2024)127}{{\em JHEP}
  {\bfseries 03} (2024) 127}, \href{http://arxiv.org/abs/2311.17044}{{\ttfamily
  arXiv:2311.17044 [hep-th]}}.

\bibitem{Cherman:2024onj}
A.~Cherman and M.~Neuzil, ``{Beta functions of 2D adjoint QCD},''
  \href{http://dx.doi.org/10.1103/PhysRevD.109.105014}{{\em Phys. Rev. D}
  {\bfseries 109} no.~10, (2024) 105014},
  \href{http://arxiv.org/abs/2401.16604}{{\ttfamily arXiv:2401.16604
  [hep-th]}}.

\bibitem{DiFrancesco:1997nk}
P.~Di~Francesco, P.~Mathieu, and D.~Senechal,
  \href{http://dx.doi.org/10.1007/978-1-4612-2256-9}{{\em {Conformal Field
  Theory}}}.
\newblock Graduate Texts in Contemporary Physics. Springer-Verlag, New York,
  1997.

\bibitem{EberhardtWZW}
L.~Eberhardt, ``{Wess-Zumino-Witten Models, Lectures Given at YRISW 2019},''.

\bibitem{Schellekens:1989am}
A.~N. Schellekens and S.~Yankielowicz, ``{Extended Chiral Algebras and Modular
  Invariant Partition Functions},''
  \href{http://dx.doi.org/10.1016/0550-3213(89)90310-6}{{\em Nucl. Phys. B}
  {\bfseries 327} (1989) 673--703}.

\bibitem{Lin:2016gcl}
Y.-H. Lin, S.-H. Shao, Y.~Wang, and X.~Yin, ``{(2, 2) superconformal bootstrap
  in two dimensions},'' \href{http://dx.doi.org/10.1007/JHEP05(2017)112}{{\em
  JHEP} {\bfseries 05} (2017) 112},
  \href{http://arxiv.org/abs/1610.05371}{{\ttfamily arXiv:1610.05371
  [hep-th]}}.

\bibitem{Waterson:1986ru}
G.~Waterson, ``{Bosonic Construction of an $N=2$ Extended Superconformal Theory
  in Two-dimensions},''
  \href{http://dx.doi.org/10.1016/0370-2693(86)91002-6}{{\em Phys. Lett. B}
  {\bfseries 171} (1986) 77--80}.

\end{thebibliography}\endgroup

\end{document}